\documentclass[letter,onecolumn,12pt]{IEEEtran}
\usepackage{cite,graphicx,amsmath,amssymb}
\usepackage{subfigure}
\usepackage{citesort}
\usepackage{fancyhdr}
\usepackage{mathrsfs}
\usepackage{url}

\usepackage[dvips]{color}

\usepackage{setspace}
\newtheorem{theorem}{Theorem}

\newtheorem{proposition}{Proposition}

\newtheorem{lemma}{Lemma}

\newtheorem{corollary}{Corollary}

\newtheorem{remark}{Remark}

\newcommand{\be}{\beta}
\newcommand{\lb}{\left}
\newcommand{\rb}{\right}

\begin{document}

\title{Mutual Information Distribution of MIMO Channels: A Unified Characterization via Painlev\'{e} Transcendents}

\title{On the Distribution of MIMO Mutual Information: An In-Depth Painlev\'{e} Based Characterization}

%\author{Yang Chen${}^{\dagger}$ and Matthew R.\ McKay${}^{*}$
%\\
%{\small \vspace*{0.2cm} ${}^\dagger$  Department of Mathematics,
%Imperial College, 180 Queen's Gate, London SW7 2BZ, UK \\
%${}^*$ Department of Electronic and Computer Engineering, Hong Kong
%Uni.\ of Science \& Tech., Hong Kong
%\\ }}

\author{ {\large Shang Li$^{\dagger}$, Matthew~R.~McKay$^{\dagger}$, Yang Chen$^{*}$} \\ \vspace{0.5cm}
{\small $^{\dagger}$Department of Electronic and Computer Engineering, \\ \vspace{-0.2cm}
Hong Kong University of Science and Technology, Kowloon, Hong Kong\\
} \vspace{0.3cm}
{\small $^{*}$Department of Mathematics, Imperial College, London, UK}
%\\$^{\ddag}$Department of Mathematics, University of Macau, Macau}
\thanks{This work was supported by the Hong Kong Research Grants Council under grant number 616911.}
 }

\IEEEaftertitletext{\vspace{-2\baselineskip}} \maketitle

\setcounter{page}{1}

{\singlespace
\begin{abstract}
This paper builds upon our recent work which computed the moment
generating function of the MIMO mutual information exactly in terms
of a Painlev\'{e} V differential equation.  By exploiting this key
analytical tool, we provide an in-depth characterization of the
mutual information distribution for sufficiently large (but finite)
antenna numbers. In particular, we derive systematic closed-form
expansions for the high order cumulants. These results yield
considerable new insight, such as providing a technical explanation
as to why the well known Gaussian approximation is quite robust to
large SNR for the case of unequal antenna arrays, whilst it deviates
strongly for equal antenna arrays.  In addition, by drawing upon our
high order cumulant expansions, we employ the Edgeworth expansion
technique to propose a \emph{refined} Gaussian approximation which
is shown to give a very accurate closed-form characterization of the
mutual information distribution, both around the mean and for
moderate deviations into the tails (where the Gaussian approximation
fails remarkably). For stronger deviations where the Edgeworth
expansion becomes unwieldy, we employ the saddle point method and
asymptotic integration tools to establish new analytical
characterizations which are shown to be very simple and accurate.
Based on these results we also recover key well established
properties of the tail distribution, including the
diversity-multiplexing-tradeoff.
\end{abstract}
}

\newpage

\section{Introduction}

Multiple-input multiple-output (MIMO) technologies form a key
component of emerging broadband wireless communication systems due
to their ability to provide substantial capacity growth over power
and bandwidth constrained channels. Such technologies have received
huge attention for over a decade now, with recent trends focusing
mainly on incorporating MIMO into complicated system configurations;
e.g., those employing relaying
\cite{Helmut06,Yijia07,Wagner08,PD10,Shin10}, cooperative multi-cell processing
\cite{Melda07,Gesbert10,Dahrouj10}, information-theoretically secure
systems \cite{Tie09,Mukherjee11,Frederique11}, and ad-hoc networking
\cite{Bchen06,Hunter08,Ray11,yueping11}.
However, despite the huge progress, some fundamental questions
regarding the information-theoretic limits of MIMO systems still
remain unclear, even for the simplest point-to-point communication
scenarios. Of these, one the most important is the characterization
of the \emph{outage capacity}, which gives an achievable rate for
transmission over quasi-static channels.

Characterizing the outage capacity is much more difficult than the ergodic capacity, since it requires solving for the entire
distribution of the channel mutual information, rather than simply
the mean, and thus far this distribution is only partially
understood. For example, in \cite{ZhengdaoWang}, assuming
independent and identically distributed (IID) Rayleigh fading with
perfect channel state information (CSI) at the receiver, the mutual
information distribution was characterized via an exact expression
for the moment generating function (MGF). This result was given in
terms of a determinant of a certain Hankel matrix which yields
little insight and becomes unwieldy when the number of antennas are
not small. A similar determinant representation for the MGF was
adopted in \cite{Shin06} to provide a saddle point approximation for
the cumulative distribution function (CDF), however the solution was
complicated and once again revealed little insight. An alternative
result was presented in \cite{Oyman03} which considered the MGF of
the mutual information at high signal-to-noise ratios (SNR) and used
this to establish a Chernoff bound on the CDF. Whilst this approach
avoided dealing with complicated determinants, simulations
demonstrated that the bound was not particularly tight, particularly
when considering the CDF region representing outage probabilities of
practical interest. As an alternative method to characterizing the
mutual information distribution through its MGF,
\cite{smith03letter,smith04} took the more direct approach of using
classical transformation theory to derive exact expressions for the
probability density function (PDF) and CDF for MIMO systems with
small numbers of antennas. It was shown however, that even for the
simplest MIMO configuration with dual antennas (i.e., having two
transmit or two receive antennas), closed-form solutions were not
forthcoming and one must rely on numerically evaluating complicated
integrals.

To overcome the complexities of finite antenna characterizations, another major line of work has
focused on giving a large-antenna asymptotic analysis, which
provides more intuitive results; see e.g.,
\cite{Moustakas03,Hochwald04,Hachem06,Chen_McKay}. In such analyzes,
the most well known conclusion is that the mutual information
distribution approaches a Gaussian as the number of antennas become
sufficiently large. Different approaches have also been employed to
derive closed-form expressions for the asymptotic mean and variance
\cite{Chen_McKay,PBRap,Hochwald04,Moustakas03,Moustakas06,Hachem06}.
%{\bf **
%Need to add more refs -- Najim, Moustakas's repica papers, etc \\
%\emph{One question from Samuel: the references here are nearly the same as those mentioned in the first sentence of this paragraph. Are they overlapped?}**}
Quoting \cite{Chen_McKay} as an example, for a MIMO system subjected
to IID Rayleigh fading with perfect receiver CSI, with $n_t$
transmit antennas, $n_r$ receive antennas, and SNR $P$, if $n_t$ and $n_r$ are both sufficiently large then the
mutual information distribution is approximated by a Gaussian with
mean $\mu_0$ and variance $\sigma^2_0$ given as
\begin{align}
&\mu_{0}=n\Big[\frac{a+b}{2}\ln\left(\frac{\sqrt{\beta+aP}+\sqrt{\beta+bP}}{2\sqrt{\beta}}\right)\nonumber\\&\hspace{1cm}-\sqrt{ab}\ln
\left(\frac{\sqrt{a\left(\beta+bP\right)}+\sqrt{b\left(\beta+aP\right)}}{\sqrt{a\beta}+\sqrt{b\beta}}\right)-\frac{\left(\sqrt{\beta+aP}-\sqrt{\beta+bP}\right)^2}{4P}\Big],\label{eq:coulomb_mean}\\
&\sigma^2_{0}=2\ln\left[\frac{1}{2}\left(\frac{\beta+aP}{\beta+bP}\right)^{1/4}+\frac{1}{2}\left(\frac{\beta+bP}{\beta+aP}\right)^{1/4}\right]
 \label{eq:coulomb_variance}
\end{align}
where
\begin{align}
n:=\min\{n_t, n_r\}, \;m :=\max\{n_t, n_r\},\; \beta :=m/n ,\;
a:=\left(\sqrt{\beta+1}-1\right)^2,\;b:=\left(\sqrt{\beta+1}+1\right)^2
. \nonumber
\end{align}
The Gaussian approximation has been considered extensively due to its relative
simplicity compared to the exact characterizations. However, in
practice, the number of antennas in MIMO systems is typically not
huge, and it turns out that the Gaussian approximation may
sometimes be very inaccurate. These deviations have been reported in
several previous contributions \cite{Chen_McKay,Shin06,Moustakas11},
and here we give two concrete examples to demonstrate them.

\begin{figure}[!ht]
\centering \subfigure[$n_t=n_r=3$]{
\includegraphics[width=0.45\columnwidth]{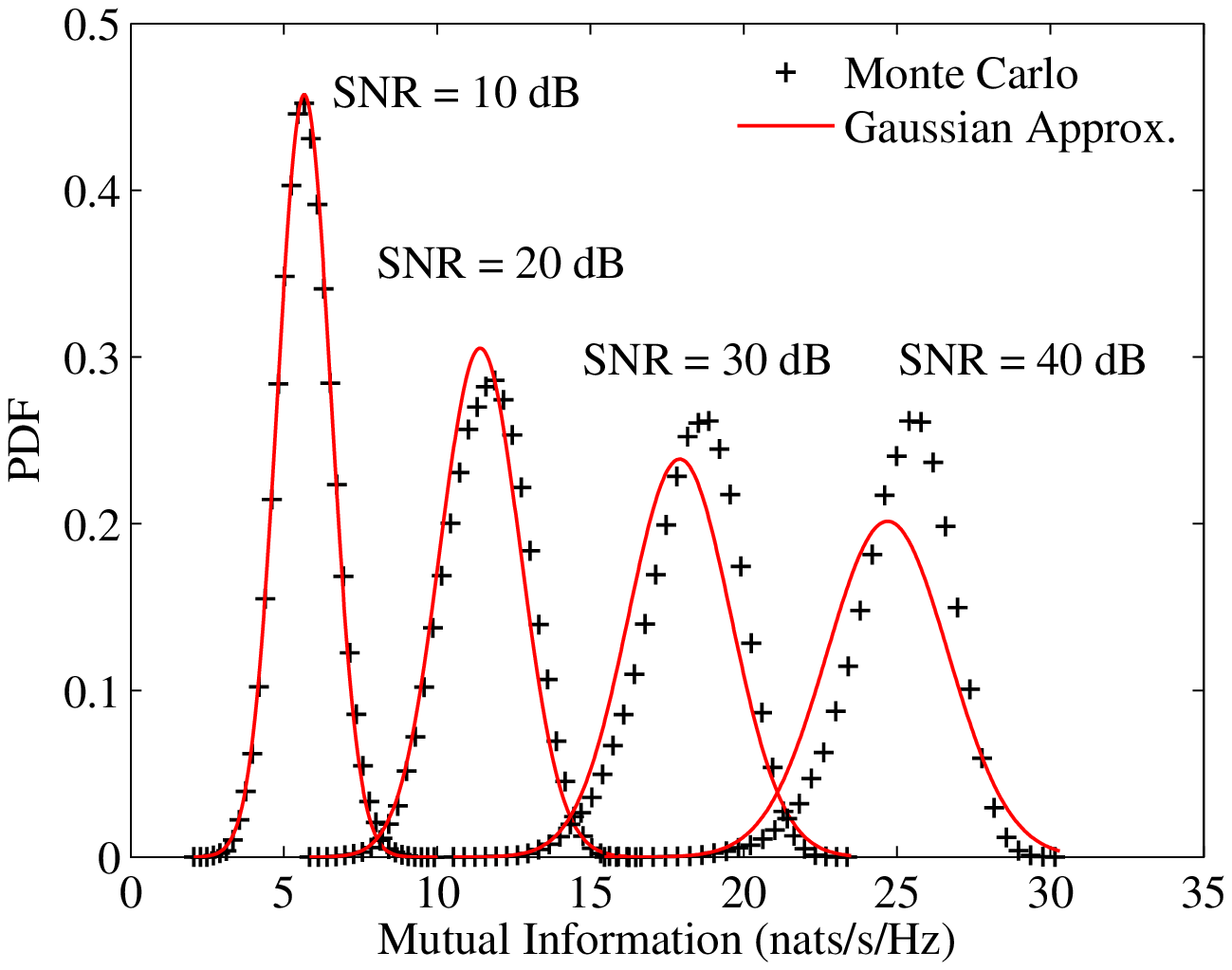}\label{Gaussian_equal}}
\subfigure[$n_t=3, \;n_r=2$]{
\includegraphics[width=0.45\columnwidth]{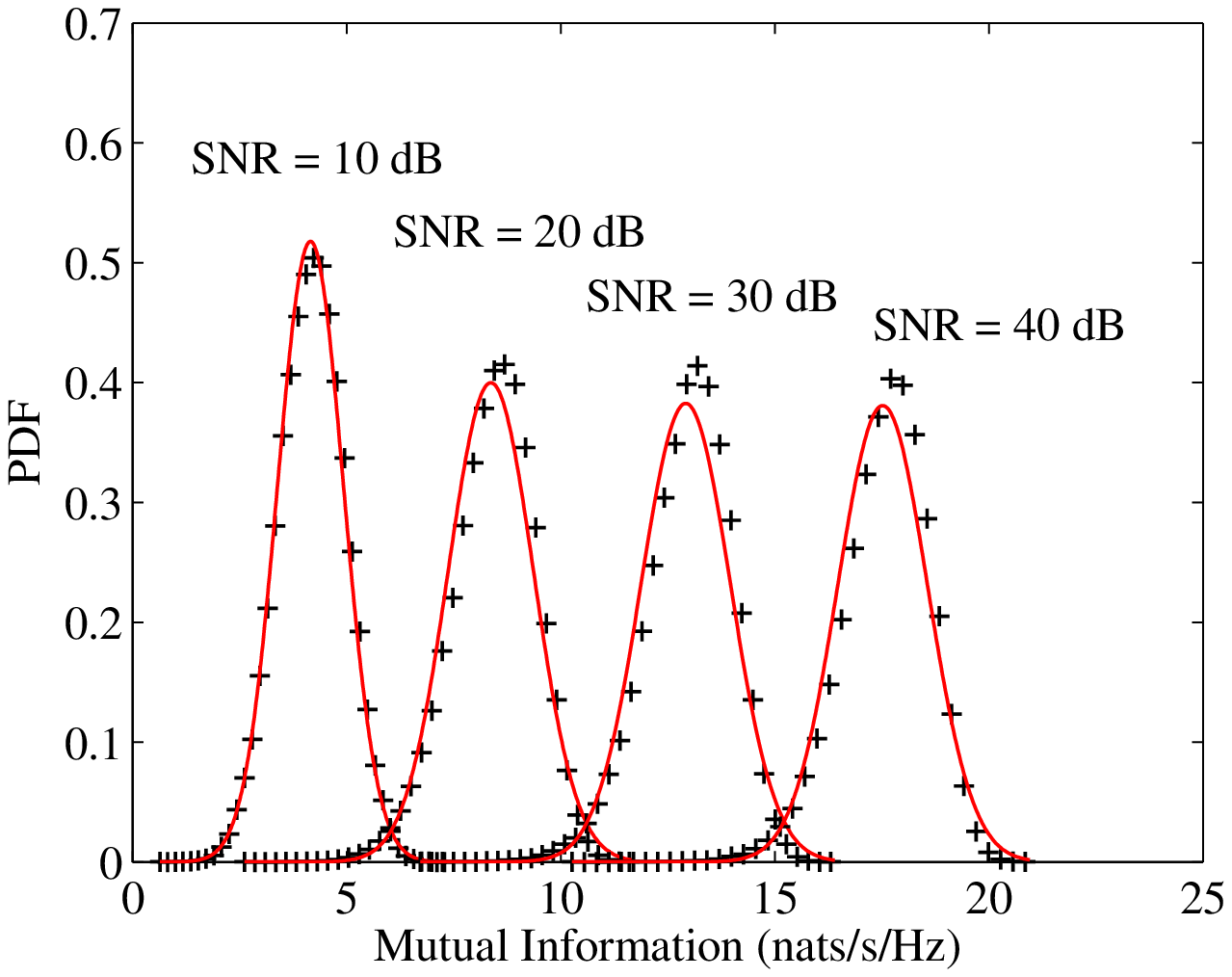}\label{Gauss_unequal}}
\caption{PDF of the mutual information of MIMO Rayleigh fading
channels for different antenna configurations and different SNRs.}
\end{figure}

As a first example, as shown in Figs. \ref{Gaussian_equal}--\ref{Gauss_unequal}, the PDF of the Gaussian approximation deviates
significantly from the true (simulated) PDF of the mutual
information for finite antenna arrays when the SNR becomes large.
This phenomena was emphasized and investigated in our recent work
\cite{Chen_McKay} for the specific case of equal-antenna arrays
(i.e., $n_t = n_r$) by looking at the cumulants of the mutual
information. Quite interestingly, the figures indicate that the
deviation from Gaussian is significantly \emph{stronger} when $n_r =
n_t$ compared with the alternative case, despite the fact that there
are more antennas. Thus, when $n_t \neq n_r$, it appears that the
Gaussian approximation is more robust to increasing SNR. This key
observation is unexpected, and to the best of our knowledge it
hitherto lacks any rigorous explanation.

\begin{figure}[!ht]
\centering \subfigure[$n_t=4,\; n_r=2$]{
\includegraphics[width=0.45\columnwidth]{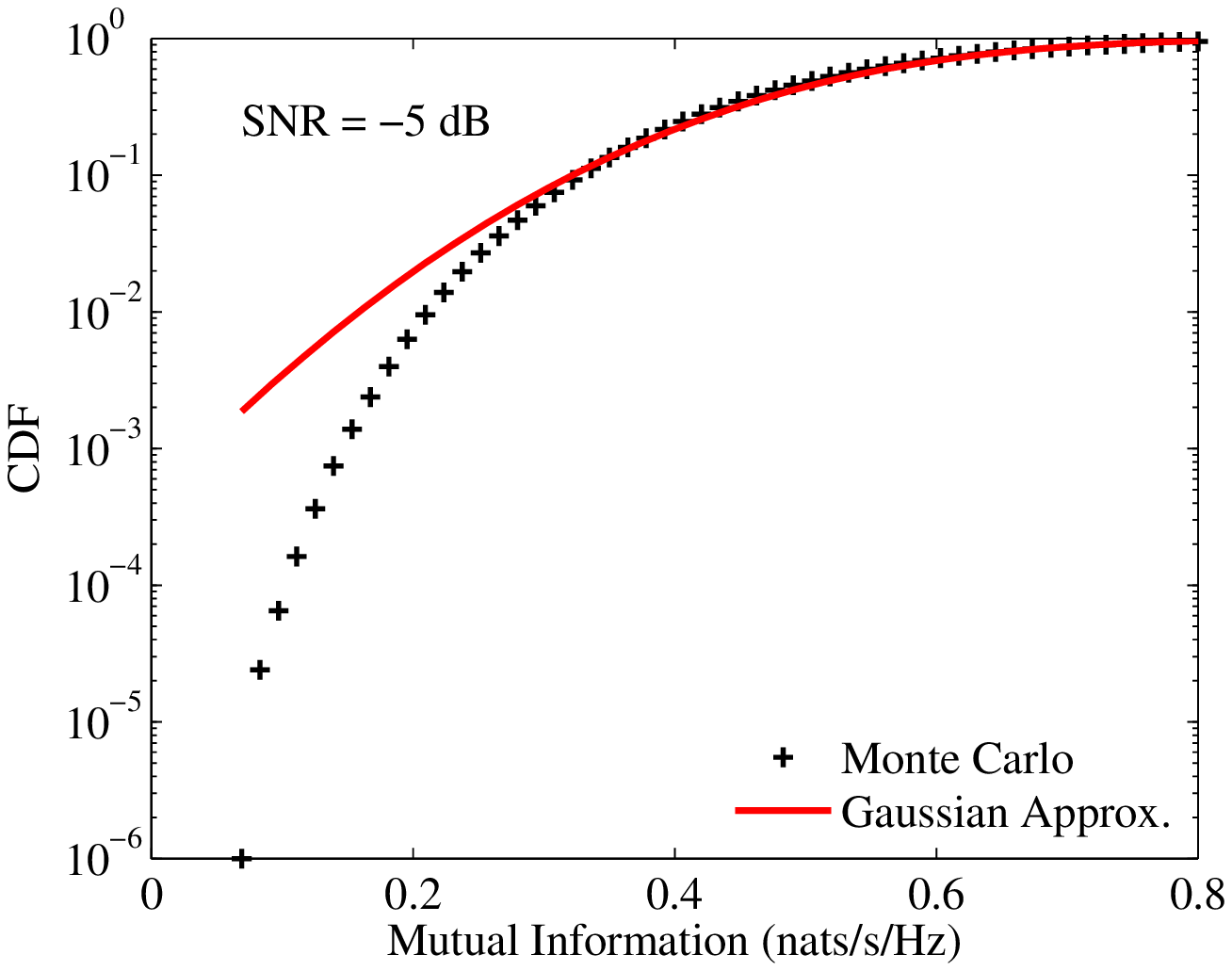}\label{cdf_Gaussian_smallP}}
\subfigure[$n_t=4,\; n_r=2$]{
\includegraphics[width=0.45\columnwidth]{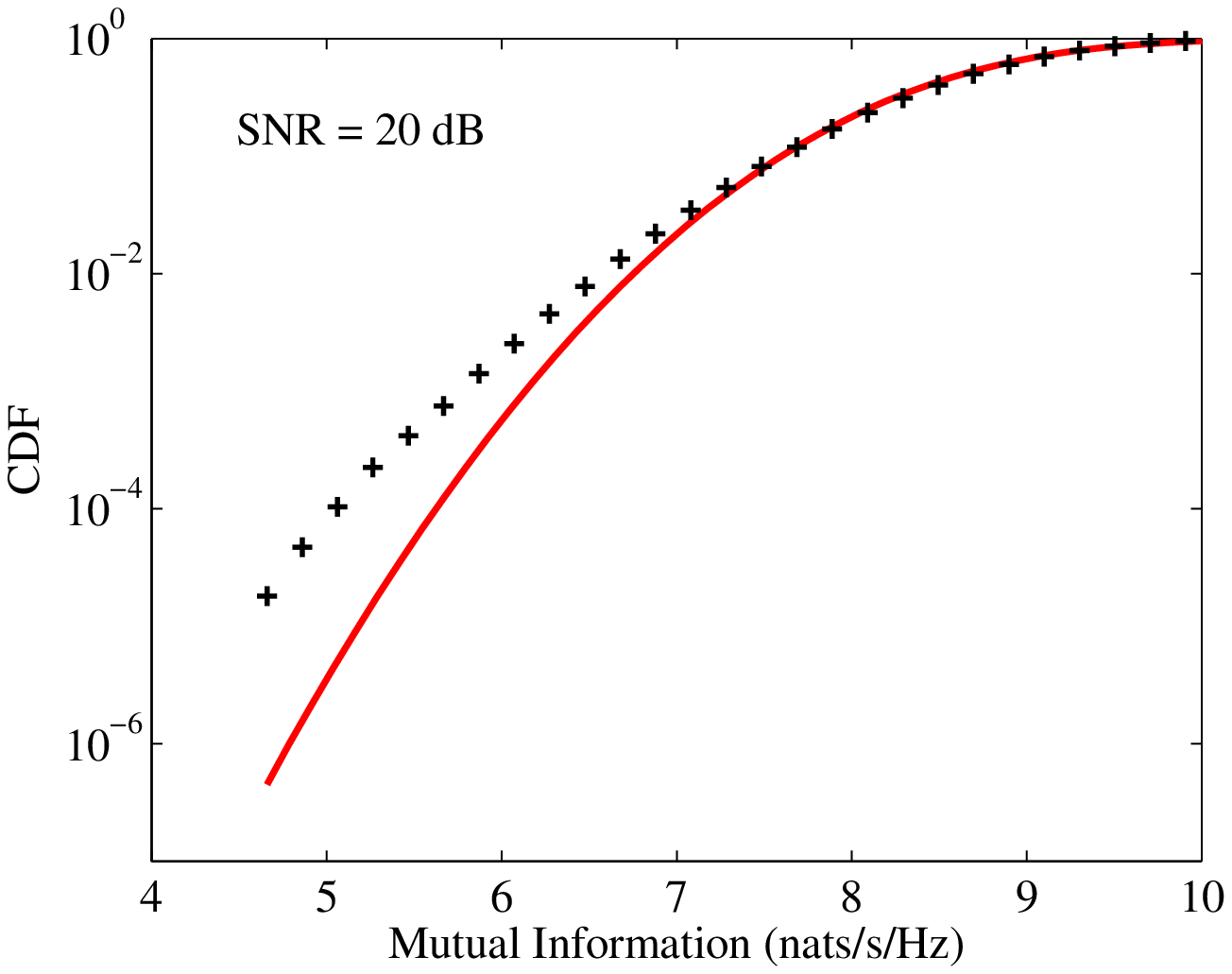}\label{cdf_Gaussian_largeP}}
\caption{CDF of mutual information of MIMO Rayleigh fading channels
for small and large SNRs.}
\end{figure}

As a second example, aside from the deviation observed in the bulk
of the distribution for large SNR, it turns out that the Gaussian
approximation is typically very inaccurate in the tail of the
distribution when $n$ is finite and the SNR is either small or
large. This is observed in Figs. \ref{cdf_Gaussian_smallP}--\ref{cdf_Gaussian_largeP}, where in both cases the Gaussian curve
fails markedly in tracking the simulations for small but practical
outage probabilities. This strong deviation from Gaussian was also
discussed in \cite{Moustakas11}, where a refined approximation was
presented based on adopting an intuitive large-$n$ Coulomb fluid
interpretation from statistical physics, leading to a set of coupled
non-linear equations requiring numerical computation. As discussed
therein, the main utility of the large deviations approach is that
it allows one to capture the tail behavior in the regime of $O(n)$
deviations away from the mean, whilst the Gaussian is restricted to
capturing deviations which are close to $O(1)$.

Obtaining a clear and rigorous understanding of the distributional
behavior indicated above appears difficult with random matrix
theoretic tools which are currently well known to information and
communication theorists, such as those based on the Stieltjes
transform \cite{Hachem06} and the replica method
\cite{Moustakas03,Moustakas06}. To address this problem, in our
recent work \cite{Chen_McKay} we introduced a powerful methodology
involving orthogonal polynomials and their so-called ``ladder
operators'' which led to a new and convenient exact
characterization of the MGF of the mutual information in terms of a
Painlev\'{e} differential equation. This valuable representation
provides the machinery for systematically capturing the finite-$n$
corrections to the asymptotic Gaussian results, and thereby
investigating deviations from Gaussian and correcting for them. Some
related subsequent work for interference-limited multi-user MIMO
systems was also presented in \cite{Li_Chen_McKay}. However, whilst
the Painlev\'{e} MGF representation in \cite{Chen_McKay} laid the
platform for further analysis, the main focus of the analysis
presented therein was restricted to the case $n_t = n_r$. Moreover,
even for this case, a rigorous treatment of the large-deviations
region considered in \cite{Moustakas11} was not pursued.

%In this paper, combined with the Edgeworth expansion and saddle point method, we are able to establish a unified understanding and effective calculation of the mutual information distribution. For example, we can consider the various regimes of the mutual information with their valid approximations. Specifically, the mutual information approaches a Gaussian
%distribution within the regime of $O(1)\sigma$ ($\sigma$ denotes the
%standard deviation) away from the mean. This regime is adequate in
%the situation where $n$ is large such that the $O(1)\sigma$
%deviation gives sufficiently low outage probabilities (e.g., in most
%cases the target outage level is around $P_{\rm out}\sim 10^{-6}$).
%On the contrary, in the case where $n$ is finite, deviation of the
%order $O(n^\epsilon)\sigma, 0<\epsilon\leq 1$ becomes relevant.
%Based on the discussion in \cite{Moustakas11} and \cite{Chen_McKay},
%the distribution further into the tail with $O(n^\epsilon)\sigma$
%away from the mean involves the effects of higher cumulants as
%$\epsilon$ increases. Eventually for $\epsilon=1$, in theory, we
%should take all the cumulants into account and Gaussian
%approximation obviously fails in this region. A more rigorous
%discussion about the effects of higher cumulants is given in Section
%\ref{sec:SPA}.

In this paper, we significantly expand upon our existing studies in
\cite{Chen_McKay} to provide a much broader characterization and
understanding of the MIMO mutual information distribution.  Starting
with the exact Painlev\'{e} representation for the mutual
information MGF derived in \cite{Chen_McKay}, for sufficiently large
$n$ (but not assuming that $n_t = n_r$), we systematically
compute new closed-form expansions for the higher order cumulants of
the mutual information distribution.  These series expansions reveal
interesting fundamental differences between the two cases, $n_t \neq
n_r$ and $n_t = n_r$.  Notably, it is demonstrated that the typical
approach of considering only the leading-order terms in the
large-$n$ series expansions for each cumulant (e.g.,
(\ref{eq:coulomb_mean}) and (\ref{eq:coulomb_variance}) respectively
for the mean and variance) is relatively stable for the asymmetric
case $n_t \neq n_r$ compared with the symmetric case $n_t = n_r$,
when the SNR $P$ becomes large. This is because the correction
series for each cumulant (i.e., comprising all terms other than the
leading $n$ term) is shown to converge to a bounded constant as $P$
increases, and therefore becomes quite small relative to the leading
term when $P$ is large. For the symmetric case, on the other hand,
the situation is very different---the correction series for each
cumulant not only grows with $P$, but also at a faster rate than the
leading term; a phenomenon which was discussed at length in
\cite{Chen_McKay}. These results allow us to provide a technical
explanation for the intriguing large-SNR phenomena observed in Figs.
\ref{Gaussian_equal} and \ref{Gauss_unequal}.

In addition to gaining fundamental insight into the behavior of the
mutual information distribution, we also provide new accurate analytical characterizations for the distribution by
employing two different approaches, each being useful in their own
region of interest. First, we draw upon the Edgeworth expansion
technique along with our derived cumulant expressions to give a
simple refined closed-form distribution approximation which is shown
to be very accurate around the mean and also for certain moderate
deviations into the tails. These results generalize previous
Edgeworth expansions which were derived for the specific case of
$n_t = n_r$ in \cite{Chen_McKay}.  It is shown that in contrast to
the Gaussian approximation, which is only capable of successfully
characterizing near $O(1)$ deviations around the mean (or the
\emph{bulk}) as $n$ increases, the Edgeworth expansion captures
deviations of up to $O(n^\epsilon)$, where $0<\epsilon\leq 1$. Here,
$\epsilon$ is a key parameter which, as it increases, allows the
approximation to capture the correct distribution further into the
tails, however it also requires the addition of more cumulants to be
included in the Edgeworth series, thereby increasing the complexity.
In the extreme case, as $\epsilon \to 1$, \emph{all} cumulants must
be included; thus, the Edgeworth technique becomes unwieldy and
alternative methods are required.  In this scenario, in order to
capture such ``large deviations'' into the tail, we exploit the
saddle point method along with asymptotic integration tools to give
further analytical representations.  Very simple formulas are
obtained for the cases of high and low SNR which, taken together,
are shown to be very accurate over almost the entire range of SNR
values.  We point out that our saddle point results provide, in
effect, an alternative characterization to the results proposed in
\cite{Moustakas11}, which also considered the ``large deviations''
regime, focusing on $O(n)$ deviations from the mean.  Our results,
however, are based on a more rigorous footing, stemming from the
exact Painlev\'{e} representation for the mutual information MGF, as
opposed to intuitive statistical physics analogies. Quite
surprisingly, they are also found to be simpler.

As a final step, to further emphasize the utility of our results and
methodology, we use our analytical framework to extract the
well known diversity-multiplexing-tradeoff (DMT) formula of Zheng
and Tse \cite{Zheng_Tse} which relates to the large deviations
region in the left tail at large SNR, whilst also deriving similar
results for the right tail. These results are found to be
consistent with those obtained via the Coulomb fluid analogy in
\cite{Moustakas11}.

The rest of the paper is structured as follows. Section II describes
the system model under consideration and introduces the
Painlev\'{e} representation for the MGF of the mutual information,
which is the key tool underpinning our analysis. Then, in Section
III, we provide a systematic derivation of the mutual information
cumulants to leading order in $n$, as well as finite-$n$ correction
terms, giving closed-form expressions in both cases. Based on these
new results, Section IV analyzes the $n$-asymptotic Gaussian
approximation at large SNR, revealing fundamental differences
between the two scenarios, $n_t = n_r$ and $n_t \neq n_r$. In
Section V, we draw upon our cumulant expressions and the Edgeworth
expansion technique to provide a refined approximation to the mutual
information distribution, both around the mean and for moderate
deviations into the tails. Subsequently, Section VI exploits the
saddle point method and asymptotic integration tools to characterize
the ``large deviations'' region, extracting key behavior such as
the DMT. In the Appendix, correspondences are drawn between our
analytical framework and the Coulomb fluid large deviations method
used in \cite{Moustakas11}.

\section{System Model}
Considering a point-to-point communication system with $n_t$
transmit and $n_r$ receive antennas, under flat-fading, the
linear MIMO channel model takes the form:
\begin{align}
{\bf y}={\bf Hx}+{\bf n}
\end{align}
where ${\bf y} \in \mathbb{C}^{n_r}$ and ${\bf x} \in
\mathbb{C}^{n_t}$ denote the received and transmitted vector
respectively, whilst ${\bf H}\in \mathbb{C}^{n_r\times n_t}$
represents the channel matrix, and ${\bf n}_{n_r\times 1}\in
\mathbb{C}^{n_r}$ represents noise. Assuming rich scattering, ${\bf
H}$ is modeled as Rayleigh fading, having IID entries $h_{i,j}\sim
\mathcal{CN}(0,1)$, known to the receiver only. The noise is assumed
${\bf n}\sim \mathcal{CN}({\bf 0, I}_{n_r})$. The input is selected
to be the ergodic-capacity-achieving input distribution ${\bf x}\sim
\mathcal{CN}\left(0,\frac{P}{n_t}{\bf I}_{n_t}\right)$, where $P$ is
the transmitted power constraint and also represents the SNR due to
the normalized noise. For the model under
consideration, the mutual information between
the channel input and output is \cite{teletar99}:
\begin{align}\label{MI}
\mathcal{I}\left({\bf x};{\bf y}\right)= \left\{
\begin{array}{ll}
 \ln \det \left({\bf
I}_{n}+\frac{P}{m}{\bf HH^\dagger}\right), \quad n_t\geq n_r \\[1mm]
\ln \det \left({\bf
I}_{n}+\frac{P}{n}{\bf H^\dagger H}\right), \quad n_t<n_r\;
\end{array}.\right.
\end{align}
According to (\ref{MI}), we can assume $n_t\geq
n_r$ without loss of generality; otherwise, if $n_t<n_r$, we only
need replace $P$ with $\beta P$.

Throughout the paper, we make the well known assumption that the channel
exhibits block-fading, such that the fading coefficients vary
independently from one coding block to another, but remain constant
for the duration of each block.  In this case, the outage
probability becomes an important performance indicator, and the
``capacity-versus-outage'' tradeoff comes into play \cite{Biglieri}.
Quantifying this tradeoff requires the entire
distribution of the mutual information (\ref{MI}). Consequently, a common
approach is to investigate the MGF of $\mathcal{I}({\bf x;y})$:
\begin{align}\label{MGF}
\mathcal{M}(\lambda)&:=E\left[\exp\left(\lambda
\mathcal{I}(\mathbf{x;y})\right)\right]\nonumber\\&=E_{\bf
H}\left\{\left[\det\left({\bf I}_{n}+\frac{P}{m} \bf HH^\dagger
\right)\right]^\lambda\right\}.
\end{align}
With this, the cumulant generating function (CGF) can be expressed
as a power series about $\lambda=0$:
\begin{align}\label{CGF_expansion}
\mathcal{K}(\lambda)&:=\ln\mathcal{M}(\lambda)=\sum_{\ell=1}^\infty\kappa_\ell\frac{\lambda^\ell}{\ell!}
\end{align}
where the coefficient $\kappa_\ell$ is the $\ell$-th cumulant of
$\mathcal{I}(\mathbf{x;y})$.

To lay the foundation of our analysis, we first quote the following exact representation for the MGF
(\ref{MGF}) (or equivalently, the CGF) in \cite{Chen_McKay}; a key result derived by the
authors by drawing upon methods from random matrix theory (see e.g.,
\cite{Magnus,Chen_A.R._Its,Basor_Chen}).

\begin{proposition} \label{th:Laguerre}
The MGF (\ref{MGF}) admits the following compact representation:
\begin{align} \label{eq:SU_MIMO_Painleve}
{\cal M}(\lambda) = \exp \left( \int_\infty^{\beta/P} \frac{ G_n(x)}{x} d x \right) \;
\end{align}
where $G_n(x)$  satisfies a version of the Painlev\'e V continuous $\sigma$--form:
\begin{align}
 \left(xG_n''\right)^2=&n^2\left(xG_n'+G_n'\left(n+m+\lambda\right)/n-G_n\right)^2\nonumber\\&-4\left(xG_n'-G_n+nm\right)\left(G_n'^2+\lambda n
G'_n\right), \label{Fundamental}
\end{align}
with $'$ denoting the derivative with respect to (w.r.t.) $x$.
\end{proposition}

%Theorem \ref{theorem1} reveals the connection between the MGF
%(\ref{2.MGF_Jacobi}) and the Painlev\'{e} VI equation, which makes
%the MGF more tractable than the original form (\ref{2.MGF_Jacobi}).

Whilst an explicit solution to the Painlev\'{e} V differential
equation does not exist in general, we will show in the next
subsection that it can be used to great effect to extract deep
insight into the behavior of the mutual information distribution. In
particular, we will use it to provide a systematic method for
computing closed-form expressions for the leading-order and
correction terms to the mean, variance, and higher order cumulants,
as the numbers of antennas grow large. This will allow us to obtain
more accurate characterizations than the asymptotic Gaussian
approximation, which is based on only the leading order terms of the
mean and variance.

\section{Systematic Derivation of the Mutual Information Statistics}\label{sect:cumulants}
In this section, we further elaborate upon our work in \cite[Section
IV]{Chen_McKay} by making use of the Painlev\'{e} representation (i.e., Proposition \ref{th:Laguerre}) to obtain closed-form expressions for the higher
order cumulants to leading order in $n$, as well as first-order
correction terms, which apply for \emph{arbitrary} $\beta$.
Computing the mutual information distribution in terms of its cumulants
enables us to obtain insights into the shape of the distribution,
for example, studying its ``Gaussianity''. Furthermore, by
invoking an Edgeworth expansion approach, we will evaluate
approximations for the outage probability in {\it closed-form},
avoiding explicit computation of the inverse MGF (i.e., an inverse
Laplace Transform). A key contribution in this section is the
generalization of the results in \cite[Section IV]{Chen_McKay}
beyond the case $\beta = 1$, and, based on these results, we will
see that the mutual information distribution exhibits some key
differences in the two scenarios, $\beta = 1$ and $\beta > 1$.

\subsection{Power Series Method: Valid for Small P}\label{sec:smallPexpansion}
For a preliminary study, we use the power series method to seek
solutions to (\ref{Fundamental}). Noting that $\lim_{P\to 0}\ln
\mathcal{M}(\lambda)=0$, we assume that for sufficiently small $P$, and thus
sufficiently large $x$, that $G_n(x)$ admits an expansion of the
form:
\begin{align}\label{series}
G_n(x)=\sum_{k=1}^\infty\frac{b_k}{x^k}\;
\end{align}
for some $b_1, b_2, \ldots$.  Substituting this power series into
(\ref{Fundamental}) and solving for the $b_k$'s by matching the
coefficients of $1/x^k$ on the left and right-hand sides, we write
the first few $b_k$'s as follows:{\allowdisplaybreaks
\begin{align*}
&b_1=-n^2\beta\lambda\;,\\
&b_2=n^3\beta\left(\beta+1\right)\lambda-n^2\beta \lambda^2\;,\\
&b_3=-n^2\beta\left(n^2\beta^2+3n^2\beta+n^2+1\right)\lambda+3n^3\beta\left(\beta+1\right)\lambda^2-n^2\beta\lambda^3\;,\\
&b_4=\left[5n^3\beta\left(\beta+1\right)+n^5\beta\left(\beta+1\right)\left(\beta^2+5\beta+1\right)\right]\lambda\\
&\qquad
-\left[n^4\beta\left(6\beta^2+17\beta+6\right)+5n^2\beta\right]\lambda^2+6n^3\beta\left(\beta+1\right)\lambda^3-n^2\beta\lambda^4\;,\\
&b_5=-\left[8n^2\beta+n^4\beta\left(15\beta^2+40\beta+15\right)+n^6\beta\left(\beta^4+10\beta^3+20\beta^2+10\beta+1\right)\right]\lambda\\
&\qquad
+\left[n^5\beta\left(10\beta^3+55\beta^2+55\beta+10\right)+40n^3\beta\left(\beta+1\right)\right]\lambda^2
\\&\qquad-\left[n^4\beta\left(20\beta^2+55\beta+20\right)+15n^2\beta\right]\lambda^3+10n^3\beta\left(\beta+1\right)\beta\lambda^4-n^2\beta\lambda^5\;
.
%\\&\; \vdots
\end{align*}}
Introducing a rearrangement of the power series (\ref{series})
according to the order of $\lambda$, we obtain an expression akin to
the power series representation of the CGF (\ref{CGF_expansion}):
$G_n(x)=\lambda g_1(x)+\lambda^2 g_2(x)+\lambda^3 g_3(x)+\cdots$
where $g_\ell, \ell=1, 2, \ldots$ are each a power series in $1/x$,
which we omit for the sake of conciseness. By using
(\ref{eq:SU_MIMO_Painleve}), the $\ell$-th cumulant $\kappa_\ell$ is then computed by
\begin{align}\label{integral}
\kappa_l=l!\int_\infty^{\beta/P}\frac{g_l(x)}{x}{\rm d}x.
\end{align}
As a result of this procedure, here we write the first few
$\kappa_\ell$'s, showing the leading order and first-order
correction terms in $n$: {\allowdisplaybreaks
\begin{align}
&\mu=n\left(P-\frac{\beta+1}{2\beta}P^2+\frac{\beta^2+3\beta+1}{3\beta^2}P^3-\cdots\right)\nonumber\\
&\qquad
+\frac{1}{n}\left(\frac{P^3}{3\beta^2}-\frac{5\left(\beta+1\right)}{4\beta^3}P^4+\frac{15\beta^2+40\beta+15}{5\beta^4}P^5-\cdots\right)+O\left(\frac{1}{n^3}\right),\label{series_k1}\\
\nonumber\\&\sigma^2=\left(\frac{P^2}{\beta}-\frac{2\left(\beta+1\right)}{\beta^2}P^3+\frac{6\beta^2+17\beta+6}{2\beta^3}P^4+\cdots\right)\nonumber\\
&\qquad+\frac{1}{n^2}\left(\frac{5P^4}{2\beta^3}-\frac{16\left(\beta+1\right)}{\beta^4}P^5+\frac{175\beta^2+456\beta+175}{3\beta^5}P^6-\cdots\right)+O\left(\frac{1}{n^4}\right)\label{series_k2},
\\\nonumber\\&\kappa_3=\frac{1}{n}\left(\frac{2P^3}{\beta^2}-\frac{9\left(\beta+1\right)}{\beta^3}P^4+\frac{24\beta^2+66\beta+24}{\beta^4}P^5-\cdots\right)\nonumber\\
&\qquad
+\frac{1}{n^3}\left(\frac{18P^5}{\beta^4}-\frac{175\left(\beta+1\right)}{\beta^5}P^6+\frac{30\left(30\beta^2+77\beta+30\right)}{\beta^6}P^7-\cdots\right)+O\left(\frac{1}{n^5}\right),\label{series_k3}\\
\nonumber\\&\kappa_4=\frac{1}{n^2}\left(\frac{6P^4}{\beta^3}-\frac{48\left(\beta+1\right)}{\beta^4}P^5+\frac{4\left(50\beta^2+135\beta+50\right)}{\beta^5}P^6-\cdots\right)+O\left(\frac{1}{n^4}\right),\label{series_k4}\\
\nonumber\\&\kappa_5=\frac{1}{n^3}\left(\frac{24P^5}{\beta^4}-\frac{300\left(\beta+1\right)}{\beta^5}P^6+\frac{120\left(15\beta^2+40\beta+15\right)}{\beta^6}P^7-\cdots\right)+O\left(\frac{1}{n^5}\right)
\; . \label{series_k5}
%\\&\;\vdots\nonumber
\end{align}}
%Note that expanding the Coulomb Fluid mean and variance (\ref{3.S}) about $P=0$ gives series' which
%agree precisely with the leading order (in $n$) terms of
%(\ref{series_k1}) and (\ref{series_k2}), indicating that for small
%$P$ and arbitrary $\beta$, the Coulomb Fluid linear statistics
%approximation captures the {\it exact} mean and variance to leading
%order in $n$.

%Though these power series expansions for the cumulants are only
%valid for small $P$, it is very insightful to see the {\it
%structure} of the cumulants, and we will employ this structure in
%the following subsection to present a generalized analysis which
%applies for \emph{arbitrary} $P$.

\subsection{General Method: Valid for Arbitrary P}
From the power series expressions
(\ref{series_k1})--(\ref{series_k5}), we observe the following
large $n$ series structure:
\begin{align}\label{structure}
&\mu=n\;C_{1,0}\phantom{C_{1,1} }+\frac{1}{n}C_{1,1} \phantom{+\frac{1}{n^2}C_{1,1}}+\frac{1}{n^3}C_{1,2}\phantom{+\frac{1}{n^4}C_{1,1}}+\frac{1}{n^5}C_{1,3}\phantom{+}\cdots\nonumber\\
&\sigma^2=\phantom{nC_{1,1}+}C_{2,0}\phantom{+\frac{1}{n}C_{1,1}}+\frac{1}{n^2}C_{2,1}\phantom{+\frac{1}{n^3}C_{1,1}}+\frac{1}{n^4}C_{2,2}\phantom{+\frac{1}{n^5}C_{1,1}}+\cdots\nonumber\\
&\kappa_3=\phantom{nC_{1,1}+C_{1,1}+\!\!\!}\frac{1}{n}C_{3,0}\phantom{+\frac{1}{n^2}C_{1,1}}+\frac{1}{n^3}C_{3,1}\phantom{+\frac{1}{n^4}C_{1,1}}+\frac{1}{n^5}C_{3,2}\phantom{+}\cdots\nonumber\\
&\kappa_4=\phantom{nC_{1,1}+C_{1,1}+\frac{1}{n}C_{1,1}+\!}\frac{1}{n^2}C_{4,0}\phantom{+\frac{1}{n^3}C_{1,1}\!\!}+\frac{1}{n^4}C_{4,1}\phantom{+\frac{1}{n^5}C_{1,1}}+\cdots\nonumber\\
&\;\vdots \phantom{=
nC_{1,1}+C_{1,1}+\frac{1}{n}C_{1,1}+\frac{1}{n^2}C_{1,1}}\ddots\phantom{+\frac{1}{n^4}C_{1,1}+\frac{1}{n^5}}\ddots\phantom{+++\quad}\end{align}
Note that $C_{i,j}$ is {\it independent} of $n$, but depends on $P$
and $\beta$. This structure gives an important general understanding
of the mutual information distribution in terms of cumulants, and
also paves the way for our arbitrary $P$ analysis. As seen from
(\ref{structure}), as $n$ grows large with other parameters fixed,
only the leading terms of the mean and variance survive, hence the
mutual information distribution approaches Gaussian with increasing
$n$. This agrees with previous results, see e.g.,
\cite{ZhengdaoWang,Hochwald04,Hachem06}.

In the previous subsection we solved $C_{i,j}$ for small $P$. Here,
we present a more general approach to compute these quantities in
closed-form for arbitrary $P$. First, we establish a recursive
computation for all cumulants to leading order in $n$ (i.e.,
generating $C_{\ell,0}, \ell=1, 2, \ldots$ along the leading
diagonal in (\ref{structure})). Then, employing the same machinery,
we obtain a recursion for systematically computing the first-order
correction terms for all cumulants, allowing us to derive
closed-form expressions for $C_{\ell,1}, \ell=1, 2, \ldots$ along
the second main diagonal in (\ref{structure}).  This, in turn,
allows us to capture improved accuracy for finite $n$. In addition
to computing the terms along the diagonals of (\ref{structure}), we
also demonstrate how our machinery based on the Painlev\'{e} MGF
representation can be employed to systematically compute the
horizontal power series expansion in (\ref{structure}) for a given
cumulant, to arbitrary degree of accuracy. For
this, we focus on the series computation of the mean by way of
example, with the other cumulant series being computable in a
similar way.

% each individual cumulant to recursively
%compute its large-$n$ series (e.g., generating $C_{1,\ell}, \ell=0,
%1, \ldots$), resulting in an increasingly accurate expressions for
%that cumulant when $n$ is finite.

%The results in this subsection will allow us to obtain more accurate
%characterizations than the asymptotic Gaussian approximation, whilst
%also allowing us to accurately capture the behavior of the
%distribution in the tails (something which is significantly missed
%by the Gaussian approximation). In short, with the exact
%Painlev\'{e} representation of the MGF, we are capable of obtaining
%the mutual information distribution with arbitrary accuracy.

\subsubsection{Cumulants to Leading Order in $n$}\label{sec:MVar_Gauss}
Recalling the definition of the cumulants in (\ref{CGF_expansion}) and their large-$n$ series structure in (\ref{structure}),
%we assume $G_n(x)$ takes the following form
%\begin{align}\label{structure_G}
%G_n(x)=&\lambda\left[ny_{1}(x)+\frac{z_{1}(x)}{n}+O\left(n^{-3}\right)\right]\nonumber\\&
%+\lambda^2\left[y_2(x)+\frac{z_2(x)}{n^2}+O\left(n^{-4}\right)\right]\nonumber\\&
%+\lambda^3\left[\frac{y_3(x)}{n}+\frac{z_3(x)}{n^3}+O\left(n^{-5}\right)\right]+\cdots
%\end{align}
%where the $y_k(x)$'s and $z_k(x)$'s correspond to the leading order terms and second order terms (in $n$) of the cumulants respectively.
we start by making the replacement $\lambda \to n\lambda$ and taking $n$ large, such that $G_n(x)$ admits
\begin{align}\label{Gn_G_F}
G_n(x)=n^2Y(x)+O\left(1\right)
\end{align}
where $Y(x)=\lambda y_1(x)+\lambda^2 y_2(x)+\cdots$, independent of $n$. Here, $y_\ell(x)$ corresponds to the
leading-order term (in $n$) for the $\ell$-th cumulant by the following integral:
\begin{align}
C_{\ell, 0}\!=\!\ell !\int_{\infty}^{\beta/P} \frac{y_\ell(x)}{x}{\rm d}x, \qquad
\ell = 1, 2,\ldots \label{integrate}
\end{align}
The remaining challenge is to compute closed-form formulas for
$y_\ell(x)$, $\ell = 1, 2, \ldots$, in order to obtain the
corresponding $C_{\ell, 0}$.
%\subsubsection{Cumulants to Leading Order in $n$}
Substituting (\ref{Gn_G_F}) for $G_n(x)$ into (\ref{Fundamental})
(with $\lambda\to n\lambda$) and keeping only the leading-order
terms in $n$ (i.e., $O(n^6)$), we obtain the following equation
involving\footnote{Here and henceforth, without causing confusion,
we will omit the argument $x$ when presenting differential or
difference (recursion) equations.} $Y(x)$:

\begin{align}\label{leading_Equation}
\left[\left(x+\beta+\lambda+1\right)Y'-Y\right]^2=4\left(xY'-Y+\beta
\right)\left(Y'+\lambda\right)Y'\;.
\end{align}

Note that this equation completely characterizes all of the
cumulants to leading order in $n$ (i.e., based on this, we may
compute $C_{\ell,0}, \ell=1, 2, \ldots$ in the structure
(\ref{structure})). This can be done upon substituting $Y(x)=\lambda
y_1(x) +\lambda^2 y_2(x)+\cdots$ into (\ref{leading_Equation}) and
matching the coefficients of $\lambda^k$ on the left and right-hand
side. Using this approach, we systematically obtain equations
involving $y_k(x), k=1, 2, \ldots$ which are elaborated upon below.

The case $k=1$ characterizes the mean of the mutual information. In
this case, we obtain the non-linear differential equation involving
$y_1(x)$:
\begin{align}\label{leading_mean_equation}
\left[x^2+2\left(\beta+1\right)x+\left(\beta-1\right)^2\right]\left(y_1'\right)^2-2\left(x+\beta+1\right)y_1'y_1-4\beta
y_1'+y_1^2=0\;,
\end{align}
%which exactly characterizes the mean to leading order in $n$.
%Solving this non-linear differential equation gives us:
which can be explicitly solved to give
\begin{align}
y_1(x)=\frac{\sqrt{x^2+2\left(\beta+1\right)x+\left(\beta-1\right)^2}-x-\left(\beta+1\right)}{2}\;.\label{y_1}
\end{align}
Integrating $y_1(x)$ via (\ref{integrate}), we find a closed-form
expression for $C_{1,0}$. With this expression, we find that
$nC_{1,0}$ agrees precisely with the $n$-asymptotic mean of the
mutual information (i.e., $\mu_0$ in (\ref{eq:coulomb_mean})).

%{\bf ** Samuel, you need to be careful above.  It seems that you
%need to compute $y_1$ and $y_2$ first, and these form the initial
%conditions for the recursion.  The expression above doesn't make any
%sense for $k = 1$ and $k=2$, since the upper summation limits become
%undefined...Perhaps, it may be clearer to explicitly give the mean
%and variance equations first, and then present the generalized
%recursion when looking at the higher order cumulants...***}

The case $k=2$ characterizes the variance of the mutual information.
In this case, we obtain the non-linear differential equation
involving $y_1(x)$ and $y_2(x)$:
\begin{align}\label{y_2}
&\left\{-2\left[x^2+2\left(\beta+1\right)x+\left(\beta-1\right)^2\right]y_1'+2\left(x+\beta+1\right)y_1+4\beta\right\}y_2'\nonumber\\
&+2\left[\left(x+\beta+1\right)y_1'-y_1\right]y_2+y_1'\left\{4x\left(y_1'\right)^2+2\left(x-\beta-1\right)y_1'-4y_1'y_1-2y_1\right\}=0.
\end{align}
Whilst seemingly complicated, quite remarkably, once we substitute
$y_1(x)$ with (\ref{y_1}), the coefficient of $y_2'(x)$ vanishes, i.e.,
\begin{align} \label{eq:Collapse}
-2\left[x^2+2\left(\beta+1\right)x+\left(\beta-1\right)^2\right]y_1'+2\left(x+\beta+1\right)y_1+4\beta
= 0 \; ,
\end{align}
and the differential equation (\ref{y_2}) collapses to a simple {\it
algebraic equation} in $y_2(x)$. The solution is then easily obtained
as:
\begin{align} \label{eq:y2Expr}
y_2(x)=\frac{x}{2\sqrt{x^2+2\left(\beta+1\right)x+\left(\beta-1\right)^2}}-\frac{1}{2}\;\frac{x^2+\left(\beta+1\right)x}{x^2+2\left(\beta+1\right)x+\left(\beta-1\right)^2}\;.
\end{align}
Integrating $y_2(x)$ via (\ref{integrate}), we obtain a closed-form
expression for $C_{2,0}$ which is found to  agree precisely with the
approximation for the $n$-asymptotic variance of the mutual information (i.e., $\sigma^2$ in
(\ref{eq:coulomb_variance})).

%The results above, indicating that the higher order cumulants beyond
%the mean and variance all vanish with increasing $n$ on the order of
%$O(1/n)$ or faster, recover the well-known fact that as $n$ grows
%towards infinity (with all other parameters kept fixed), the
%Gaussian approximation with $\mu_0$ and $\sigma^2_0$ approaches the
%exact mutual information distribution.

The higher order cumulants, beyond the mean and variance, are
characterized by the case $k > 2$. Evaluating and investigating
these cumulants is important, since it allows one to study
deviations from Gaussian for finite dimensions. As we will see,
these higher order cumulants may also be used to ``refine'' the
Gaussian approximation to provide increased accuracy.

For $k > 2$, $y_k(x)$ are found to satisfy the following recursion:
{\allowdisplaybreaks
\begin{align}\label{recursion}
&\sum_{i=1}^{k}y_i'\left\{\left[x^2\!+\!2\left(\beta\!+\!1\right)x\!+\!\left(\beta\!-\!1\right)^2\right]y_{k-i+1}'\!-\!2\left(x\!+\!\beta\!+\!1\right)y_{k-i+1}\right\}\!+\!\sum_{i=1}^{k}
y_iy_{k-i+1}\!-\!4\beta
y_{k}'\nonumber\\&=2\sum_{i=1}^{k-1}y_i'\left[\sum_{j=1}^{k-i}\left(2xy_j'y_{k-i-j+1}'
\!-\!2y_j'y_{k-i-j+1}\right)\!+\!\left(x\!-\!\beta\!-\!1\right)y_{k-i}'\!-\!y_{k-i}\right]\!-\!\sum_{i=1}^{k-2}y_i'y_{k-i-1}'\;
\end{align}
with initial conditions $y_1(x)$ and $y_2(x)$ given in (\ref{y_1}) and
(\ref{eq:y2Expr}) respectively.  In theory, the recursive equation
(\ref{recursion}) enables us to systematically compute the leading
order expressions (in $n$) in closed-form for any desired number of
higher order cumulants, in sequence. In its current form, however,
(\ref{recursion}) appears very complicated. Fortunately, this
expression can be simplified considerably by observing that the only
term in (\ref{recursion}) which involves $y_k(x)$ (or its derivative)
is:
\begin{align} \label{eq:LeadTerm}
\left\{\left[x^2+2\left(\beta+1\right)x+\left(\beta-1\right)^2\right]y_1'-\left(x+\beta+1\right)y_1-2\beta\right\}y_k'
- \left[\left(x+\beta+1\right)y_1'-y_1\right]y_k\;,
\end{align}
with all other terms involving the previously computed $y_i(x), i<k$.
As for the variance, quite remarkably, the term (\ref{eq:LeadTerm})
simplifies considerably upon noting that the coefficient of $y_k'(x)$
is precisely zero, by virtue of (\ref{eq:Collapse}). This
interesting observation indicates that the computation of the higher
cumulants, for $k > 2$, only involves solving simple algebraic
equations in $y_k(x)$, rather than non-linear differential equations
involving $y_k(x)$ and $y_k'(x)$. More specifically, the recursive
equation (\ref{recursion}) collapses to the following:
\begin{align}
y_k =
\frac{R_k\!-\!\sum_{i=2}^{k-1}\left\{\left[x^2\!+\!2\left(\beta\!+\!1\right)x\!+\!\left(\beta\!-\!1\right)^2\right]y_i'y_{k-i+1}'\!-\!2\left(x+\beta+1\right)y_i'y_{k-i+1}+
y_iy_{k-i+1}\right\}}{2\left[\left(x+\beta+1\right)y_1'-y_1\right]}
\end{align}
for $k = 3, 4, \ldots$, where $R_k$ represents the right-hand side
of (\ref{recursion}), which depends only on the previously computed
$y_i(x), i<k$. With this result, $y_k(x)$ can be easily computed,
systematically in closed-form, for any value of $k$. We simply write
down $y_3(x)$ and $y_4(x)$ here by way of example:{\allowdisplaybreaks
\begin{align}
&y_3(x)=-\frac{1}{2}\;x\left(x+1-\beta\right)\left(x-1+\beta\right)\left(\frac{x+\beta+1-\sqrt{x^2+2\left(\beta+1\right)x+\left(\beta-1\right)^2}}{\left(x^2+2\left(\beta+1\right)x+\left(\beta-1\right)^2\right)^{5/2}}\right)\;,\label{y_3}\\
&y_4(x)=\frac{x}{2}\frac{3x^4+\left(\beta+1\right)x^3-\left(\beta-1\right)^2\left[6x^2+3\left(\beta+1\right)x-\left(\beta-1\right)^2\right]}{\left(x^2+2\left(\beta+1\right)x+\left(\beta-1\right)^2\right)^{7/2}}\nonumber\\&
\quad+\frac{x}{2}\frac{-3x^5\!-\!4\left(1\!+\!\beta\right)x^4\!+\!\left(\beta\!-\!1\right)^2\left[5x^3\!+\!9\left(\beta\!+\!1\right)x^2\!+\!2\left(\beta\!+\!1\right)^2x\!-\!\left(\beta\!+\!1\right)\left(\beta\!-\!1\right)^2\right]}{\left(x^2+2\left(\beta+1\right)x+\left(\beta-1\right)^2\right)^4}\;.\label{y_4}
\end{align}}
Integrating $y_k(x)$ via (\ref{integrate}), we obtain $C_{3, 0}$ and
$C_{4, 0}$ in (\ref{structure}) in closed form:
%\begin{align}
%\kappa_{l, {\rm
%leading}}=-\frac{1}{n^{l-2}}\int^P_0\frac{y_l\left(x\right)}{x}\;{\rm
%d}x\;.
%\end{align}
%We compute the first four cumulants:
{\allowdisplaybreaks
\begin{align}
%&\kappa_{1, {\rm
%Leading}}=n\Bigg[\ln\left(\frac{\sqrt{\be^2+2\lb(\be+1\rb)\be
%P+\lb(\be-1\rb)^2P^2}+\lb(\be-1\rb)P+\be}{2\beta}\right)\nonumber\\&\qquad\qquad\qquad+\beta\ln\left(\frac{\sqrt{\beta^2+2\left(\beta+1\right)\beta
%P+\left(\beta-1\right)^2P^2}-\left(\beta-1\right)P+\beta}{2\beta}
%\right)\nonumber\\&\qquad\qquad\qquad\qquad\qquad\quad-\frac{\beta+\left(\beta+1\right)P-\sqrt{\beta^2+2\left(\beta+1\right)\beta
%P+\left(\beta-1\right)^2P^2}}{2P}\Bigg]\;,\label{k1leading}\\\nonumber \\
%&\kappa_{2,{\rm
%Leading}}=\ln\left[\frac{\beta+\left(\beta+1\right)P}{2\sqrt{\beta^2+2\left(\beta+1\right)\beta
%P+\left(\beta-1\right)^2P^2}}+\frac{1}{2}\right]\;,\label{k2leading}\\\nonumber
%\\
&C_{3,  0}=\frac{\beta+1}{2\beta}-\frac{3\beta
P}{\beta^2+2\left(\beta+1\right)\beta
P+\left(\beta-1\right)^2P^2}\nonumber\\&\qquad-\frac{1}{2}\frac{\left(\beta-1\right)^2\left[\left(\beta^2+1\right)P^3+3\beta\left(\beta+1\right)P^2\right]
+3\beta^2\left(\beta^2+1\right)P+\beta^3\left(\beta+1\right)}{\beta\left(\beta^2+2\left(\beta+1\right)\beta
P+\left(\beta-1\right)^2P^2\right)^{3/2}}\;,\label{k3leading}
\\\nonumber \\&C_{4, 0}=-\!\frac{\beta^2\!+\!1}{2\beta^2}\!+\!\frac{18\beta^4\!+\!28\beta^3\left(\beta\!+\!1\right)P\!+\!\left(\beta\!-\!1\right)^2\left[3\beta^2P^2\!-\!6\beta\left(\beta\!+\!1\right)P^3\!+\!\left(\beta\!-\!1\right)^2P^4\right]}{\left(\beta^2+2\left(\beta+1\right)\beta
P+\left(\beta-1\right)^2P^2\right)^3}P^2\nonumber\\&\qquad\qquad
+\frac{1}{2\beta^2\left(\beta^2+2\left(\beta+1\right)\beta
P+\left(\beta-1\right)^2P^2\right)^{5/2}}\bigg[\left(\beta^2+1\right)\beta^5+5\beta^4\left(\beta+1\right)\left(\beta^2+1\right)P\nonumber\\&\qquad\qquad+\left(10\beta^4+10\beta^3-16\beta^2+10\beta+10\right)\beta^3P^2+10\beta^2\left(\beta+1\right)\left(\beta^2+\beta+1\right)\left(\beta-1\right)^2P^3\nonumber\\&\qquad\qquad
+\beta\left(5\beta^4+14\beta^2+5\right)\left(\beta-1\right)^2P^4+\left(\beta+1\right)\left(\beta-1\right)^6P^5\bigg]
\;.\label{k4leading}
\end{align}}
As expected, expanding (\ref{k3leading}) and (\ref{k4leading}) about
$P=0$ gives exactly the same power series as the leading terms in
(\ref{series_k3}) and (\ref{series_k4}).

\subsubsection{Cumulants to Second Order in $n$}\label{sect:z}

In addition to computing the cumulants to leading order in $n$, it
is also of interest to compute the correction terms (in $n$), to
capture deviations and achieve higher accuracy at finite $n$. To
this end, similar to before, we consider
%the higher-order term in (\ref{Gn_G_F}) and assume
\begin{align}\label{Gn_G_F_2}
G_n(x)=n^2Y(x)+Z(x)+O\left(n^{-2}\right)
\end{align}
where $Y(x)$ is defined as in (\ref{Gn_G_F}) whilst $Z(x)=\lambda
z_1(x)+\lambda^2 z_2(x)+\cdots$, independent of $n$. Here,
$z_\ell(x)$ corresponds to the first-order correction term for the
$\ell$-th cumulant (i.e., characterized by $C_{\ell, 1}, \ell=1,
2, \ldots$ in (\ref{structure})) by the following integral:
\begin{align}
C_{\ell, 1}\!=\!\ell !\int_{\infty}^{\beta/P} \frac{z_\ell(x)}{x}{\rm d}x, \qquad
\ell = 1, 2,\ldots \label{integrate_2}
\end{align}

%Later by comparing the results with the small $P$ series in
%section IV(B), we can check the validity of our results.

Substituting (\ref{Gn_G_F_2}) into (\ref{Fundamental}) (with $\lambda\to n\lambda$)
and setting the second leading-order terms (in $n$) to be zero, we obtain the simple {\it
algebraic} equation involving $Z(x)$:
\begin{align}\label{second_Equation}
&\Big\{2\beta \lambda
-\left[\lambda^2+2(1+\beta-x)\lambda+x^2+2(1+\beta)x+(\beta-1)^2\right]Y'+(x-\lambda+\beta+1)Y\nonumber\\&-4Y'Y+6xY'^2\Big\}Z'
=\lb(2\lb(Y'\rb)^2-\lb(x+\be+1-\lambda\rb)Y'-Y\rb)Z-\frac{x^2}{2}\lb(Y''\rb)^2
\; .
\end{align}
This equation captures the {\it exact} first-order corrections to
all leading order cumulant approximations. Substituting $Y(x)=\lambda y_1(x)
+\lambda^2 y_2(x)+\cdots$ and $Z(x)=\lambda z_1(x) +\lambda^2
z_2(x)+\cdots$ into (\ref{second_Equation}) and matching the
coefficients of $\lambda^k$, we are able compute the $z_k(x)$'s
systematically.

The case $k=1$ corresponds to the correction term for the mean. In
this case, we obtain the equation involving $z_1(x)$:
\begin{align}\label{eq:z_1}
&\left\{2\beta-[x^2+2(\beta+1)x+(\beta-1)^2]y_1'+(x+\beta+1)y_1\right\}z_1'\nonumber\\&\qquad\qquad\qquad\qquad+\left[\left(x+\beta+1\right)y_1'-y_1\right]z_1+\frac{x^2}{2}
y_1''^2=0\;.
\end{align}
Again, once we substitute $y_1(x)$ with (\ref{y_1}), the coefficient of
the $z_1'(x)$ vanishes and (\ref{eq:z_1}) collapses to an algebraic
equation whose solution is
\begin{align}
&z_1(x)=\frac{-\beta
x^2}{\left(x^2+2\left(\beta+1\right)x+\left(\beta-1\right)^2\right)^{5/2}}\;.
\end{align}
The case $k=2$ corresponds to the correction term for the variance.
In this case, we obtain the equation involving $z_2(x)$:
\begin{align}\label{eq:z_2}
&\left\{2\beta-[x^2+2(\beta+1)x+(\beta-1)^2]y_1'+(x+\beta+1)y_1\right\}z_2'+\left[(x+\beta+1)y_1'-y_1\right]z_2\nonumber\\
&\!+\!\left\{2\beta\!-\![x^2+2(\beta+1)x\!+\!(\beta\!-\!1)^2]y_2'\!+\!(x+\beta+1)y_2\!+\!6x(y_1')^2\!-\!2(1+\beta-x)y_1'\!-\!4y_1'y_1\!-\!y_1\right\}z_1'\nonumber\\&+\left[(x+\beta+1)y_1'-y_1\right]z_1
+\left[(x+\beta+1)y_2'-2(y_1')^2-y_2-y_1'\right]=0\;.
\end{align}
The coefficient of $z_2'(x)$ is the same as in (\ref{eq:z_1}) and
vanishes, which means (\ref{eq:z_2}) is an algebraic equation
for $z_2(x)$. More interestingly, after substituting $y_1(x)$ with
(\ref{y_1}) and $y_2(x)$ with (\ref{y_2}), the coefficient of $z_1'(x)$
also becomes zero, which further simplifies the computation. Thus we
obtain $z_2(x)$ as follows:
\begin{align}
z_2(x)&=\frac{\left[(x+\beta+1)y_1'-y_1\right]z_1
+\left[(x+\beta+1)y_2'-2(y_1')^2-y_2-y_1'\right]}{(x+\beta+1)y_1'-y_1}\nonumber\\&=-\frac{x^2}{2}\frac{\left(\beta+1\right)x^2+2\left(\beta^2-3\beta+1\right)x+\left(\beta+1\right)\left(\beta-1\right)^2}{\left(x^2+2\left(\beta+1\right)x+\left(\beta-1\right)^2\right)^{7/2}}\nonumber\\&
\quad
+\frac{x^2}{2}\frac{\left(\beta\!+\!1\right)x^3\!+\!\left(3\beta^2\!-\!14\beta\!+\!3\right)x^2\!+\!\left(3\beta\!-\!1\right)\left(\beta\!-\!3\right)\left(\beta\!+\!1\right)x\!+\!
\left(\beta^2\!+\!10\beta\!+\!1\right)\left(\beta\!-\!1\right)^2}{\left(x^2\!+\!2\left(\beta\!+\!1\right)x\!+\!\left(\beta\!-\!1\right)^2\right)^4}\;.
\end{align}
For the case $k> 2$, we can derive a recursive equation as follows:
\begin{align}\label{recursion2}
&\sum_{i=1}^kz_i'\left\{[x^2+2(\beta+1)x+(\beta-1)^2]y_{k-i+1}'+(x+\beta+1)y_{k-i+1}\right\}+2\beta
z_k'\nonumber\\&=\sum_{i=1}^{k-1}z_i'\left\{y_{k-i}+2(1+\beta-x)y_{k-i}'+4\sum_{j=1}^{k-i}y_{k-i-j+1}'y_j-6x\sum_{j=1}^{k-j}y_{k-i-j+1}'y_j'\right\}+\sum_{i=1}^{k-2}z_i'y_{k-i-1}'\nonumber\\&-\sum_{i=1}^{k}z_i\left[\left(x\!+\!\beta\!+\!1\right)y_{k-i+1}'\!-\!y_{k-i+1}\right]+\sum_{i=1}^{k-1}z_i\left[2\sum_{j=1}^{k-i}y_j'y_{k-i-j+1}'\!+\!y_{k-i}'\right]\!-\!\frac{x^2}{2}\sum_{i=1}^{k}y_i''y_{k-i+1}''\;.
\end{align}
The only term in (\ref{recursion2}) which involves $z_k(x)$ (or its
derivative) is:
\begin{align}
\left\{-\left[x^2+2(\beta+1)x+(\beta-1)^2\right]y_1'+(x+\beta+1)y_1+2\beta\right\}z_k'+[(x+\beta+1)y_1'-y_1]z_k
\end{align}
where the coefficient of $z_k'(x)$ is exactly the same as the one of
$y_k'(x)$ in (\ref{eq:LeadTerm}), which has been shown to be zero. This
observation indicates that we only need to solve an algebraic
equation for $z_k(x)$. Interestingly, apart from $z_k'(x)$, it is found
that the coefficients of $z_i'(x), i=1, 2, \dots, k-1$ in
(\ref{recursion2}) are also identically equal to zero.
% (though
%lacking a rigorous justification at this moment, examples up to
%$k=4$ have been tested).

%{\bf \noindent*********************************\\
%{\it Prof. McKay}: Now, a general question --- do the coefficients of $y_k'$ vanish also?  If so, then the recursion would be very very simply...\\
%{\it Response}: Genrerally, the coefficient of $y_k'$ does not vanish, thus (\ref{eq:recursive_z_k}) is the simplest version of the recursion.\\
%*************************************}

Given that this vanishing property holds for any $k$, eliminating
terms involving $z_i'(x), i=1, 2, \dots, k$ in
%\begin{align}
%\left(x^2+2(\beta+1)x+(\beta-1)^2\right)y_1'+(x+\beta+1)y_1+2\beta
%\end{align}
(\ref{recursion2}) immediately leads to the general recursive
solution:
\begin{align}\label{eq:recursive_z_k}
z_k =
\frac{\sum_{i=1}^{k-1}z_i\left[2\sum_{j=1}^{k-i}y_j'y_{k-i-j+1}'\!+\!y_{k-i}'\!-\!\left(x\!+\!\beta\!+\!1\right)y_{k-i+1}'+y_{k-i+1}\right]\!-\!\frac{x^2}{2}\sum_{i=1}^{k}y_i''y_{k-i+1}''}{\left(x\!+\!\beta\!+\!1\right)y_{1}'\!-\!y_{1}}\;.
\end{align}
With this result, we can compute $z_k(x)$ in closed-form for any value
of $k$. For example, $z_3(x)$ is evaluated as
follows:{\allowdisplaybreaks
\begin{align}
&z_3(x)=-\frac{x^2}{2\left(x^2+2\left(\beta+1\right)x+\left(\beta-1\right)^2\right)^6}\Bigg[-x^7+8\left(\beta+1\right)x^6+\left(49\beta^2-38\beta+49\right)x^5\nonumber\\&+6\left(\beta+1\right)\left(15\beta^2-38\beta+15\right)x^4+\left(65\beta^4-76\beta^3-42\beta^2-76\beta+65\right)x^3\nonumber\\
&+4\left(\beta+1\right)\left(\beta^2+26\beta+1\right)\left(\beta-1\right)^2x^2-\left(17\beta^2-46\beta+17\right)\left(\beta-1\right)^4x-6\left(\beta+1\right)\left(\beta-1\right)^6\Bigg]\nonumber\\
&-\frac{x^2}{2\left(x^2+2\left(\beta+1\right)x+\left(\beta-1\right)^2\right)^{11/2}}\Bigg[x^6-9\left(\beta+1\right)x^5-8\left(5\beta^2-11\beta+5\right)x^4\nonumber\\&
-10\left(\beta+1\right)\left(5\beta^2-13\beta+5\right)x^3-\left(15\beta^4+134\beta^3-330\beta^2+134\beta+15\right)x^2\nonumber\\&+\left(\beta+1\right)\left(11\beta^2-100\beta+11\right)\left(\beta-1\right)^2x+6\left(\beta^2+5\beta+1\right)\left(\beta-1\right)^4\Bigg].
\end{align}}

%\noindent*********************************************************************\\
%{\bf Note that given the property that the coefficient of $z_i',
%i=1, 2, \dots, k$ in (\ref{recursion2}) all identically vanish holds
%for any $k$, it naturally implies that the coefficient of $Z'$ in
%(\ref{second_Equation}) vanishes. Therefore, we write
%\begin{align}\label{eq:LeadingEq_V2}
%2\beta \lambda
%&-\left[\lambda^2+2(1+\beta-x)\lambda+x^2+2(1+\beta)x+(\beta-1)^2\right]Y'\nonumber\\&\qquad\qquad\qquad+(x-\lambda+\beta+1)Y-4Y'Y+6xY'^2=0\;.
%\end{align}
%which is seemingly another version of (\ref{leading_Equation}).
%Substituting $Y=\lambda y_1+\lambda^2 y_2 +\cdots$ into
%(\ref{eq:LeadingEq_V2}), we can recover exactly the same $y_k$ as
%those derived from (\ref{leading_Equation}), which in return
%confirms the vanishing property of $z_i', i=1, 2, \ldots, k$. The
%examples of $y_1, y_2, y_3$ and $y_4$ (e.g., (\ref{y_1}) -
%(\ref{y_4})) have been tested.}\\
%*********************************************************************

Integrating $z_k(x)$ via (\ref{integrate_2}), we obtain the first-order
correction terms in closed-form. For example, $C_{1, 1}$ and $C_{2,
1}$ are given as follows:{\allowdisplaybreaks
\begin{align}
&C_{1, 1}=-\frac{\beta\!+\!1}{24\beta}\!+\!\frac{\beta^3\left(\beta\!+\!1\right)\!+\!3\beta^2\left(\beta\!+\!1\right)^2P\!+\!3\beta\left(\beta\!+\!1\right)\left(\beta^2\!+\!1\right)P^2\!+\!\left(\beta^2\!+\!1\right)\left(\beta\!-\!1\right)^2P^3}{24\beta\left(\beta^2+2\beta\left(\beta+1\right)P+\left(\beta-1\right)^2P^2\right)^{3/2}}\label{mean_correction}\;,\\&
C_{2, 1}=\frac{\be^2\!+\!1}{24\be^2}\!-\!\frac{4\left(\beta\!+\!1\right)\beta^3P^3\!+\!\left(9\beta^2\!-\!42\beta\!+\!9\right)\beta^2P^4\!+\!6\beta\left(\beta\!+\!1\right)\left(\beta\!-\!1\right)^2P^5\!+\!\left(\beta\!-\!1\right)^4P^6}{12\left(\beta^2+2\beta\left(\beta+1\right)P+\left(\beta-1\right)^2P^2\right)^{3}}\nonumber\\&
-\frac{1}{24\beta^2\left(\beta^2+2\beta\left(\beta+1\right)P+\left(\beta-1\right)^2P^2\right)^{7/2}}\Bigg[\lb(\be^2+1\rb)\be^7+7\be^6\lb(\be+1\rb)\lb(\be^2+1\rb)P\nonumber\\&
+7\be^5\lb(\be^2+1\rb)\lb(3\be^2+4\be+3\rb)P^2+\lb(\be+1\rb)\lb(35\be^4+62\be^2+35\rb)\be^4P^3\nonumber\\&
+\lb(35\be^6\!+\!9\be^4\!+\!8\be^3\!+\!9\be^2\!+\!35\rb)\be^3P^4\!+\!\lb(\be\!+\!1\rb)\lb(\be\!-\!1\rb)^2\lb(21\be^4\!-\!14\be^3\!-\!2\be^2\!-\!14\be\!+\!21\rb)\be^2P^5\nonumber\\&
+7\lb(\be+1\rb)^2\lb(\be-1\rb)^6\be
P^6+\lb(\be+1\rb)\lb(\be-1\rb)^8P^7\Bigg].\label{var_correction}
%&\kappa_{3, {\rm Correction}}=
\end{align}}
The first-order correction terms to the higher cumulants $\kappa_3,
\kappa_4, \ldots$ are omitted here in order to keep the presentation
concise, though they follow trivially. Again, expanding
(\ref{mean_correction}) and (\ref{var_correction}) about $P=0$
yields exactly the same power series as the $O(n^{-1})$ term in
(\ref{series_k1}) and the $O(n^{-2})$ term in
(\ref{var_correction}). Note that if we aim to compute further high
order correction terms (i.e., $C_{i, j}, i=1, 2, \ldots,  j \ge 2$)
for each cumulant, we can invoke the same procedure as what has been
used for computing the two leading-order terms, which only takes
more algebraic effort.

\subsubsection{Large-$n$ Series Computation for a Given Cumulant}
In addition to the machinery used in the previous subsection for
systematically computing the high order cumulants one after the
other for a given order in $n$, we can also
%also employ the one used in \cite{Chen_McKay} and
%\cite{Li_Chen_McKay} to
derive recursions for systematically computing the large-$n$ series
for a \emph{given} cumulant.
%To this end, instead of arranging (\ref{Fundamental}) (with $G_n(x)$
%represented in power series of $\lambda$) by the order of $n$ and
%derive, in sequence, the recursion for $C_{\ell,0}, \ell=1, 2,
%\ldots$, $C_{\ell,1}, \ell=1, 2, \ldots$, etc., we rearrange the
%equations by the order of $\lambda$ and recursively compute
%$C_{0,j}, j=1, 2, \ldots$, $C_{1,j}, j=1, 2, \ldots$, etc.
%Essentially, both machineries aim to extract equations for the
%corresponding terms of $C_{i,j}$'s from (\ref{Fundamental}), by
%substituting $G_n$ with power series shown in (\ref{structure}) and
%matching the coefficients of $n$ and $\lambda$ respectively. With
%these equations, $C_{\ell,j}$'s can be obtained in closed-form
%expressions with the integral (\ref{integral}).
Here we give the computation of the mean as an example, with the
corresponding computations for the higher cumulants  following
similarly. More details for the specific case $n_t = n_r$ can be
found in \cite{Chen_McKay}.

Based on the power series representation of the CGF (in $\lambda$),
substituting
\begin{align}
G_n(x)=\lambda g_1(x) +\lambda^2 g_2(x) +\lambda^3 g_3(x)+\cdots
\end{align}
into (\ref{Fundamental}) and matching the coefficients of
$\lambda^k$, we obtain the equations for $g_i(x),i=1, 2, \ldots$.
For the mean we are interested in $g_1(x)$, which takes the form:
%Specifically, the exact equation for the corresponding term of the
%mean is as follows:
\begin{align}\label{meanEq}
x^2\left(g_1''\right)^2-&n^2\left(x^2+2\left(\beta+1\right)x+\left(\beta-1\right)^2\right)\left(g_1'\right)^2\nonumber\\&+2n^2\left[\left(x+\beta+1\right)g_1+2\beta
n\right]g_1'-n^2g_1^2=0\;.
\end{align}
Making use of the large-$n$ series structure in (\ref{structure}), we assume
\begin{align}\label{eq:mean_series}
g_1(x)=n Q_0(x) +\frac{1}{n} Q_1(x) + \frac{1}{n^3} Q_2(x)+\ldots
\end{align}
where $Q_{\ell}(x), \ell=0, 1, \ldots$ are independent of $n$ and are
related to the large-$n$ series expansion for the mean via the
following integral:
\begin{align}\label{integrate_3}
C_{1,\ell}=\int_\infty^{\beta/P}\frac{Q_\ell(x)}{x}{\rm d}x,\qquad \ell=1, 2, \ldots
\end{align}
Substituting (\ref{eq:mean_series}) into (\ref{meanEq}) and matching
the coefficients of $n^k$, equations involving $Q_\ell(x), \ell=0, 1, \ldots$
are obtained successively. Considering the highest order in $n$, we
have
\begin{align}
\left[x^2+2\left(\beta+1\right)+\left(\beta-1\right)^2\right]\left(Q_0'\right)^2-2\left[\left(x+\beta+1\right)Q_1+2\beta\right]Q_0'+Q_0^2=0\;.
\end{align}
Note that this equation is exactly the same as
(\ref{leading_mean_equation}), because they both characterize the
leading order term (in $n$) of the mean and lead to the same
solution (\ref{y_1}).
%To compute correction terms, we continue to consider the coefficients to next order in $n$
%and obtain
%\begin{align}
%&\left\{2\left[x^2+2\left(\beta+1\right)x+\left(\beta-1\right)^2\right]Y_1'-\left(x+\beta+1\right)Y_1-2\beta\right\}Y_2'\nonumber\\&\qquad\qquad\qquad\qquad+2\left[Y_1-\left(x+\beta+1\right)Y_1'\right]Y_2-x^2\left(Y_1''\right)^2=0
%\end{align}
For $k\geq 1$, we derive the following recursion
\begin{align}\label{eq:recursive_Y_k}
4\beta &Q_k'+2(x+\beta+1)\sum_{i=0}^kQ_{k-i}Q_{i}'+x^2\sum_{i=0}^{k-1}Q_{k-i-1}''Q_{i}''\nonumber\\&=\sum_{i=0}^{k}Q_{k-i}Q_i+\sum_{i=0}^{k}\left[x^2+2(\beta+1)x+(\beta-1)^2\right]Q_{k-i}'Q_i'.
\end{align}
Interestingly enough, we find again that in
(\ref{eq:recursive_Y_k}), the only term which involves $Q_k(x)$ (or its
derivative) is
\begin{align*}
&\left\{-2\left[x^2+2(\beta+1)x+(\beta-1)^2\right]Q_0'+2(x+\beta+1)Q_0+4\beta\right\}Q_k'+2\left[(x+\beta+1)Q_0'-Q_0\right]Q_k
\end{align*}
where the coefficient of $Q_k'(x), k=1, 2, \ldots$ vanishes
identically. Setting $k=1$ results in exactly the same equation as
(\ref{eq:z_2}), since they both correspond to $C_{1,1}$ in
(\ref{structure}). For $k\geq 2$ we obtain the general recursion of
$Q_k(x)$:
\begin{align}
&Q_k =
\nonumber\\&\frac{\sum_{i=1}^{k-1}\!\left[Q_{k-i}Q_i\!+\!\left(x^2\!+\!2(\beta\!+\!1)x\!+\!(\beta\!-\!1)^2\right)Q_{k-i}'Q_i'\!-\!2(x\!+\!\beta\!+\!1)Q_{k-i}Q_i'\right]\!-\!x^2\!\sum_{i=0}^{k-1}\!Q_{k-i-1}''Q_{i}''}{2\left[(x+\beta+1)Q_0'-Q_0\right]}.
\end{align}
Note that this recursion is a generalization of \cite[Eq. (132)]{Chen_McKay} which was derived for $\beta=1$.
%\begin{align}
%Y_3=-{\frac { \left( 8{x}^{4}+ 4\left(\beta +1\right) {x}^{3}-
% \left(15{\beta}^{2}-39\beta+15 \right) {x}^{2}-10\left( \beta
%+1 \right)  \left( \beta-1 \right) ^{2}x+ \left( \beta-1 \right) ^{4}
% \right) {x}^{2}\beta}{ \left( {x}^{2}+2(\beta+1)x+{\beta}^{2}-2\,
%\beta+1 \right) ^{11/2}}}
%\end{align}
Integrating $Q_k(x)$ via (\ref{integrate_3}) gives us closed-form
expressions for $C_{1,\ell}, \ell=0, 1, \ldots$. We write $C_{1,2}$
here as an example, and omit the presentation of the higher order
correction formulas for the sake of conciseness: {\footnotesize
\begin{align}\label{eq:C_1_3}
&C_{1,2}=\frac{1+\beta^3}{240\beta^3}-\frac{1}{240\beta^3\left((\beta-1)^2P^2+2(\beta+1)P+\beta^2\right)^{9/2}}\Big[\left( 1+{\beta}^{3} \right) {\beta}^{9}+9 \left( {\beta}^{2}-\beta
+1 \right)  \left( \beta+1 \right) ^{2}{\beta}^{8}P\nonumber\\&+18 \left( \beta+
1 \right)  \left( 2{\beta}^{2}+3\beta+2 \right)  \left( {\beta}^{2
}-\beta+1 \right) {\beta}^{7}{P}^{2}+42\, \left( 2{\beta}^{2}+\beta+
2 \right)  \left( {\beta}^{2}-\beta+1 \right)  \left( \beta+1 \right)
^{2}{\beta}^{6}{P}^{3}\nonumber\\&+42 \left( 2{\beta}^{2}+\beta+2 \right)
 \left( {\beta}^{2}-\beta+1 \right)  \left( \beta+1 \right) ^{2}{\beta
}^{5}{P}^{4}+ \left( 126+126{\beta}^{8}-384{\beta}^{4}+126{\beta
}^{5}+126{\beta}^{3} \right) {\beta}^{4}{P}^{5}\nonumber\\&+6\left( \beta+1
 \right)  \left( 14{\beta}^{8}-35{\beta}^{7}+35{\beta}^{6}-21{
\beta}^{5}-16{\beta}^{4}-21{\beta}^{3}+35{\beta}^{2}-35\beta+
14 \right) {\beta}^{3}{P}^{6}\nonumber\\&+6 \left( 6{\beta}^{8}-9{\beta}^{7}
-3{\beta}^{6}+9{\beta}^{5}+56{\beta}^{4}+9{\beta}^{3}-3{
\beta}^{2}-9\beta+6 \right)  \left( \beta-1 \right) ^{2}{\beta}^{2}{
P}^{7}\nonumber\\&+9 \left( \beta+1 \right)  \left( {\beta}^{6}-3{\beta}^{5}+3
{\beta}^{4}+3\,{\beta}^{2}-3\beta+1 \right)  \left( \beta-1
 \right) ^{4}\beta\,{P}^{8}+ \left( {\beta}^{6}-3{\beta}^{5}+3{
\beta}^{4}+3{\beta}^{2}-3\beta+1 \right) \left( \beta-1 \right) ^
{6}{P}^{9}
\Big].
\end{align}}
Systematic derivations of the series expansions for the variance and
higher order cumulants follow similarly, using the same approach. In
this way, one can obtain the large-$n$ series for each cumulant to
arbitrary degree of accuracy.
%when $n$ is finite.
%More details about this alternative
%cumulants derivation can be found in
%\cite{Chen_McKay,Li_Chen_McKay}.

%\\*********************************\;\;\;{\bf Main Discussion}\;\;\;*************************** \\
%Seen from the first three cumulants with first order corrections,
%there involve three parameters to affect the mutual information
%distribution, referred to as ($n, \beta, P$) or ($n, m, P$)
%parameter space. It is well-know that for sufficiently large $n$,
%the mutual information becomes Gaussian asymptotically.
%Nevertheless, it is still not clear that how large $n$ needs to be
%for any given value of $\beta$ and $P$, in order to make the
%large-dimension-based results accurate. Some one sees from the
%simulation that the large $n$ approximation in fact works well for
%even small $n$, whilst others observe strong deviation. Here we can
%achieve a full understanding of how ($n,\beta,P$) play the role in
%the mutual information distribution.

\section{Large SNR Analysis of the Gaussian Deviation}\label{sect:largeP}

The systematic high order cumulant expansions derived in the
previous section allow us to closely investigate the behavior of the
mutual information distribution under various conditions.  For
example, our analysis in Section \ref{sec:MVar_Gauss} recovered the
fact that as the number of antennas $n$ grows large, with the SNR
$P$ kept fixed, the distribution approaches a Gaussian.
However, in practice, the number of antennas is finite and is not
typically huge (though some recent trends have considered such
systems \cite{Marzetta10,Ngo11,Ngo12,Rusek12,Gop11}), whilst the SNR
may vary significantly depending on the application.  Thus, a
natural question is, for asymptotically high SNR, if and by how much
the mutual information distribution deviates from Gaussian.

In this section, we focus on the high SNR regime of the mutual
information. In this case, as shown in Figs.
\ref{Gaussian_equal}--\ref{Gauss_unequal}, the distribution appears
to deviate from Gaussian as the SNR becomes large, whilst the
deviation appears to be much \emph{stronger} for the case of equal
numbers of transmit and receive antennas (i.e., $n = m$) compared
with the case $n \neq m$.  This is an interesting observation which
thus far has resisted theoretical validation or explanation. Here
we draw insight into this phenomenon by employing our new
closed-form high order cumulant expansions given in the previous
section.

Interestingly, by taking $P$ large in (\ref{eq:coulomb_mean}),
(\ref{eq:coulomb_variance}), (\ref{k3leading}),
(\ref{mean_correction}) and (\ref{var_correction}), it turns out
that we obtain very different limiting results depending on whether
$\beta = 1$ (i.e., $n_t = n_r$) or not. For $\beta=1$, as $P$ grows
large we obtain\footnote{Note that the correction term for the third
cumulant $\kappa_3$ (i.e., $C_{3, 1}$ in (\ref{structure})) was
obtained in the derivation described in Section \ref{sect:z}, but
was not presented explicitly in this paper for the sake of
conciseness.}:
\begin{align}
&\mu\sim n\ln P +\frac{1}{16}\frac{\sqrt{P}}{n}+\frac{3}{1024}\left(\frac{\sqrt{P}}{n}\right)^3+\frac{45}{32768}\left(\frac{\sqrt{P}}{n}\right)^5+\cdots\; \label{beta1k1}\\
&\sigma^2\sim \frac{1}{2}\ln\left(\frac{P}{2}\right)+\frac{1}{32}\left(\frac{\sqrt{P}}{n}\right)^2+\cdots\; \label{beta1k2}\\
&\kappa_3\sim
\frac{1}{4n}+\frac{1}{32}\left(\frac{\sqrt{P}}{n}\right)^3+\cdots\label{beta1k3}
\end{align}
These large-$n$---large-$P$ series expansions were also documented
in \cite{Chen_McKay}, which focused primarily on the case $n_t =
n_r$. From these expressions, we see that for each cumulant the
correction terms grow with $P$, and indeed grow \emph{faster} than
the leading order terms.  As such, approximating each cumulant via their
leading order term will be inaccurate when $P$ is large.
%As
%such, it is clear that the commonly-assumed Gaussian approximation
%(based on the leading order terms of $\mu$ and $\sigma$ only), will
%become inaccurate for sufficiently large $P$.

Now, considering $\beta \neq 1$, we obtain:
\begin{align}
&\mu \sim
n\left[\ln P-\left(\beta-1\right)\ln\left(\frac{\beta-1}{\beta}\right)-1\right]+\frac{1}{n}\left[\frac{1}{12\left(\beta-1\right)}-\frac{1}{12\beta}\right]\nonumber\\&\qquad+\frac{1}{n^3}\left[-\frac{1}{120(\beta-1)^3}+\frac{1}{120\beta^3}\right]+\frac{1}{n^5}\left[-\frac{1}{252(\beta-1)^5}+\frac{1}{252\beta^5}\right]+\cdots\;,\label{k1largeP}\\
&\sigma^2 \sim
\ln\left(\frac{\beta}{\beta-1}\right)+\frac{1}{n^2}\left[-\frac{1}{12\left(\beta-1\right)^2}+\frac{1}{12\beta^2}\right]+\cdots\;,\label{k2largeP}\\
&\kappa_3 \sim
\frac{1}{n}\left[-\frac{1}{\left(\beta-1\right)}+\frac{1}{\beta}\right]+\frac{1}{n^3}\left[\frac{1}{6\left(\beta-1\right)^3}-\frac{1}{6\beta^3}\right]+\cdots\;\label{k3largeP}
\; .
%&\kappa_4 \sim
%\frac{1}{n^2}\left[\frac{1}{\left(\beta-1\right)^2}-\frac{1}{\beta^2}\right]+\frac{1}{n^4}\left[-\frac{1}{2\left(\beta-1\right)^4}+\frac{1}{2\beta^4}\right]+\cdots\;\label{k4largeP}
\end{align}
From these we can make some key observations, which we summarize in
the following remarks:
\begin{remark}
In contrast to the results in (\ref{beta1k1})--(\ref{beta1k3}), for
$n_t \neq n_r$, all terms other than the leading term of $\mu$ are
\emph{strictly bounded} as $P$ increases, converging to constants
depending on $\beta$.  Moreover, as $n$ or $\beta$ increase, the
correction terms have less effect, eventually becoming negligible,
even when $P$ is very large.
\end{remark}
\begin{remark} \label{th:pro3}
For $n_t \neq n_r$, the $n$-asymptotic power series representations
for the cumulants remain valid for \emph{arbitrary} SNR $P$; whilst
in contrast, for $n_t = n_r$ they break down for sufficiently large
$P$.  This clearly indicates that the commonly-assumed Gaussian
approximation (based on the leading order terms of $\mu$ and
$\sigma$ only) is quite robust at high SNRs for the case $n_t \neq
n_r$, but not for the case $n_t = n_r$.
\end{remark}

These remarks are illustrated in Figs.
\ref{fig:ka_equal}--\ref{fig:ka_unequal2}, showing $\sigma^2$ and
$\kappa_3$. Fig. \ref{fig:ka_equal} shows that when $n_t=n_r$, the
leading order term of $\sigma^2$ (i.e., representing the variance of the
Gaussian approximation) and the leading order term of $\kappa_3$ both
deviate strongly from simulations when $P$ is sufficiently large. In
contrast, when $n_t\neq n_r$ (i.e., Figs.
\ref{fig:ka_unequal}--\ref{fig:ka_unequal2}), the leading order
cumulants capture the simulations accurately for arbitrary $P$, even
with smaller $n$. We also see that increasing $\beta$ enhances the
accuracy of the leading order results, whilst including the
first-order correction term provides improved accuracy as well.
These observations explain technically why the Gaussian
approximation, based purely on the leading order terms of $\mu$ and
$\sigma^2$, is relatively more robust to increasing SNRs in Fig.
\ref{Gauss_unequal}, compared with the results in Fig.
\ref{Gaussian_equal}. Nevertheless, even in the former case, the
higher cumulants (e.g., of order $O(1/n)$) still contribute to some
deviations from Gaussian. These deviations become particularly
significant in the tail region, yielding the
Gaussian approximation unsuitable for capturing low outage
probabilities of practical interest.  This will be considered
further in the subsequent sections, where we will make use of our
closed-form cumulant expansions to ``correct'' for these deviations,
both in the tail and around the mean, and thereby refine the
Gaussian approximation.

\begin{figure}[htbp]
\centering \subfigure[$n_t=4, n_r=4$]{
\includegraphics[width=0.485\columnwidth]{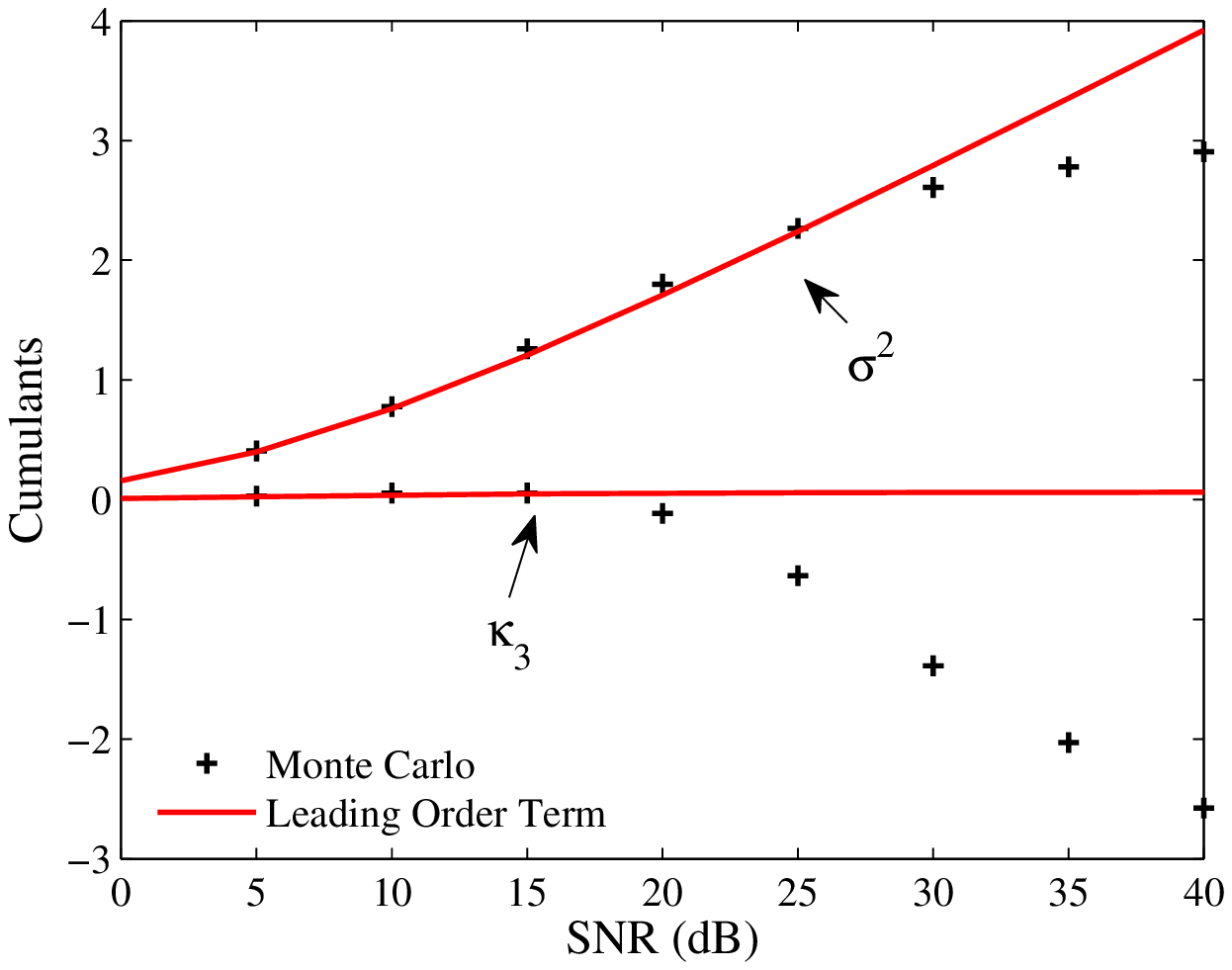}\label{fig:ka_equal}}
\subfigure[$n_t=4, n_r=3$]{
\includegraphics[width=0.485\columnwidth]{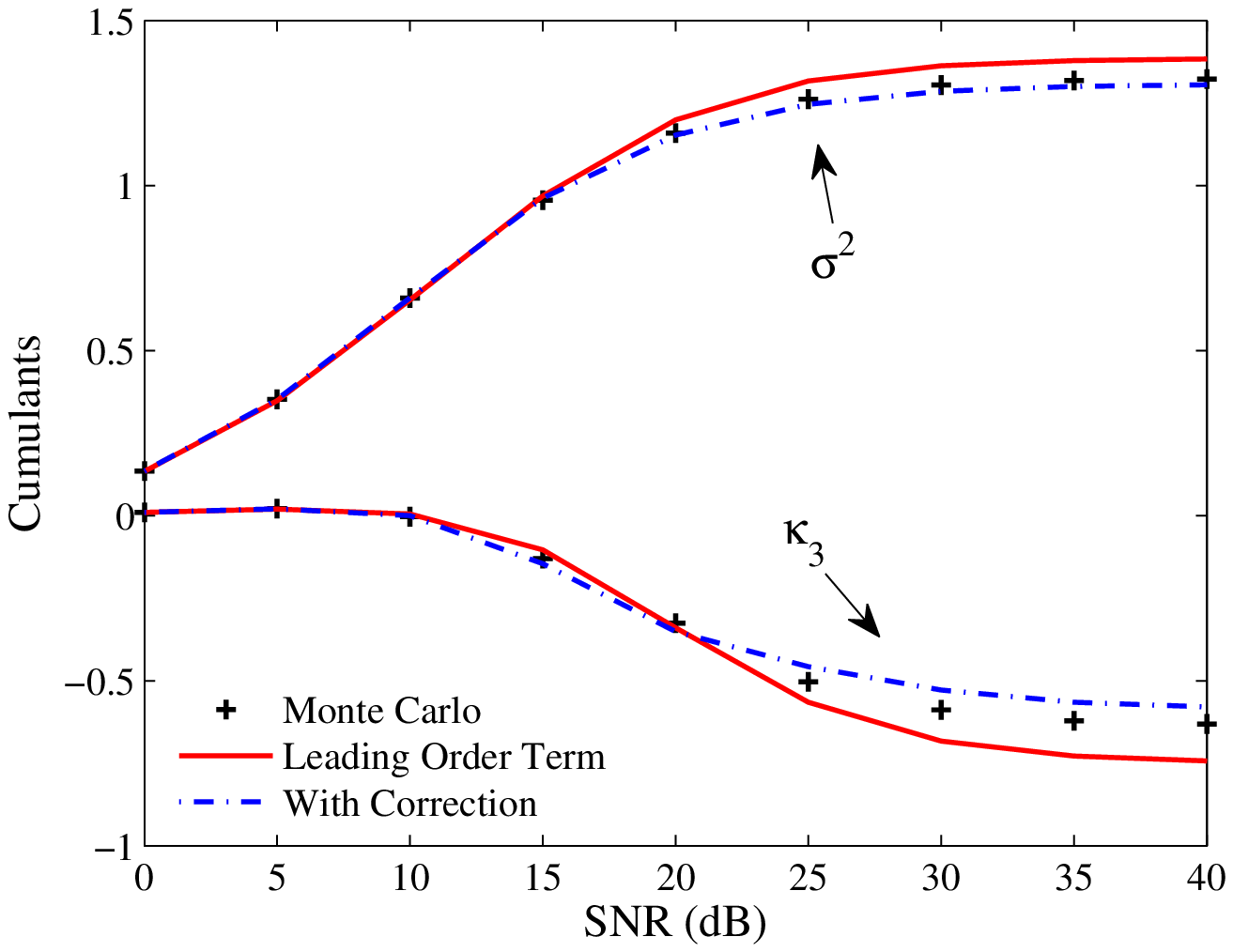}\label{fig:ka_unequal}}
\subfigure[$n_t=5, n_r=3$]{
\includegraphics[width=0.485\columnwidth]{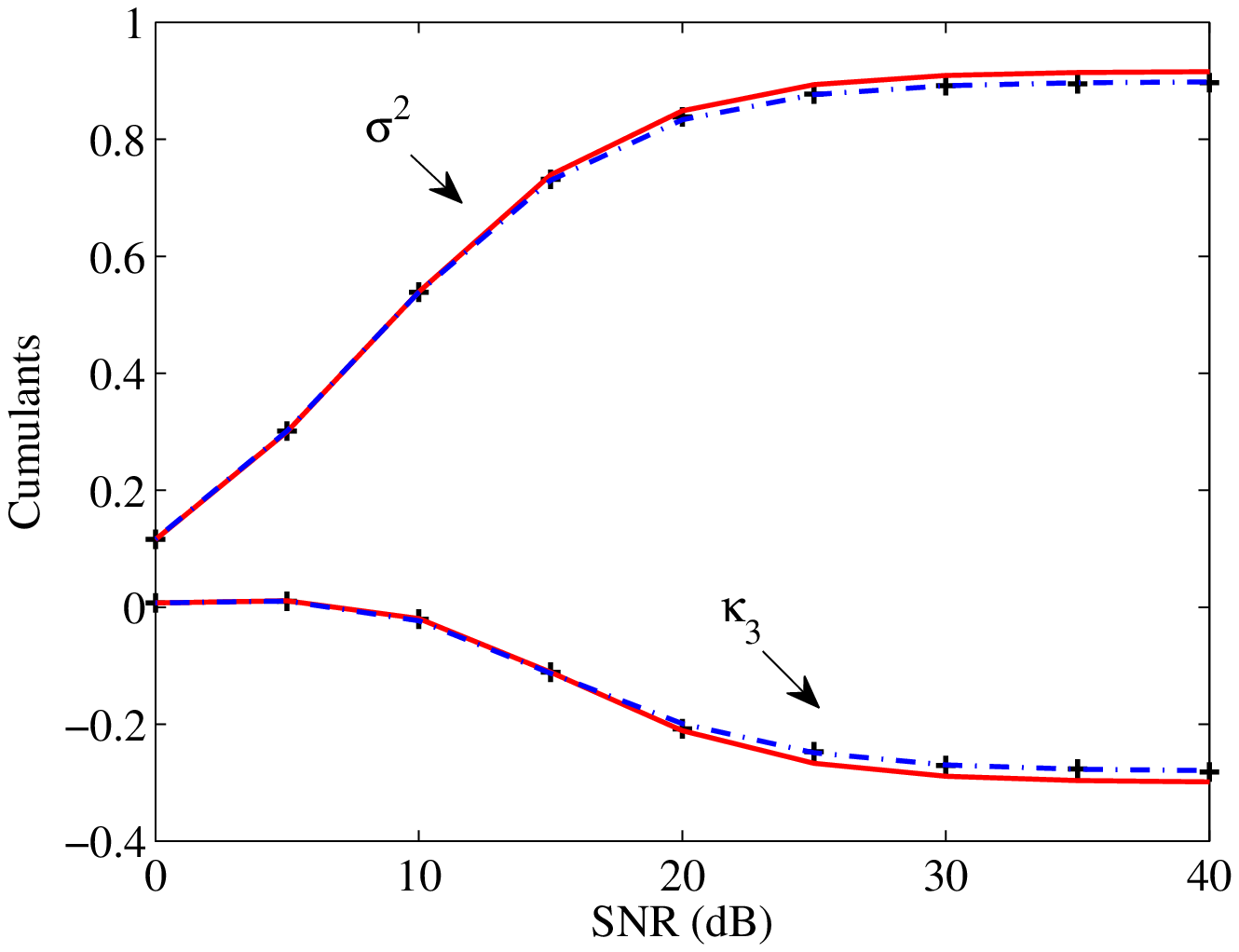}\label{fig:ka_unequal2}}
%\subfigure[$n_t=6, n_r=3$]{
%\includegraphics[width=0.485\columnwidth]{SU_k234_nt_6_nr_3.eps}}
\caption{Comparison of $\sigma^2$ to leading order (in $n$) and with
first-order correction, $\kappa_3$ to leading order and with
first-order correction, and Monte Carlo simulations. Results are
shown for different antenna configurations.}
\end{figure}

Before presenting these refinements, we would like to briefly
connect our large-$n$--large-$P$ cumulant expansions with existing
results, which were obtained via different means by considering $n$
fixed and taking $P$ large at the beginning. This approach was used
to obtain the leading-order terms of $\mu$ and $\sigma^2$ in
\cite{Hochwald04}, and these results coincide exactly with the
leading-order terms in (\ref{beta1k1}) and (\ref{beta1k2}).  In
addition, the first few terms of the mean and variance expansions (i.e., (\ref{k1largeP}) and (\ref{k2largeP}) respectively)
match exactly with the results in \cite[Eq.
(134)]{Lozano2005} and \cite[Lemma A.2]{Alex02}. For example,
\cite{Lozano2005} has the large-$P$ mean representation:
\begin{align}
\mu&\sim n\ln \left(\frac{P}{m}\right)-n+m\psi (m)-(m-n)\psi (m-n),
\quad P\to \infty\;
\end{align}
which gives (\ref{k1largeP}) by expanding the digamma function $\psi (x)$ at large $x$ as
%where $\psi (x)$ denotes the digamma function, admitting the large
%$x$ expansion:
\begin{align}\label{eq:psi_expansion}
\psi (x) =
\ln{x}-\frac{1}{2x}-\frac{1}{12x^2}+\frac{1}{120x^4}-\frac{1}{252x^6}+O\left(\frac{1}{x^8}\right), \quad x\to \infty.
\end{align}
Our results can also be shown to be consistent with the large-$P$
MGF representation derived in \cite{Oyman03}.

%In addition, recalling the large-$P$ characteristic function (CF) of
%mutual information reported in \cite[Eq. (27)]{Oyman03} (based on
%the result of \cite{Goodman63}), we find our cumulants can also be
%recovered by expanding the asymptotic CF there, considering the
%expansion for $\psi(x)$ in (\ref{eq:psi_expansion}). It is quite
%interesting that our large-$n$--large-$P$ procedure has arrived to
%the same results as those in \cite{Oyman03}, which build on taking
%$P$ large at the very beginning.

%%%%%%%%%%%%%%%%%%%%%%%%%%% Connection with Oyman %%%%%%%%%%%%%%%%%%%%%%%%%%%%
%In fact, the $P$-asymptotic cgf has been computed by Oyman \cite{Oyman03} using Gamma function as follows:
%\begin{align}\label{Oyman}
%\ln \mathcal{M}(\lambda)=n\lambda \ln \left(\frac{P}{m}\right)+\sum^n_{j=1} \left[\ln\Gamma(m+1-j+\lambda)-\ln \Gamma(m+1-j)\right],
%\end{align}
%through which the cumulants can be computed as
%\begin{align}
%\kappa_{\ell}=\sum_{j=1}^n\psi^{(\ell-1)}(m+1-j).\label{cumu_finite_n}
%%\left[x\psi (x)\right]^{(\ell-1)}\big| ^m_{m-n}.
%\end{align}
%We can see that $\ell$th cumulant is connected to the
%$(\ell-1)$-order derivative of digamma function (i.e. polygamma
%function). Note that expanding (\ref{cumu_finite_n}) in large-$n$
%power series leads to our results (e.g. (\ref{k1largeP}) -
%(\ref{k4largeP})) exactly. This indicates that interchanging the
%order of limits $P\to \infty$ and $n\to \infty$ should not change
%the asymptotic CGF.
%%%%%%%%%%%%%%%%%%%%%%%%%%%%%%%%%%%%%%%%%%%%%%%%%%%%%%%%%%%%%%%%%%%%%%%%

%\subsection{Large Deviation Theory and Refined Tail Distribution}
\section{Characterization with the Edgeworth Expansion}
Armed with closed-form expressions for $C_{i,j}$ in
(\ref{structure}), in this section we draw upon the Edgeworth
expansion technique. This approach allows us to start with a
Gaussian distribution and to systematically correct this by
including higher cumulant effects (i.e., other than the mean and
variance), giving an explicit expression for the corrected PDF.
Moreover, the CDF, which directly defines the outage probability,
can be obtained explicitly through a straightforward integration.

We first write the MGF in the following form
\begin{align}\label{origin}
\mathcal{M}(\lambda)=\exp\left(\sum_{\ell=3}^\infty\frac{\kappa_\ell}{\ell
!}\lambda^\ell\right)\mathcal{M}^{(G)}(\lambda)
\end{align}
where $\mathcal{M}^{(G)}(\lambda)$ represents the MGF of a Gaussian
distribution with mean $\mu$ and variance $\sigma^2$. Note that the
MGF is, in effect, the Laplace transform of a PDF (evaluated along the real axis), thus in the PDF domain (\ref{origin}) is equivalent to
\begin{align}\label{origin_edgeworth_expansion}
p_{\mathcal{I}(\mathbf{x};\mathbf{y})}(t)=\exp
\left(\sum_{\ell=3}^\infty\frac{\kappa_\ell}{\ell
!}\left(-\partial_t\right)^\ell\right)p_{\mathcal{I}(\mathbf{x};\mathbf{y})}^{(G)}(t)
\end{align}
where
$p_{\mathcal{I}(\mathbf{x};\mathbf{y})}^{(G)}(t)=\frac{1}{\sqrt{2\pi}\sigma}e^{-z^2/2},\; z:=(t-\mu)/\sigma$ and $\partial_t:={\rm d}/{\rm d}t$.
Expanding the exponent in (\ref{origin_edgeworth_expansion}) and
collecting terms according to the powers of $\sigma$, the PDF of the mutual information takes the form \cite[Eq.43]{Blinnikov}:
\begin{align}\label{GC_Complete}
p_{\mathcal{I}(\mathbf{x}; \mathbf{y})}(t) \approx \frac{1}{{\sqrt{2\pi}\sigma}}{\rm
e}^{-z^2/2} \left[1+ \mathcal{D}(z) \right]
\end{align}
where
\begin{align} \label{eq:Dx}
\mathcal{D}(z)=\sum_{s=1}^L \sum_{ \{ k \}}
\frac{{\rm He}_{s + 2r}\left( z \right)}{\sigma^{s+2r}}
\prod_{\ell=1}^s \frac{1}{k_\ell!} \left(
\frac{\kappa_{\ell+2}}{(\ell+2)!} \right)^{k_\ell}
%
%&= \sum_{s=1}^\infty \frac{1}{n^s } \sum_{ \{ k \}}
%\frac{1}{\sigma_0^{s+2r}} {\rm He}_{s + 2r}\left( \frac{x -
%n\mu_0}{\sigma_0} \right) \prod_{\ell=1}^s \frac{1}{k_\ell!} \left(
%\frac{C_{\ell+2}}{(\ell+2)!} \right)^{k_\ell}
\end{align}
is the quantity which determines any deviation from Gaussian. Here,
$L$ is a positive integer characterizing how many cumulants are
included in the corrected PDF (i.e., $\kappa_\ell, \ell=3, 4, \ldots, L+2$ are involved).
% (i.e., considering $\kappa_3 \to
%\kappa_{L+2}$ for refinement).
Note that the second summation enumerates all sets $k = \{ k_1, k_2,
\cdots, k_s \}$ containing the non-negative integer solutions of the
Diophantine equation $k_1 + 2 k_2 + \cdots + s k_s = s$. For each of
these solutions, a corresponding constant $r$ is defined as $r = k_1
+ k_2 + \cdots + k_s$. In \cite{Blinnikov}, a practical algorithm
for computing the $\{k\}$ solutions is proposed in general. ${\rm He}_\ell (z)$ is the $\ell$-th Chebyshev-Hermite polynomial,
with the explicit form \cite[Eq. (13)]{Blinnikov}
\begin{align}
{\rm He}_\ell (z) = \ell !  \sum_{k=0}^{\lfloor \ell/2 \rfloor}
\frac{ (-1)^k z^{\ell - 2 k}}{k! (\ell - 2 k)! 2^k} \;
\end{align}
where $\lfloor \cdot \rfloor$ denotes the floor function.
These are generated by differentiating the standard normal
distribution:
\begin{align}\label{hermite_def}
{\rm
He}_\ell\left(z\right)=\left(-1\right)^\ell g_0^{\left(\ell\right)}(z)/g_0(z)
\end{align}
where $g_0(z)=\exp\left(-z^2/2\right)/\sqrt{2\pi}$ and the
superscript $(\ell)$ denotes the $\ell$-th derivative w.r.t. $z$.
Here we list some specific polynomials that we are going to use
subsequently:
\begin{align}
{\rm He}_3(z)=z^3-3z , \quad {\rm He}_4(z)=z^4-6z^2+3 , \quad {\rm
He}_6(z) = z^6 - 15 z^4 + 45 z^2 - 15   \; .
\end{align}
In theory, we can approximate the mutual information distribution
with arbitrary accuracy, by taking $L$ sufficiently large in
$\mathcal{D}(z)$. We first set $L=2$ (i.e., including $\kappa_3$ and
$\kappa_4$):
\begin{align}\label{corrected_Gauss_s_2}
p_{\mathcal{I}(\mathbf{x}; \mathbf{y})}(t) \approx
 \frac{p_{I(\mathbf{x},\mathbf{y})}^{(G)}(t)}{\sigma}
\biggl[1+\frac{\kappa_{3}} {6 \sigma^3} {\rm He}_3\left(z\right) +
\frac{\kappa_{4}}{24 \sigma^4} {\rm He}_4\left(z \right) + \frac{
\kappa_{3}^2 }{72 \sigma^6} {\rm He}_6\left(z \right) \biggr]
\end{align}
and compare the Edgeworth expansion with the Monte Carlo simulations and the Gaussian approximation in
Fig. \ref{Gauss_correction}. Note
that for the Edgeworth expansion curves we only use the leading order terms of the cumulants (i.e., $C_{\ell,0}/n^{2-\ell}$ in (\ref{structure})), since as we
have shown, the leading order terms of the first few cumulants give valid approximations for arbitrary $P$ if $\beta \neq 1$ (c.f. \emph{Remark
\ref{th:pro3}}).
%In contrast to the $\beta=1$ case where the leading
%terms (i.e., $C_{\ell,0}/n^{2-\ell}$) become inaccurate for large SNR,
%the Edgeworth expansion with leading order contribution of cumulants in $\beta \neq 1$ case is
%shown to capture the correct behavior in the whole region of SNR
%(see \cite[Fig. 7]{Chen_McKay}) for comparison).

\begin{figure}[ht]
\centering
\includegraphics[width=0.485\columnwidth]{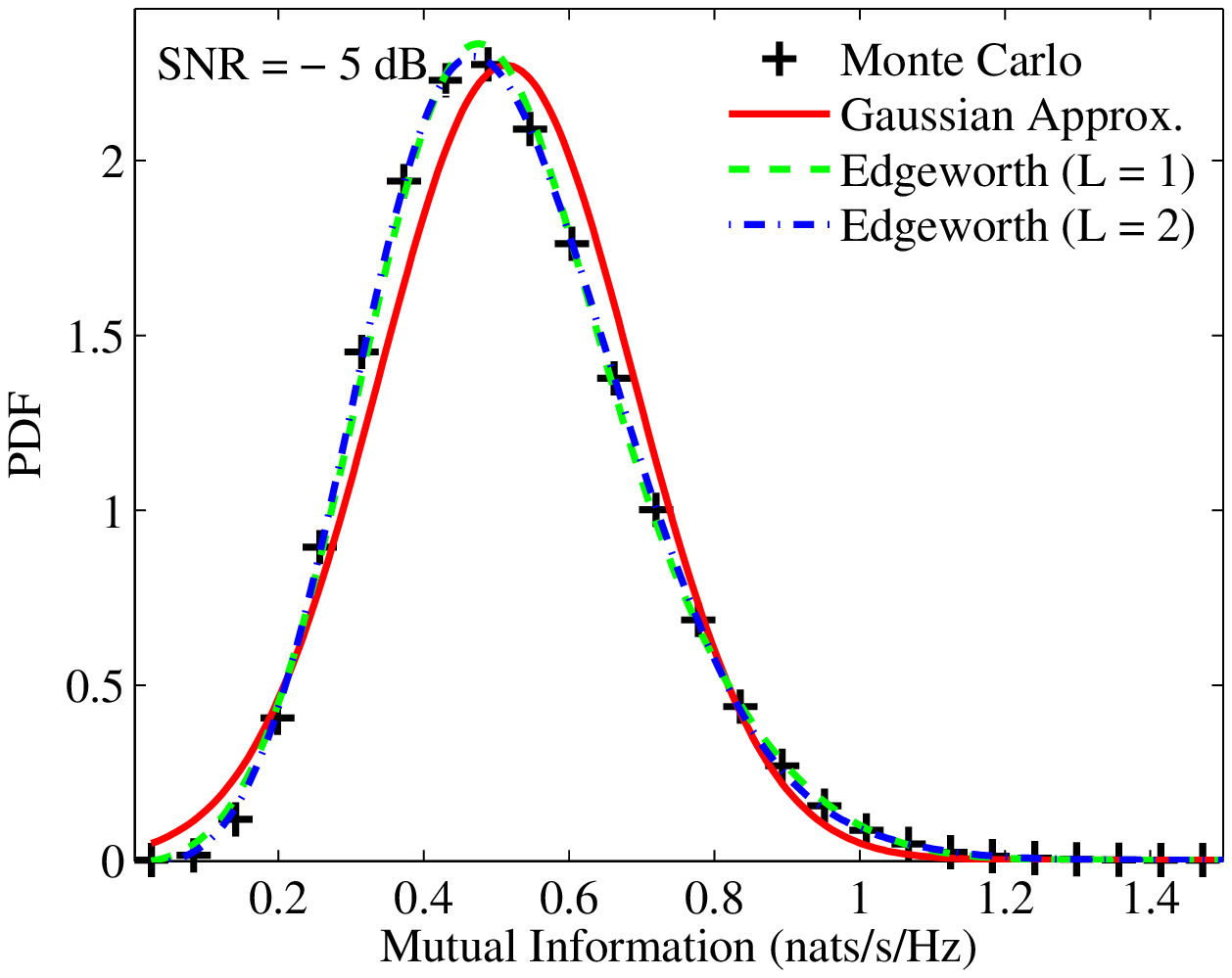}
\includegraphics[width=0.485\columnwidth]{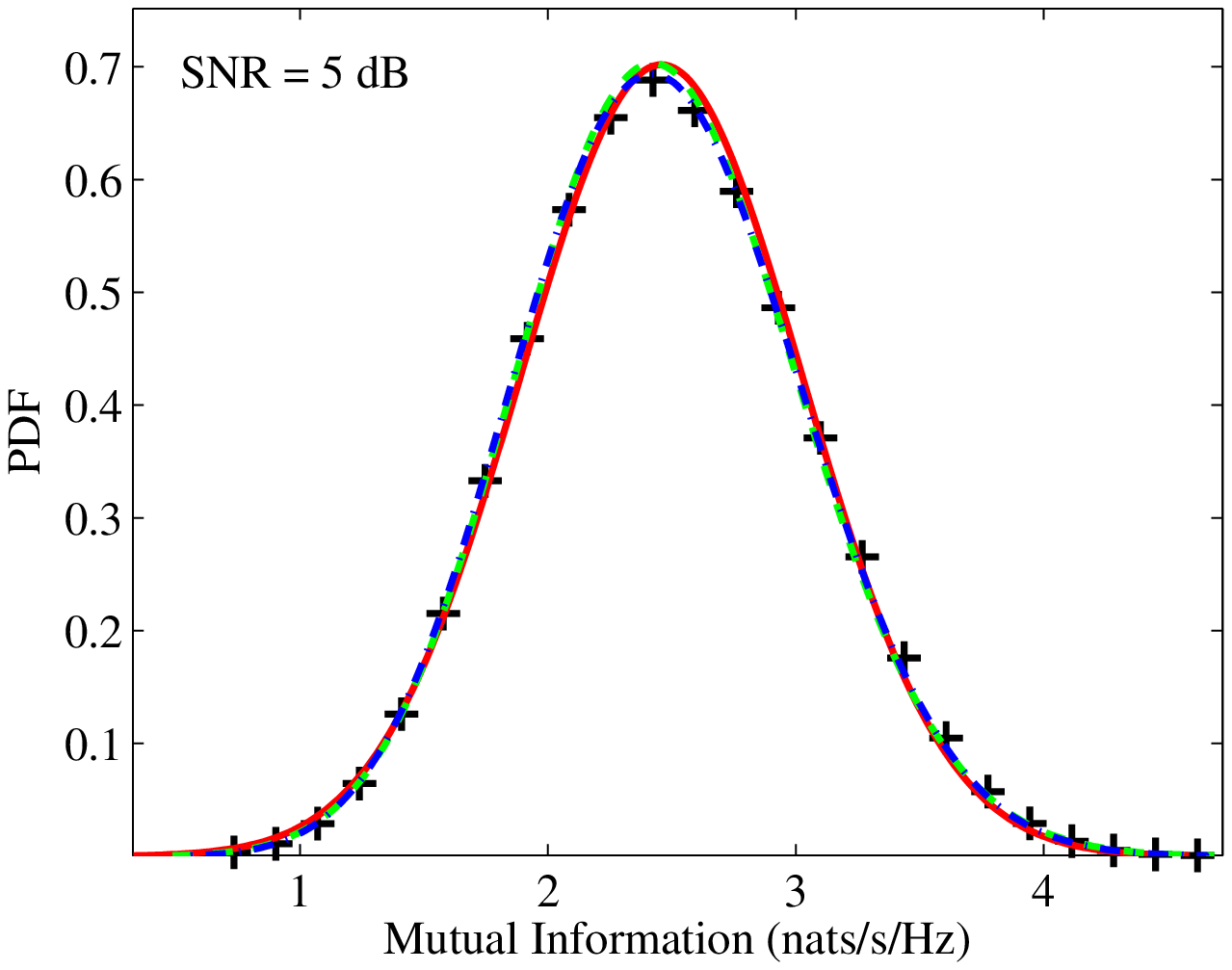}
\includegraphics[width=0.485\columnwidth]{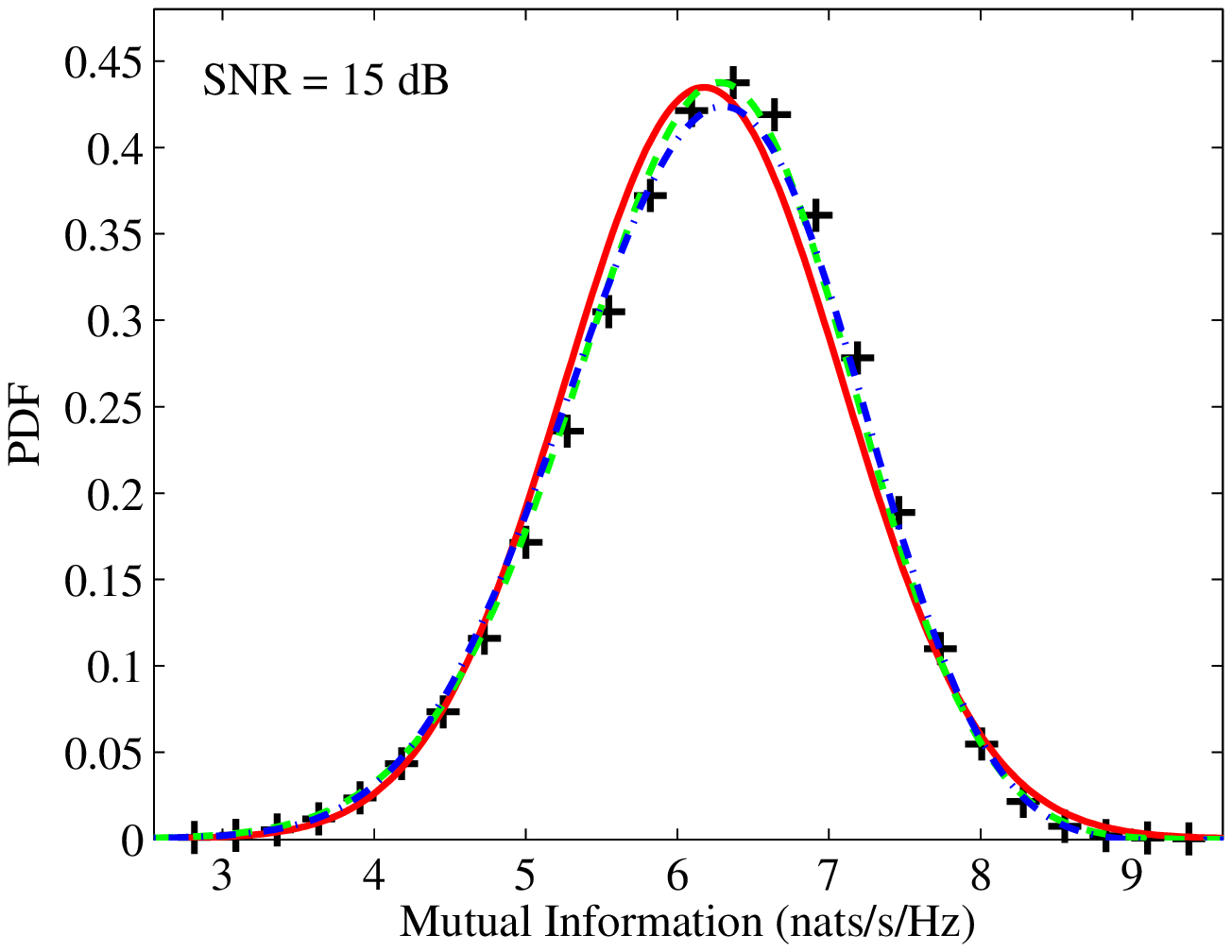}
\includegraphics[width=0.485\columnwidth]{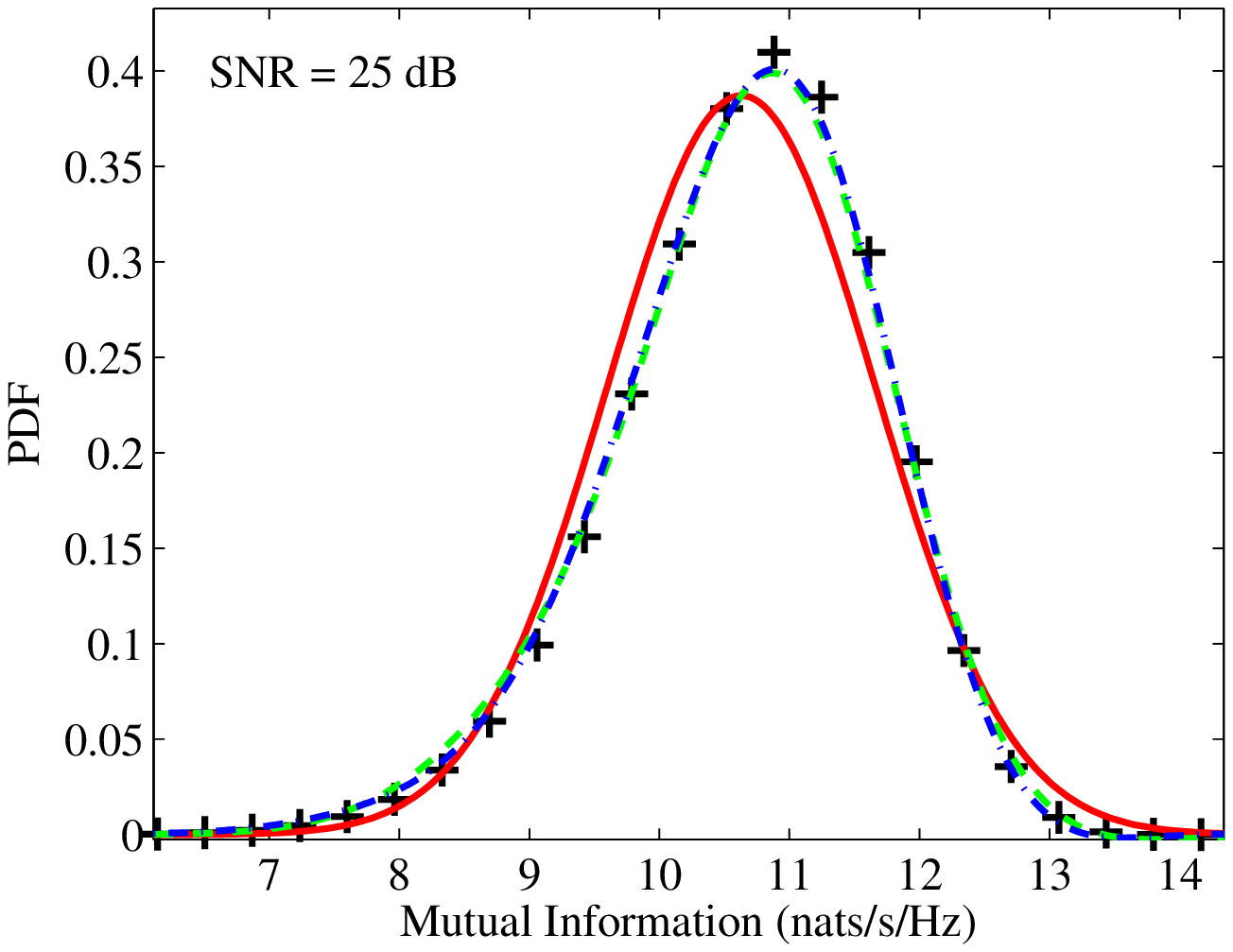}
\caption{PDF of mutual information, comparing the Gaussian
approximation, Edgeworth expansions (with $L=1$ and $L=2$), and
Monte Carlo simulations. Results are shown for $n_t=3, n_r=2$ and
for different SNRs. }\label{Gauss_correction}
\end{figure}
To examine the outage probability, we need to derive the CDF of the mutual information. %Recalling that
%\begin{align}\label{hermite_def}
%{\rm
%He}_n\left(z\right)=\left(-1\right)^ng_0^{-1}g_0^{\left(n\right)}(z)
%\end{align}
%where $g_0(z)=\exp\left(-z^2/2\right)/\sqrt{2\pi}$.
Recalling (\ref{hermite_def}), we
have
\begin{align}
\int_{-\mu/\sigma}^z {\rm He}_\ell\left(u\right)g_0\left(u\right) {\rm d}u\approx\int_{-\infty}^z {\rm He}_\ell\left(u\right)g_0\left(u\right) {\rm d}u
&=-{\rm
He}_{\ell-1}(z)g_0(z).
\end{align}
Therefore, the Edgeworth PDF formula
(\ref{GC_Complete}) can be immediately integrated to give the CDF:
\begin{align}\label{CDF}
F_{\mathcal{I}(\mathbf{x};
\mathbf{y})}(t)\approx 1-Q\left(z\right)-g_0(z)\sum_{s=1}^L \sum_{ \{
k \}} \frac{{\rm He}_{s + 2r-1}\left( z \right)}{\sigma^{s+2r}}
\prod_{\ell=1}^s \frac{1}{k_\ell!} \left(
\frac{\kappa_{\ell+2}}{(\ell+2)!} \right)^{k_\ell}
\end{align}
where $Q(z):=\frac{1}{\sqrt{2\pi}}\int^\infty_z e^{-u^2/2}{\rm d}u$.
Comparisons between the CDF curves computed by the Edgeworth
expansion and the Gaussian approximation are made in Fig.
\ref{CDF_plot}. We see that in the tail region, the Edgeworth
expansions with higher cumulants nicely approach the simulations (in
this case, $L=6$ was required to achieve good accuracy), whilst the
Gaussian curve strongly deviates. This confirms that, by the virtue
of our new cumulant expressions and the Edgeworth expansion, we can
compute the outage probability of {MIMO} systems with high accuracy
in closed-form for arbitrary SNR.
\begin{figure}[ht]
\centering
\includegraphics[width=0.48\columnwidth]{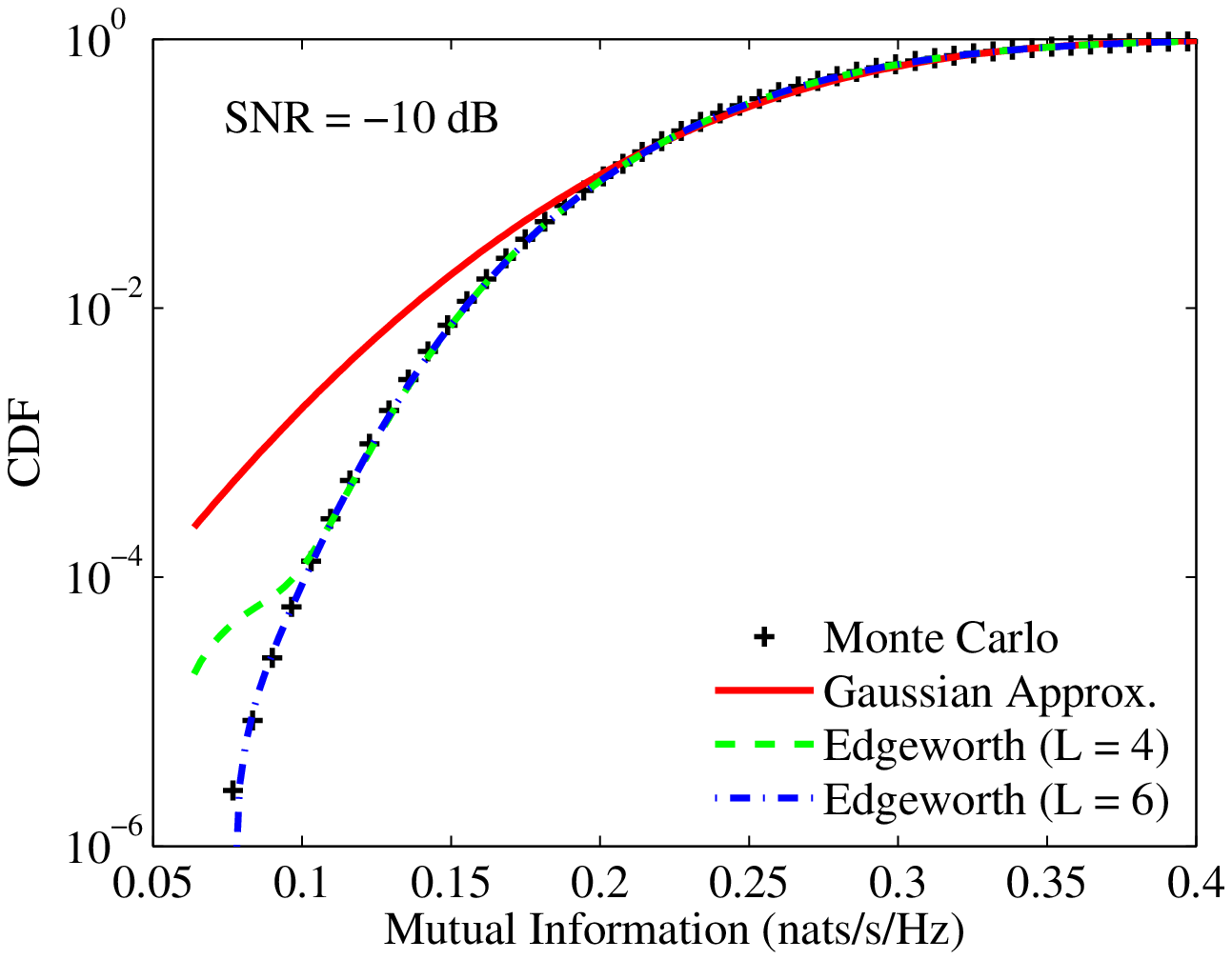}
\includegraphics[width=0.48\columnwidth]{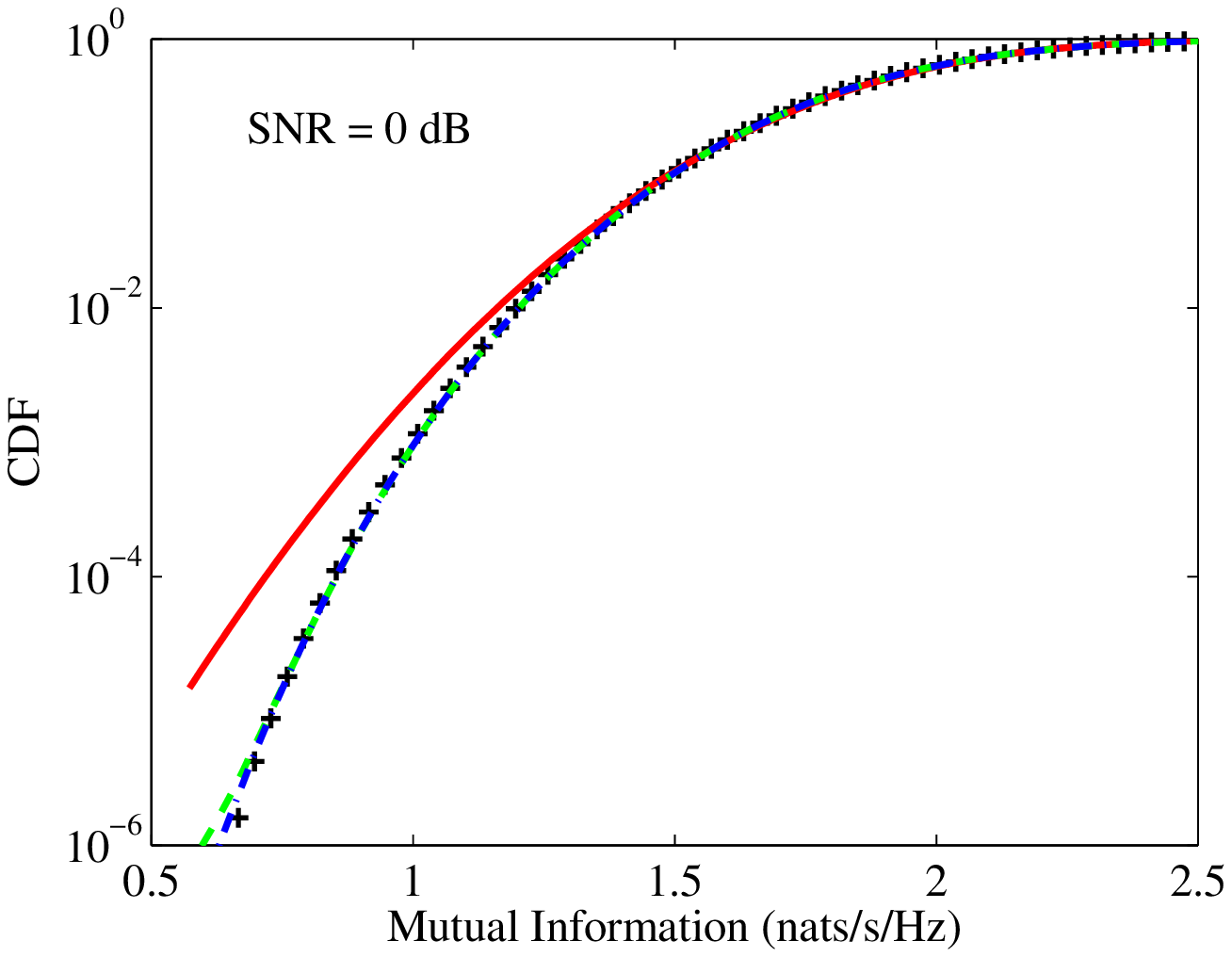}
\includegraphics[width=0.48\columnwidth]{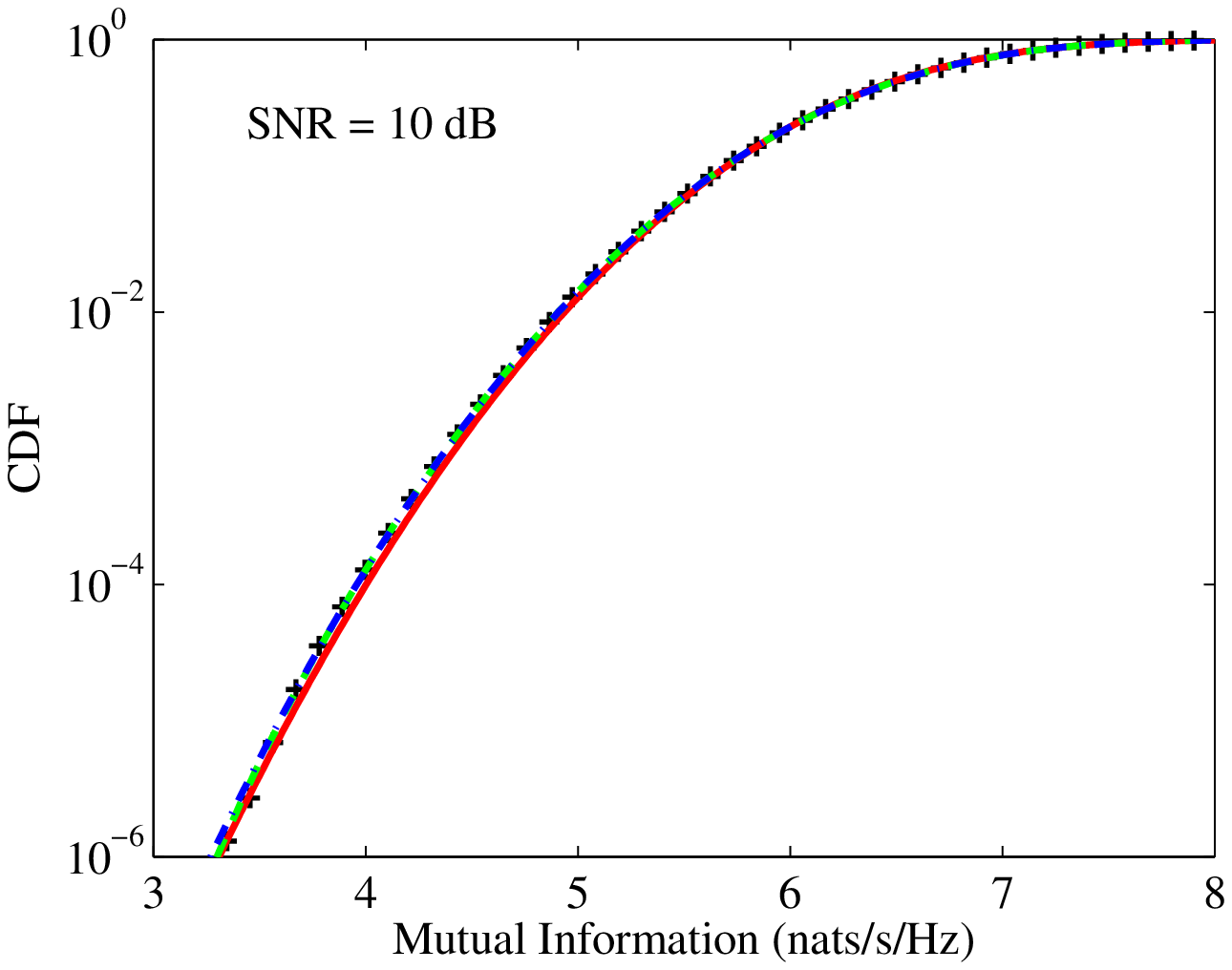}
\includegraphics[width=0.48\columnwidth]{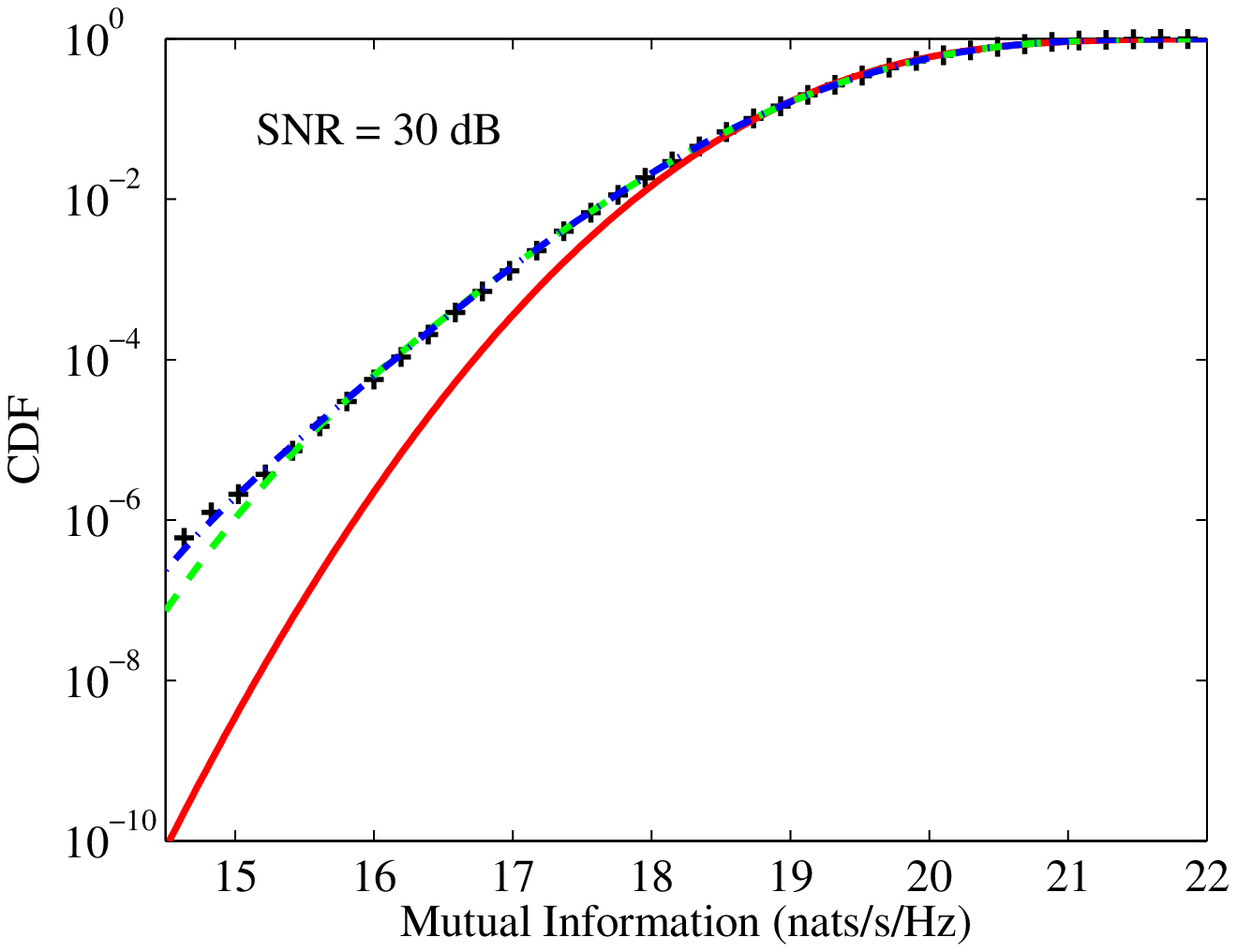}
\caption{CDF of mutual information, comparing the Gaussian
approximation, Edgeworth expansions (with $L=4$ and $L=6$), and
Monte Carlo simulations. Results are shown for $n_t=6, n_r=3$ and
for different SNRs. }\label{CDF_plot}
\end{figure}

Moreover, from these numerical tests, for small SNR (i.e., ${\rm SNR} = -10\, {\rm dB}, 0\, {\rm dB}$), we
see that the Gaussian approximation overestimates the outage
probability in the tail, whilst for large SNR, the outage
probability is underestimated. These tail deviations can be analyzed
as follows. For points far away from the mean (i.e., large $z$), we
approximate
\begin{align}
{\rm He}_{\ell}\left(z\right)\sim z^{\ell},
\end{align}
and the Edgeworth expansion (\ref{eq:Dx}) becomes
\begin{align}
\mathcal{D}(z) &\approx \sum_{s=1}^{L}\sum_{ \{ k \}}
\left(\frac{z}{\sigma}\right)^{s+2r} \prod_{\ell=1}^s
\frac{1}{k_\ell!} \left( \frac{\kappa_{\ell+2}}{ (\ell+2)!}
\right)^{k_\ell}.
\end{align}
For large $z$, the highest order of $(z/\sigma)^{s+2r}$ will
dominate the lower order terms. Since $r$ is maximized for $r=s$
with $\{k_1, k_2, \cdots, k_s\}=\{s,0,\cdots,0\}$, letting $L\to
\infty$,
 $\mathcal{D}(z)$ can be further estimated as
\begin{align}
\mathcal{D}(z) &\approx \sum_{s=1}^\infty
 \frac{1}{s!} \left( \frac{\kappa_{3}}{6
\sigma^{3}}\; z^3\right)^{s}\nonumber\\&=
\exp\left(\frac{\kappa_3}{6\sigma^3}\;z^3\right)-1\;.
\end{align}
With this, the PDF of the mutual information becomes:
\begin{align}\label{tail_refinement}
p_{\mathcal{I}(\mathbf{x}; \mathbf{y})}(t)\approx
\frac{1}{{\sqrt{2\pi}\sigma}}
e^{-\frac{\left(t-\mu\right)^2}{2\sigma^2}+\frac{\kappa_3}{6\sigma^6}\left(t-\mu\right)^3}
\end{align}
which gives a more accurate distribution than the Gaussian in the tails for finite
$n$, since the effect of $\kappa_3$ is considered. The same formula
was presented in \cite{Chen_McKay}, which focused on the case $n_t =
n_r$. Here, we provide more general insights with our new cumulant
expressions which apply for both $n_t = n_r$ and $n_t \neq n_r$.

%{\bf ** Samuel, the discussion below is EXTREMELY unclear.  The
%deviations analysis, for example, can only become clear when you
%identify or recall how the individual cumulants scale with $n$. This
%is ``hidden'' in (74).  Also, the logical flow of the explanations
%in this section is bad.  What is the key message?  The Gaussianity?
%The behavior in the tails?  The high/low SNR analysis? Connecting
%with prior work in Moustakas? Currently everything is mixed up---a
%clear and more logical discussion is needed. **}

First, (\ref{tail_refinement}) indicates that the Gaussian
approximation is accurate under the condition:
\begin{align}
|t-\mu|<<\sqrt[3]{\left|\frac{6\sigma^6}{\kappa_3}\right|}\approx \sqrt[3]{\left|\frac{6\sigma_0^6}{C_{3,0}}\right|}\;n^{1/3}\;.
\end{align}
Thus, the large-$n$ Gaussian approximation is valid for deviations
of $|t-\mu|\sim O(n^{1/3})$ or less from the mean. Outside this
regime, the second term of the exponent in (\ref{tail_refinement})
becomes non-negligible.  In this case, by examining the factor
$\kappa_3 / \sigma^6$ in (\ref{tail_refinement}), we can further
understand how the distribution behaves compared with the Gaussian
approximation. To this end, considering the variance and third
cumulant to leading order in $n$ (i.e., $\sigma_0^2$ given in
(\ref{eq:coulomb_variance}), and $C_{3,0}/n$ with $C_{3,0}$ given in
(\ref{k3leading})), for small $P$,
\begin{align}
\frac{\kappa_3}{\sigma^6}\approx \frac{2\beta}{n\,P^3}\;, \qquad P\to 0\;,
\end{align}
whilst for large $P$,
\begin{align}
\frac{\kappa_3}{\sigma^6}\approx \left\{
\begin{array}{ll}
\frac{2}{n\,\left(\ln P\right)^3}  \;,  \; \; \,  &  \beta = 1 \\
\frac{1}{n\,\beta\left(\beta-1\right)\left(\ln\left(1-\beta^{-1}\right)\right)^3}
\;, \; \; \; \; \; \; \; \, & \beta
>1
\end{array}
\right.\;,\qquad P\to \infty \; . \label{eq:HighSNRResult}
\end{align}
These results are in perfect agreement with \cite[Eq.
(70)]{Moustakas11} obtained via Coulomb fluid arguments, and also
\cite[Eq. (218)]{Chen_McKay} obtained for the special case $\beta =
1$. Therefore, seen from (\ref{tail_refinement}), for small SNR,
$\kappa_3/\sigma^6>0$, and the left tail of the PDF should be always
above the Gaussian approximation (similarly for the CDF in the left
tail). For large SNR and $\beta>1$ on the other hand, $\kappa_3/\sigma^6<0$, and
the situation is the opposite.

Note that the interpretation of the large SNR results above for the
particular case $\beta=1$ should be taken with caution since, as
discussed in Section \ref{sect:largeP} and also in
\cite{Chen_McKay}, the leading order expressions for $\sigma^2$ and
$\kappa_3$ (upon which the arguments are based) become inaccurate in
that scenario, unless $n$ is also very large. For the case $\beta
\neq 1$ however, there is no such problem.

%{\bf **Prof. McKay: I think that at this point you should mention about ``when''
%this expansion is valid, and when other terms come into the picture.
%We did this in the previous paper also. We should also give some
%relevant citations to our previous paper; otherwise the reviewer may
%have the impression that we are copying (the derivation is basically
%the same, and it looks very similar visually) and will not know what
%is new here. Samuel: revised**}
%
%{\bf ** Also, need to be consistent.  Some places you use $\mu$
%and $\sigma^2$, other places you use $\mu$ and $\sigma^2$. Check
%carefully throughout paper for consistency. Samuel: checked**}

\section{Refining the Tail Distribution via the Saddle Point Method}\label{sect:SPA}
Whilst the Edgeworth expansion technique provides an accurate
closed-form characterization of the mutual information distribution,
it becomes unwieldy when too many cumulants are needed for obtaining
the desired accuracy. Particularly, in the case when one is
interested in the tail region of $O(n)$ away from the mean (i.e.,
the ``large deviation'' region discussed in \cite{Moustakas11}), we
need to consider the effect of \emph{all} cumulants to obtain high
accuracy. Therefore, to supplement the Edgeworth expansion results,
in this section we will draw upon the saddle point method and the
cumulant results from Section \ref{sect:cumulants} to further
investigate the large deviation scenario.

\subsection{Saddle Point Method}\label{sec:SPA}
Assuming that the MGF (equivalently, the CGF) is known, the PDF of the
mutual information can be derived through the inversion formula:
\begin{align}\label{eq:inverse_Fourier}
p_{\bf \mathcal{I}(x;y)}(t)=\frac{1}{2\pi i}\int_{\tau-i\infty}^{\tau+i\infty} e^{ \mathcal{K}(\lambda)-\lambda t} {\rm d}\lambda ,
\end{align}
where $\mathcal{K}(\lambda)$ is the CGF and $\tau$ is a real number
defining the integration path. In most cases, it is infeasible to
evaluate (\ref{eq:inverse_Fourier}) in closed-form. However, noting
that $\mathcal{K}(\lambda)-\lambda t \sim O(n)$ (given $\mu\sim
O(n)$, thus
$\mathcal{K}(\lambda)=\mu\lambda+\frac{\sigma^2}{2}\lambda^2+\cdots\sim
O(n)$ and $t=\mu+O(n)\sigma\sim O(n)$), if $n$ is large, the saddle
point method in \cite{Daniels54} can be used to provide an accurate
approximation for this integral. This is done by choosing the path
of integration to pass through a saddle point $\lambda^\star$ such
that the integrand is negligible outside the neighborhood of this
point. More specifically, for a given $t$, $\lambda^\star$ is
computed as the real-valued root of
\begin{align}\label{eq:saddlepoint}
t=\mathcal{K}'(\lambda)\;,
\end{align}
where $'$ denotes derivative w.r.t. $\lambda$. Defining the
so-called \emph{rate function},
\begin{align}\label{eq:I}
I(t):=t\lambda^\star(t)-\mathcal{K}(\lambda^\star(t)) \; ,
\end{align}
the PDF to leading order in $n$ admits \cite{Daniels54}
\begin{align}\label{SPA_PDF}
p_{\bf \mathcal{I}(x;y)}(t)\approx\frac{e^{-I(t)}}{\sqrt{2\pi \mathcal{K}''(\lambda^\star(t))}}\;.
\end{align}

By the virtue of the saddle point equation (\ref{eq:saddlepoint}),
we can examine how the large-$n$ behavior of the mutual information
varies in different regimes of $|t-\mu|\sim O(n^\epsilon),
0\leq\epsilon\leq 1$. Substituting the cumulant power series
expansion of the CGF (\ref{CGF_expansion}) into both $I(t)$ and
(\ref{eq:saddlepoint}) gives
\begin{align}\label{eq:how_ka_involved_1}
I(t)=(t-\mu)\lambda^\star(t)-\frac{\sigma^2}{2}(\lambda^\star(t))^2-\frac{\kappa_3}{3!}(\lambda^\star(t))^3+O((\lambda^\star(t))^4),
\end{align}
with $\lambda^\star(t)$ the solution to the saddle point equation
\begin{align}\label{eq:how_ka_involved_2}
t=\mu+\sigma^2 \lambda^\star(t) +\kappa_3 \frac{
(\lambda^\star(t))^2}{2}+\kappa_4\frac{ (\lambda^\star(t))^3}{3!}+O(
(\lambda^\star(t))^4).
\end{align}
Based on (\ref{eq:how_ka_involved_1}) and (\ref{eq:how_ka_involved_2}), we draw the following remarks:
\begin{itemize}
\item In the region $|t-\mu|\sim O(1)$, as $n\to \infty$, recalling that $\mu = O(n)$,
$\sigma^2 = O(1)$, whilst $\kappa_\ell = O(n^{2-\ell})$ for $\ell \geq 3$, (\ref{eq:how_ka_involved_2}) results in
$\lambda^\star(t) \approx (t-\mu)/\sigma^2$ which yields the
Gaussian exponent $I(t) \approx (t-\mu)^2/(2\sigma^2)$ in
(\ref{eq:how_ka_involved_1}). We see that the higher cumulants
(e.g., $\kappa_3, \kappa_4,$ etc.) vanish for large $n$ in
(\ref{eq:how_ka_involved_2}) and (\ref{eq:how_ka_involved_1}). This
is the region where, as argued in \cite{Moustakas11}, the central
limit theorem is valid, and thus the Gaussian approximation is
asymptotically accurate.

\item Looking at further (sub-linear) deviations from the mean, in the region $|t-\mu| \sim O(n^\epsilon), \epsilon<1$, (\ref{eq:how_ka_involved_2}) generates
the same saddle point $\lambda^\star(t)=(t-\mu)/\sigma^2$ as $n\to
\infty$. However, since $\lambda^\star(t) \sim O(n^\epsilon)$, in
this case some of the terms in (\ref{eq:how_ka_involved_1})
involving the higher cumulants ($\kappa_\ell$, $\ell \geq 3$) do
\emph{not} vanish. More specifically, this includes all cumulants
$\kappa_\ell$ for which $(\ell-2)/\ell  \leq \epsilon$.  For
example, for large $n$, $\kappa_3$ becomes effective (provides a
non-negligible contribution) for deviations of $O(n^{1/3})$ or more,
$\kappa_4$ is effective for deviations of $O(n^{1/2})$ or more,
$\kappa_5$ is effective for deviations of $O(n^{3/5})$ or more, and
so on. This behavior is consistent with the discussion in the
previous section and in \cite{Chen_McKay}. In these scenarios, our
Edgeworth expansion results presented in the previous section
provide an accurate closed-form approximation, by accounting for a
fixed number of higher cumulant effects. However, when the
deviations become stronger (e.g., as $\epsilon$ increases towards
$1$), more and more terms are required in the Edgeworth expansion in
order to account for the increasing number of non-negligible high
order cumulants, thereby significantly increasing the computational
complexity.

\item Looking at even further (linear) deviations from the mean, in the region $|t-\mu| \sim O(n)$, as $n\to \infty$, (\ref{eq:how_ka_involved_2})
indicates $\lambda^\star(t) \sim O(n)$. In this case, \emph{all}
cumulants to leading order in $n$ (i.e., an infinite number) contribute in
(\ref{eq:how_ka_involved_2}) and (\ref{eq:how_ka_involved_1}). In
addition, we find that $I(t)\sim O(n^2)$. In this ``large
deviation'' region which is sufficiently deep in the tails of the
distribution, the Gaussian approximation strongly misses the correct
behavior, whilst the Edgeworth expansion in (\ref{CDF}) also becomes
intractable.
%Therefore, we require an alternative method for
%capturing the correct behavior in this region.  This will be
%considered below.
\end{itemize}
Importantly, the above discussions provide a unified picture of the
mutual information distribution for large-$n$. Specifically, they
show that the well known Gaussian approximation, our Edgeworth
expansion approximation, and the large deviation results (also
considered in \cite{Moustakas11}) have their own region of validity,
depending on how far one looks into the tail of the distribution as
$n$ increases. Whilst the Gaussian approximation and the Edgeworth
expansion have been well characterized in the previous sections,
further work is required in relation to the large deviation region,
which is now considered. First, we note that a key problem
encountered with (\ref{SPA_PDF}) is how to compute $I(t)$.  In
addition, whilst in principle the CDF of the mutual information can
be obtained by integrating the PDF expression (\ref{SPA_PDF}) (i.e., $F_{\mathcal{I}\bf (x;y)}(t)=\int_0^t p_{\mathcal{I}\bf (x;y)}(u){\rm d}u$), this
is difficult and generally does not yield a closed-form expression.
We will first address this integration problem, then deal with the
problem of computing the rate function $I(t)$.

%Fig. \ref{fig:PDF_SPA} indicates the accuracy of the Saddle point approximated pdf (\ref{SPA_PDF}).
%\begin{figure}[ht]
%\centering
%\subfigure[$n_t=6,\;n_r=2, \; P = 30\; {\rm dB}$]{
%\includegraphics[width=0.48\columnwidth]{PDF_SPA_6_2_30dB.eps}}
%\caption{The PDF computed by Saddle point approximation (\ref{SPA_PDF}).}\label{fig:PDF_SPA}
%\end{figure}

In order to integrate (\ref{SPA_PDF}), we can derive an asymptotic
expression for large $n$ via Laplace's method (see e.g.,
\cite[Chapter 2]{Avramidi00}). Recall that, as previously described,
$I(u)\sim O(n^2)$, whilst $\mathcal{K}''(\lambda^\star)\sim O(1)$.
Since $-I(u)$ is an increasing function for $u\leq \mu$ (decreasing
function for $u>\mu$), to leading order in $n$, the integration of (\ref{SPA_PDF}) is dominated by the region in the
neighborhood of the upper limit $t$ for $u\leq \mu$ (lower limit $t$
for $u>\mu$). Thus we expand $I(u)$ about $t$ as
\begin{align}
I(u)=I(t)+(u-t)I'(t)+\frac{(u-t)^2}{2}I''(t)+O((u-t)^3).
\end{align}
Additionally, in the neighborhood of $t$, the function
$\mathcal{K}''(\lambda^\star(u))$ is nearly constant and can be
approximated by its value at $t$. Based on these arguments (see
\cite[Chapter 2]{Avramidi00} for more details), we obtain
\begin{align}\label{eq:PDF2CDF_1}
\int_0^t \!\!\!\frac{e^{-I(u)}}{\sqrt{2\pi \mathcal{K}''(\lambda^\star (u))}} {\rm d}u \approx \left\{
\begin{array}{ll}
\frac{e^{-I(t)}}{\sqrt{2\pi \mathcal{K}''(\lambda^\star (t))}}\int_0^t e^{-I'(t)(u-t)-I''(t)(u-t)^2/2+O((u-t)^3)} \;{\rm d}u,\quad t<\mu\\
1-\frac{e^{-I(t)}}{\sqrt{2\pi \mathcal{K}''(\lambda^\star (t))}}\int_t^\infty e^{-I'(t)(u-t)-I''(t)(u-t)^2/2+O((u-t)^3)} \;{\rm d}u,\quad t>\mu
\end{array}
\right.\;.
\end{align}
By neglecting the terms in the exponent of order $(u-t)^2$ or
higher, we have
\begin{align}\label{eq:laplace_1}
F_{\mathcal{I}({\bf x;y})}(t)&\approx \left\{
\begin{array}{ll}
 -\frac{e^{-I(t)}}{I'(t)\sqrt{2\pi \mathcal{K}''(\lambda^\star (t))}}\;, \quad &t\leq \mu  \\
1+\frac{e^{-I(t)}}{I'(t)\sqrt{2\pi \mathcal{K}''(\lambda^\star (t))}}\;,\quad &t> \mu
\end{array}
\right.\;.
\end{align}
If one were to instead neglect the terms of order $(u-t)^3$ or
higher in (\ref{eq:PDF2CDF_1}), then the (less) asymptotic formula
proposed in \cite[Eq. (63)--(64)]{Moustakas11} results. However, it
is important to point out that (\ref{eq:laplace_1}) and \cite[Eq.
(63)--(64)]{Moustakas11} each involve the derivatives of $I(t)$
(e.g., $I'(t)$ and $I''(t)$), which are difficult to handle, both
analytically and computationally. Thus, it is of interest to derive
a more manageable representation, which is now pursued.

Recall that the Gaussian approximation is accurate around the bulk
of the distribution (which captures the vast majority of the area
under the PDF curve), whilst the absolute difference between the
true PDF and the Gaussian approximation is typically small outside
of the bulk (since the PDF naturally takes extremely small values in
this region).  Thus, the Gaussian exponent (i.e., $(t-\mu)^2/(2
\sigma^2 )$) should capture the leading effect in $I(t)$. As such,
we expand the integrand around the mean (equivalently, around
$\lambda^\star(t)=0$), giving:
\begin{align}\label{eq:Laplace_integral}
F_{\bf \mathcal{I}(x;y)}(t)\approx\int_0^t \frac{1}{\sqrt{2\pi \mathcal{K}''(0)}}e^{-\frac{(u-\mu)^2}{2\sigma^2}-O((u-\mu)^3)}{\rm d}u=\frac{1}{\sqrt{2\pi}\sigma}\int_0^t e^{-\widetilde{I}(u)}e^{-\frac{(u-\mu)^2}{2\sigma^2}}{\rm d}u
\end{align}
where
\begin{align} \label{eq:ITilde}
\widetilde{I}(u):=I(u)- \frac{ (u-\mu)^2} {2 \sigma^2 } \; \; .
\end{align}
Noting that $|\widetilde{I}(u)| \ll (u-\mu)^2/(2 \sigma^2)$, we adopt
Laplace's expansion method again as follows. Since
$(u-\mu)^2/\sigma^2$ is a decreasing function for $u\leq \mu$
(increasing function for $u>\mu$), to leading order in $n$, the
integral (\ref{eq:Laplace_integral}) is dominated by the region in
the neighborhood of the upper limit $t$ for $u\leq \mu$ (lower limit
$t$ for $u>\mu$). In this neighborhood, the function
$\widetilde{I}(u)$ can be regarded as effectively a constant and can
be approximated by its value at $t$, which leads to the following:
\begin{align}
&\frac{1}{\sqrt{2\pi}\sigma}\int_0^t e^{-\widetilde{I}(u)}e^{-\frac{(u-\mu)^2}{2\sigma^2}}{\rm d}u \approx\left\{
\begin{array}{ll}
e^{-\widetilde{I}(t)}\left[\frac{1}{\sqrt{2\pi}\sigma}\int^t_0e^{-\frac{(u-\mu)^2}{2\sigma^2}}{\rm d}u\right], \quad t\leq \mu\\ 1-e^{-\widetilde{I}(t)}\left[\frac{1}{\sqrt{2\pi}\sigma}\int^\infty_te^{-\frac{(u-\mu)^2}{2\sigma^2}}{\rm d}u\right], \quad t > \mu
\end{array}
\right.\;.
\end{align}
Thus the CDF of the mutual information is represented by the concise
formula
\begin{align}\label{eq:CDF_SPA}
F_{\mathcal{I}({\bf x;y})}(t)&\approx \left\{
\begin{array}{ll}
 {Q}\left(- \frac{t-\mu}{\sigma} \right)e^{-\widetilde{I}(t)}, \quad &t\leq \mu  \\
1-{Q}\left( \frac{t-\mu}{\sigma} \right)e^{-\widetilde{I}(t)},\quad
&t> \mu
\end{array}
\right. \; \; .
\end{align}
In the next subsection, we will show that this asymptotic CDF
captures the distributional behavior very accurately.
%\end{itemize}
%\begin{align}
% \frac{{Q}\left(\frac{\rvert I(x)'\lvert}{\sqrt{I(x)''}}\right)}{\sqrt{I(x)''}\sigma}e^{-I(x)+\frac{I(x)'^2}{2I(x)''}}, \quad x\leq \mu_{\rm Coulomb} \\
%&\approx 1-\frac{{Q}\left(\frac{\rvert I(x)'\lvert}{\sqrt{I(x)''}}\right)}{\sqrt{I(x)''}\sigma}e^{-I(x)+\frac{I(x)'^2}{2I(x)''}}, \quad x> \mu_{\rm Coulomb}
%\end{align}
%\begin{align}
%\Pr(\mathcal{I}({\bf x;y})<x) &\approx \frac{1}{\sqrt{2\pi}\sigma \lvert I(x)'\rvert}e^{-I(x)}\label{Qapprox}\\&\approx \frac{\sigma}{\sqrt{2\pi}\lvert x-\mu \rvert}e^{-I(x)}\label{Gauapprox}
%\end{align}

Now we address the remaining challenge: computing a manageable
expression for $I(t)$, and therefore $\widetilde{I}(t)$ via
(\ref{eq:ITilde}). Recalling the large-$n$ expansion structure in
(\ref{structure}), summing up the leading terms of the cumulants
(i.e., $C_{\ell,0}/n^{2-\ell}, \ell=1, 2, \ldots$) gives us the
$n$-asymptotic CGF. However, in general, the complexity of the
expressions for the $C_{\ell,0}$'s makes it difficult to derive a
generic closed-form formula for this asymptotic CGF. Thus, to make
solid analytical progress, we focus on the scenarios of small-$P$
and large-$P$.

%In this section, we aim to obtain the asymptotic CGF, and further employ the large deviation theory to compute the tail distribution, which is a complement to the Gaussian approximation.
\subsection{Asymptotic CGF at Low SNR}
We first consider the $n$-asymptotic cumulants for small-$P$. In
this case, (\ref{series_k1})--(\ref{series_k5}) indicate the generic
cumulant expression
\begin{align}
\kappa_\ell \sim
{\left(\ell-1\right)!}\;mn\left(\frac{P}{m}\right)^\ell\;,\quad \ell=1, 2, \ldots
\end{align}
Here $\kappa_1 = \mu$ and $\kappa_2 = \sigma^2$. Thus the CGF is
obtained as
\begin{align}\label{Small_P_CGF}
\mathcal{K}\left(\lambda\right)\sim&\sum_{\ell=1}^\infty
\frac{mn}{\ell}\left(\frac{P}{m}\lambda\right)^\ell
=-mn\ln\left(1-\frac{P}{m}\lambda\right).
\end{align}
In fact, we can also recover (\ref{Small_P_CGF}) by scaling and
solving the exact equation for the $n$-asymptotic CGF in
(\ref{leading_Equation}), where the derivatives are taken w.r.t.
$x$. To this end, knowing that $Y(x)=\lambda y_1(x)+\lambda^2 y_2(x)+\cdots$, and $y_\ell(x)\sim O(x^{-\ell})$ (indicated by the small $P$ expansion results in Section \ref{sec:smallPexpansion}), in order to keep the terms to leading order in $x$
of $Y(x)$, we introduce the following variable substitution:
%{\bf ** You need to be careful with notation in this section.  Here,
%$x$ means something very different than it does in the saddlepoint
%formula.  I suggest changing the saddlepoint variable to a new
%variable which has not been used (perhaps $t$ or $u$---but check).
%**}
\begin{align}\label{eq:substitution}
\frac{n \lambda}{x}:=y\;.
\end{align}
With this, equation (\ref{leading_Equation}) becomes
\begin{align}
\left[-y\left(1+\frac{\beta+1}{x}+y\right)Y'-Y\right]^2=4y\left(y Y'
+ Y-\beta\right)\left(-\frac{y}{x^2} Y'+y\right)Y'
\end{align}
where $'$ is the derivative w.r.t. $y$. Note that here we consider the large $x$ (equivalently, small $P$) but finite $y$ scenario. Letting $x\to \infty$, we have
\begin{align}
\left[Y'\left(y^2+y\right)+Y\right]^2=4y^2\left(yY'+Y-\beta\right)Y'
\end{align}
with the solution
\begin{align}
Y(y)=\frac{\beta y}{y-1}\;.
\end{align}
Integrating $Y(y)$, we obtain the large-$n$--small-$P$ CGF:
\begin{align}\label{Integration}
\mathcal{K}(\lambda)&\sim n^2\int_\infty^{\beta/P}\frac{Y(\frac{\lambda}{nx})}{x}{\rm d}x\nonumber\\
&=-mn\ln \left(1-\frac{P}{m}\lambda\right), \quad \lambda <\frac{m}{P}
\end{align}
in agreement with (\ref{Small_P_CGF}), which was obtained by summing
the cumulants. Note that the CGF in (\ref{Integration}) corresponds
to that of a chi-square distribution, indicating that in this
low-$P$ scenario:
\begin{align}\label{Eq:Small_P_Chisqure}
\mathcal{I}({\bf x;y}) \overset{d}{\sim} \frac{P}{2m} \chi^2\left(2mn\right).
\end{align}
This approximation is in fact quite well known, and it can be
readily established by noting that $\ln \det ({\bf
I}_n+\frac{P}{m}{\bf HH^\dag})\approx \frac{P}{m}{\rm tr}({\bf
HH^\dag})$ for small $P$ (i.e., obtained by expanding $\ln \det
({\bf I}_n+\frac{P}{m}{\bf HH^\dag})$ around $P=0$, and keeping the
first term), and the obvious fact that ${\rm tr}({\bf
HH^\dag})\overset{d}{\sim} \chi^2(2mn)/2$.

As depicted in Fig. \ref{CGF_Chi_Square}, if $P$ is very small
(e.g., $P=-5 \;{\rm dB}$), then the chi-square approximation lines
up quite well with the simulations; however beyond this very small
regime (e.g., for $P=5 \;{\rm dB}$), it is inaccurate. Thus, for
greater validity, further refinement beyond the leading-order
chi-square approximation is necessary.
\begin{figure}[ht]
\centering
\subfigure
%[$n_t = 4,\; n_r = 2$]
{
\includegraphics[width=0.48\columnwidth]{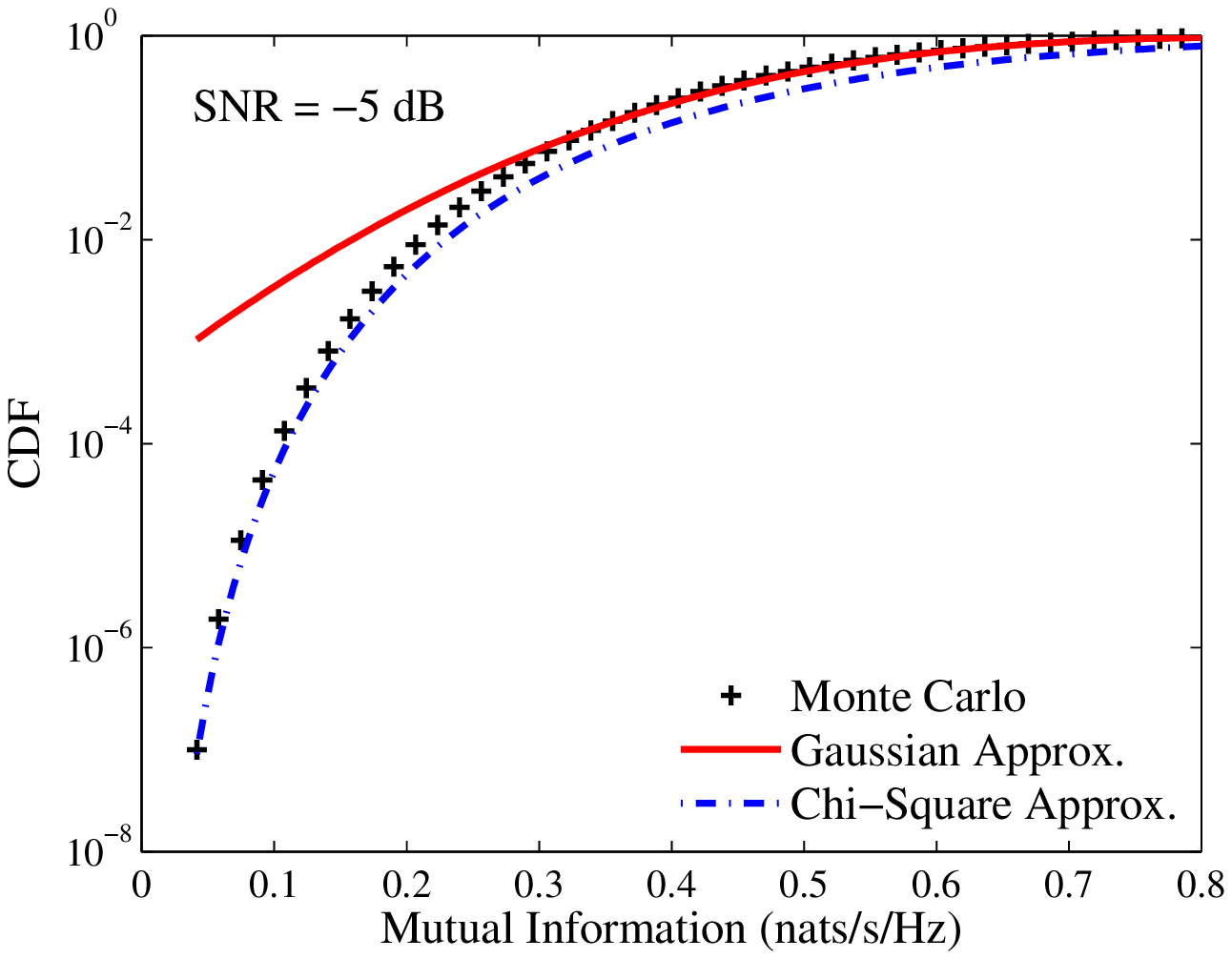}}
\subfigure
%[$n_t = 4, \;n_r = 2$]
{\includegraphics[width=0.48\columnwidth]{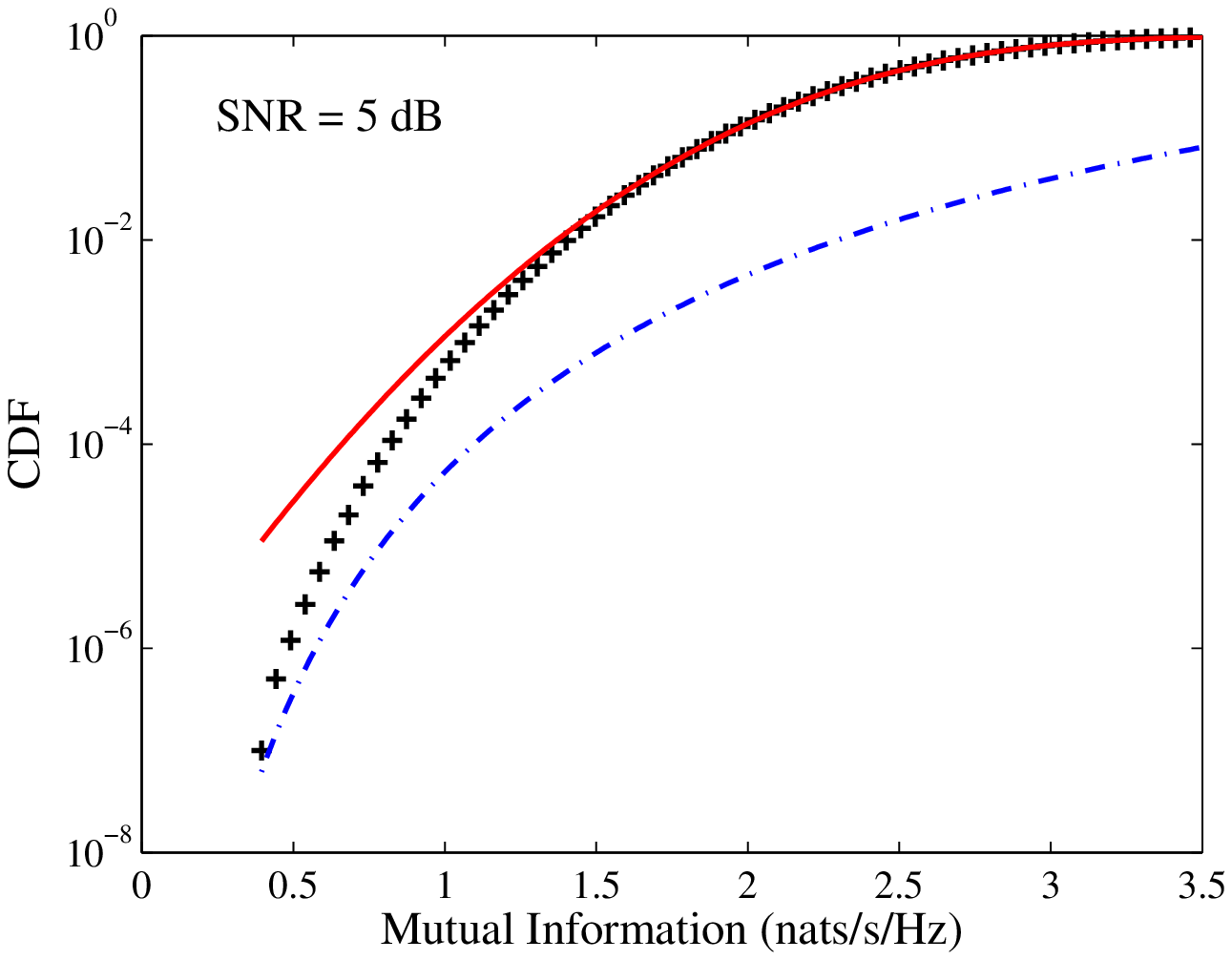}}
\caption{CDF of mutual information for small SNRs, comparing the
Gaussian approximation, chi-square approximation, and Monte Carlo
simulations.  Results are shown for $n_t = 4$, $n_r = 2$.
%It is shown
%that Gaussian approximation fails to capture the tail distribution
%in the small SNR regime. In contrast, the chi-square approximation
%lines up with the numerical results accurately for very small SNR,
%but eventually deviates as SNR increases.
}\label{CGF_Chi_Square}
\end{figure}
This requires computing the higher order correction terms (in $P$),
a task which appears difficult via direct expansion of the $\ln
\det ({\bf I}_n+\frac{P}{m}{\bf HH^\dag})$ formula, as indicated
above for the leading order chi-square approximation.  To our
knowledge, such refinement has not been computed thus far.

To develop a systematic method for solving this refinement problem,
we may once again make use of our Painlev\'{e} representation.
Noting that $x$ is large whilst the new variable $y$ is finite, we assume the following large-$x$ expansion:
%{\bf ** $x$ and $y$'s in the equation below. Does this make
%sense?  It is confusing to me; you need to clarify what you are
%doing here. ** \\ Samuel: We consider the large $x$ but finite $y$ scenario, which is mentioned before Eq. (95).}
\begin{align}\label{eq:expansion}
Y(y)=Y_0(y)+\frac{Y_1(y)}{x}+\frac{Y_2(y)}{x^2}+\cdots \;.
\end{align}
Substituting (\ref{eq:expansion}) into (\ref{leading_Equation}) and matching the
coefficient of $x^{-k}$, we can compute $Y_k(y)$ systematically in
closed-form. For example,
\begin{align}
Y_1(y)=-\frac{\beta (\beta+1)y}{(y-1)^3}\;.
\end{align}
Integrating $Y_1(y)$ through (\ref{Integration}), we obtain the CGF with
first-order correction term (in $P$):
\begin{align}\label{Small_P_CGF_corrected}
\mathcal{K}(\lambda)\approx -mn\ln \left(1-\frac{P}{m}\lambda\right)-\frac{n}{2}\frac{(1+1/\beta)\lambda P^2}{\left(\lambda P/m-1\right)^2}\;.
\end{align}
With this, the saddle point (\ref{eq:saddlepoint}) is obtained as
\begin{align}\label{equality_Small_P}
t
=\frac{mnP\left[-2\lambda^{\star 2}P^2+(nP^2+mP^2+4mP)\lambda^\star-2m^2+mnP+m^2P\right]}{2\left(\lambda^\star P-m\right)^3}\;.
\end{align}
This expression can be solved in closed-form, with the resulting
expression involving a cubic equation.  Alternatively, one may
trivially compute the solution numerically for any given value of
$t$. With $\lambda^\star$ solved for a given $t$, the value of the
rate function $I(t)$ and thus $\widetilde{I}(t)$ follow immediately
according to the definitions (\ref{eq:I}) and (\ref{eq:ITilde})
respectively (with $\mu$ and $\sigma^2$ in (\ref{eq:ITilde})
approximated via $\mu_0$ and $\sigma_0^2$). By invoking the CDF
formula (\ref{eq:CDF_SPA}), we can then compute the saddle point
approximation for the mutual information distribution. This
approximation is illustrated in Fig. \ref{fig:small_P_2_2}.
% and
%Fig. \ref{fig:small_P_4_2}.
Compared with the chi-square approximation, it is shown that this
refined CDF %(\ref{eq:CDF_SPA}) with $\widetilde{I}(x)$ (or $I(x)$)
%computed from (\ref{Small_P_CGF_corrected}) and
%(\ref{equality_Small_P})
is remarkably accurate, even for moderate $P$ values (e.g.,
$P=10\,{\rm dB}$). As $P$ further increases (beyond, for example,
$P=15 {\rm dB}$), we have found that the saddle point approximation
starts to miss the correct behavior, and higher order correction
terms (equivalently, $Y_2(y), Y_3(y), \ldots$) are needed.  These
can be systematically computed using the same procedure as before.
%was employed for $Y_1(y)$.

\begin{figure}[ht]
\centering
\subfigure[$n_t=2,\;n_r=2$]{
\includegraphics[width=0.48\columnwidth]{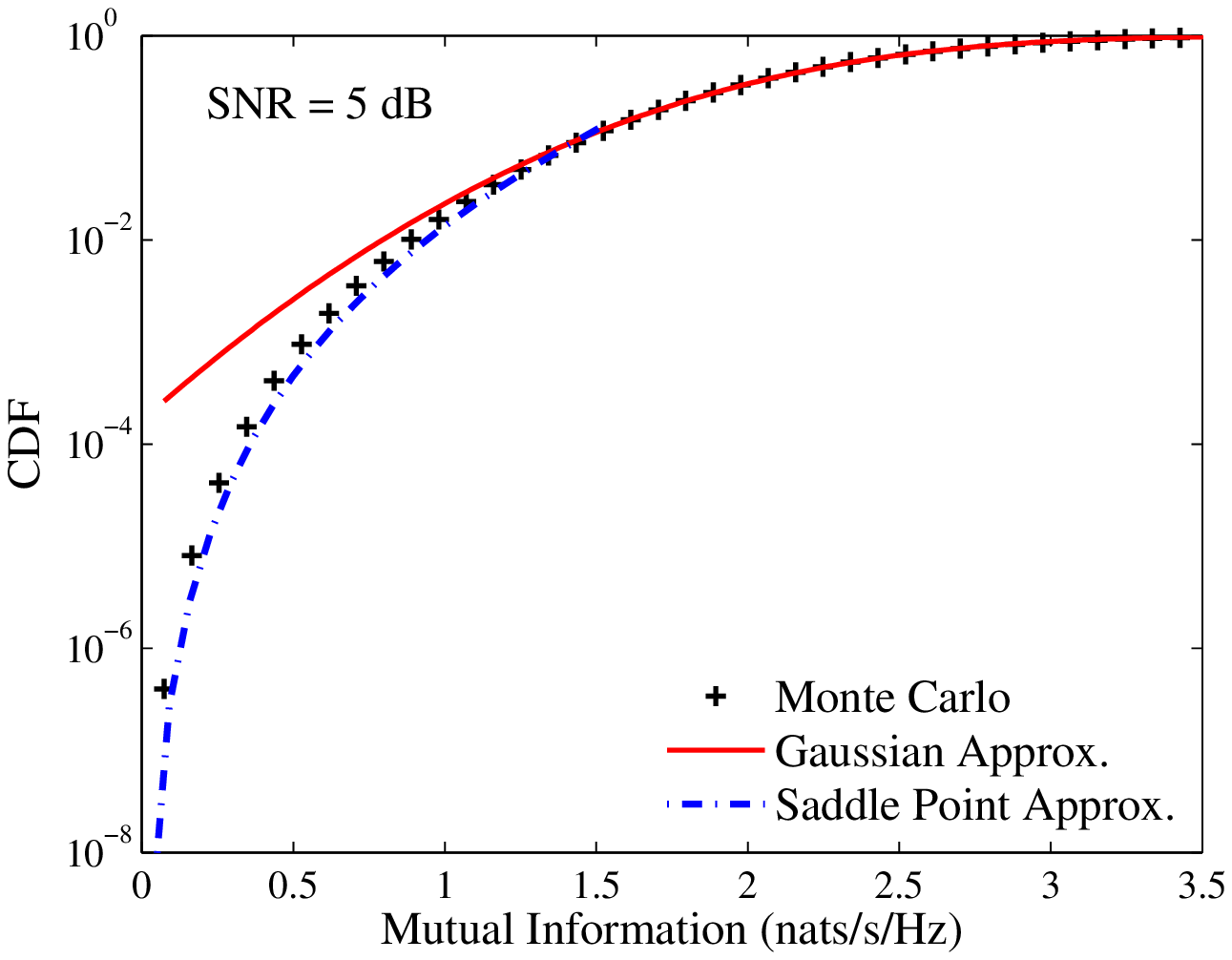}}
\subfigure[$n_t=2,\;n_r=2$]{
\includegraphics[width=0.48\columnwidth]{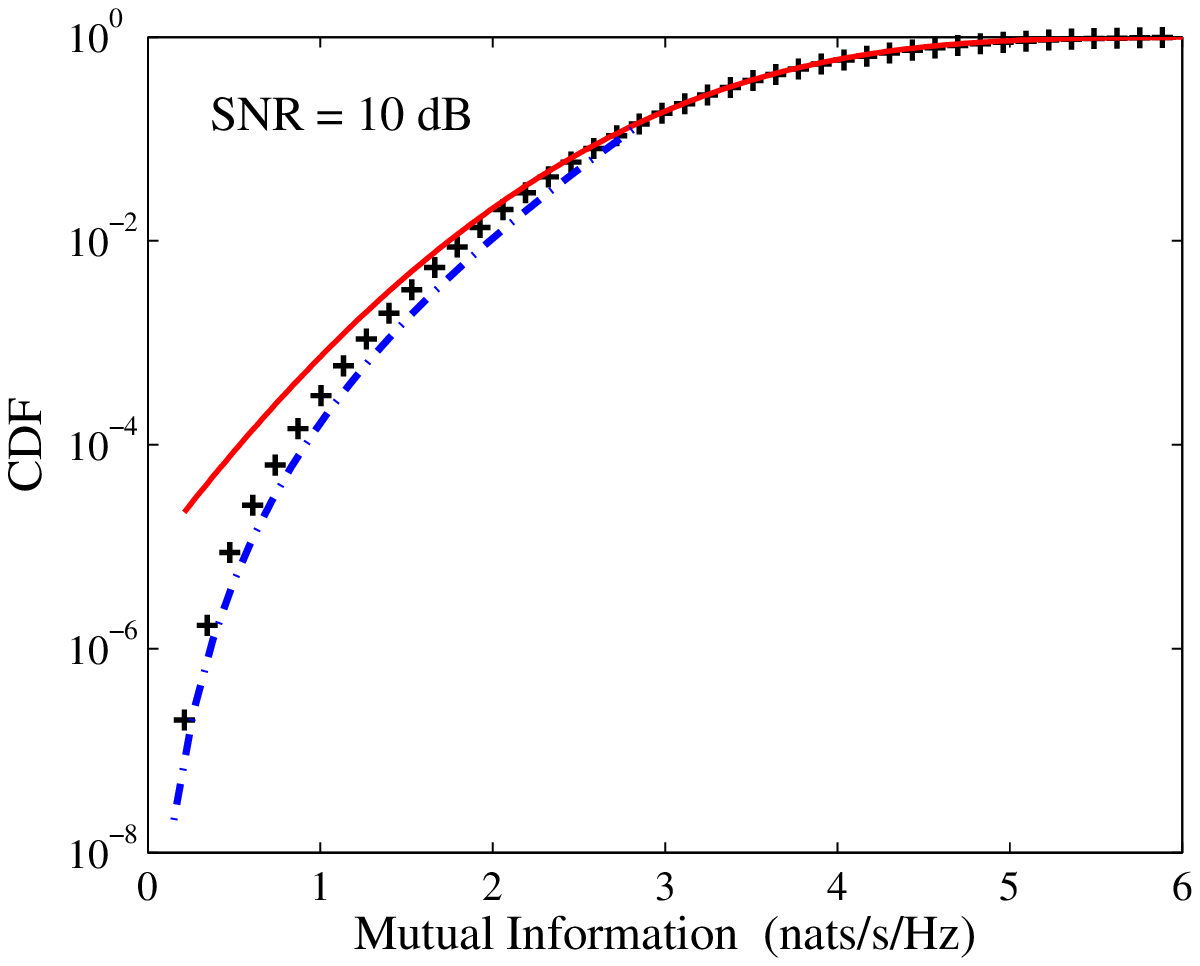}}
%\subfigure[$n_t=2,\;n_r=2, \; {\rm SNR} = 15\; {\rm dB}$]{
%\includegraphics[width=0.48\columnwidth]{Figures/CDF_2_2_15dB.eps}}
\caption{CDF of mutual information for small SNRs, comparing the
Gaussian approximation, saddle point approximation, and Monte Carlo
simulations. Results are shown for $n_t = 2$, $n_r = 2$ and for different SNRs.
%We see the saddle point approximation lines up with simulations in
%the tail region, even for moderate SNR $P=10 \;{\rm dB}$.
%But it eventually deviates for increasing $P$ and
%higher order correction terms are needed.
}
\label{fig:small_P_2_2}
\end{figure}
%\begin{figure}[ht]
%\centering
%\subfigure[$n_t=4,\;n_r=2, \; {\rm SNR} = 5 \;{\rm dB}$]{
%\includegraphics[width=0.48\columnwidth]{Figures/CDF_4_2_5dB_corrected.eps}}
%\subfigure[$n_t=4,\;n_r=2, \; {\rm SNR} = 10 \;{\rm dB}$]{
%\includegraphics[width=0.48\columnwidth]{Figures/CDF_4_2_10dB.eps}}
%%\subfigure[$n_t=4,\;n_r=2, \; {\rm SNR} = 15 \;{\rm dB}$]{
%%\includegraphics[width=0.48\columnwidth]{Figures/CDF_4_2_15dB.eps}}
%\caption{CDF of the mutual information distribution for small SNRs, computed by Gaussian approximation, saddle point approximation and Monte Carlo simulations. We see the saddle point approximation lines up with simulations in the tail region, even for moderate SNR $P=10 \;{\rm dB}$. But it eventually deviates for increasing $P$ and higher order correction terms are needed.}\label{fig:small_P_4_2}
%\end{figure}

\subsection{Asymptotic CGF at Large SNR}

% By computing
%the leading terms of the first few cumulants, and taking $P\to
%\infty$, we observe the following:{\allowdisplaybreaks
%\begin{align}\label{largeP_K}
%&\sigma^2 \sim \ln \left(\beta\right)-\ln
%\left(\beta-1\right)\;,\nonumber\\
%&\kappa_3 \sim \frac{1}{n}\left(\frac{1}{\beta}-\frac{1}{\beta-1}\right)\;,\nonumber\\
%&\kappa_4 \sim \frac{1}{n^2}
%\left(-\frac{1}{\beta^2}+\frac{1}{\left(\beta-1\right)^2}\right)\;,\nonumber\\
%&\kappa_5 \sim \frac{2}{n^3}
%\left(\frac{1}{\beta^3}-\frac{1}{\left(\beta-1\right)^3}\right)\;,\nonumber\\
%&\kappa_6 \sim
%\frac{6}{n^4}\left(-\frac{1}{\beta^4}+\frac{1}{\left(\beta-1\right)^4}\right)\;,\nonumber\\
%&\kappa_7 \sim
%\frac{24}{n^5}\left(\frac{1}{\beta^5}-\frac{1}{\left(\beta-1\right)^5}\right)\;,\\
%&\vdots\nonumber
%\end{align}}
Now we consider the large-$n$--large-$P$ scenario. Based on the
expressions which have been computed for $C_{\ell,0}, \ell=1, 2,
\ldots$ for $P\to \infty$ (e.g.,
(\ref{k1largeP})--(\ref{k3largeP})), we obtain the generic
expression for the leading term of the $\ell$-th cumulant ($\ell \geq
3$):
\begin{align}\label{ka_largeP}
\kappa_\ell \sim \left(-1\right)^\ell
\left(\ell-3\right)!\left[\frac{1}{\left(m-n\right)^{\ell-2}}-\frac{1}{m^{\ell-2}}\right]\;,
\quad \ell=1, 2, \ldots \; .
\end{align}
With this, upon summing the CGF series, the asymptotic CGF is
computed in closed form:
\begin{align}
\mathcal{K}(\lambda)&\approx \mu_0
\lambda+\frac{\sigma_0^2}{2}\lambda^2+\sum_{\ell=3}^\infty\frac{\left(\ell-3\right)!}{\ell!}
\left(-\lambda\right)^\ell\left[\frac{1}{\left(m-n\right)^{\ell-2}}-\frac{1}{m^{\ell-2}}\right]
%\nonumber\\&=\left(-\lambda\right)^3\sum_{\ell=0}^\infty\frac{\ell!}{\left(\ell+3\right)!}\left(-\lambda\right)^\ell
%\left[\frac{1}{\left(m-n\right)^{\ell+1}}-\frac{1}{m^{\ell+1}}\right]
%\nonumber\\&=
%\left(-\lambda\right)^3\left[\frac{1}{m-n}\sum_{\ell=0}^\infty
%\frac{\ell!}{\left(\ell+3\right)!}\left(-\frac{\lambda}{m-n}\right)^\ell-\frac{1}{m}\sum_{\ell=0}^\infty
%\frac{\ell!}{\left(\ell+3\right)!}\left(-\frac{\lambda}{m}\right)^\ell\right]\nonumber\\
\\&=
\left(\mu_0\!-\!\frac{n}{2}\right)
\lambda\!+\!\frac{\sigma^2_0}{2}\lambda^2\!+\!\frac{(\lambda\!+\!m)^2}{2}\ln \left(1\!+\!\frac{\lambda}{m}\right)\!-\!\frac{(\lambda\!+\!m\!-\!n)^2}{2}\ln \left(1\!+\!\frac{\lambda}{m\!-\!n}\right).
\end{align}
Now, from (\ref{eq:coulomb_mean}) and (\ref{eq:coulomb_variance}),
we have
\begin{align}
&\mu_0 \approx n\ln{P}-(m-n)\ln\left(\frac{m-n}{m}\right)-n\;,  \qquad P\to \infty,\\
&\sigma_0^2 \approx \ln\left(\frac{m}{m-n}\right)\;, \qquad P\to
\infty,
\end{align}
giving the following large-$n$--large-$P$ CGF:
\begin{align}\label{CGF_converge}
{\cal K}\left(\lambda\right)&\approx \left[n\ln
\left(\frac{P}{m}\right)-\frac{3}{2}n\right]\lambda+\frac{\left(m+\lambda\right)^2}{2}\ln\left(m+{\lambda}\right)\nonumber\\&-\frac{\left(m-n+\lambda\right)^2}{2}\ln\left(m-n+{\lambda}\right)-\frac{m^2}{2}\ln{m}+\frac{(m-n)^2}{2}\ln{(m-n)}
\; ,
\end{align}
valid for $\lambda\in (n-m, +\infty)$.

\begin{remark}
Interestingly, although (\ref{CGF_converge}) was obtained based on
the cumulant expressions which are valid only for $\beta\neq 1$
(i.e., since a singularity exists for $\beta=1$ in the
$n$-asymptotic cumulants, as seen in
(\ref{k1largeP})--(\ref{k3largeP})), by setting $\beta=1$, we have
\begin{align}\label{CGF_equal}
{\cal K}\left(\lambda\right)&\approx n\lambda\ln
\left(\frac{P}{n}\right)-\frac{3}{2}n\lambda+\frac{\left(n+\lambda\right)^2}{2}\ln\left(n+{\lambda}\right)\nonumber\\&-\frac{\lambda^2}{2}\ln{\lambda}-\frac{n^2}{2}\ln{n}
\; .
\end{align}
This will be shown to describe the correct behavior of the mutual
information.
\end{remark}

%We compare the asymptotic closed-forms of CGF (\ref{CGF_converge}) (or (\ref{CGF_equal})), Oyman's high-SNR CGF (\ref{Oyman}) and the numerical results in Fig.
%\ref{CGF_comparison}. The figures indicate the following:
%\begin{itemize}
%\item The $n$-asymptotic CGF (\ref{CGF_converge}) matches with Oyman's exact high-SNR CGF (\ref{Oyman}) perfectly.
%\item Both (\ref{CGF_converge}) (or (\ref{CGF_equal})) and (\ref{Oyman}) match with the numerical results very well for $\lambda \geq n-m$, but diverges eventually for the negative $\lambda <n-m$.  We realize that Oyman's CGF has the same singularity as (\ref{CGF_converge}) due to the singular points of $\Gamma (x)$ in negative real $x$.
%\item Both (\ref{CGF_converge}) and (\ref{Oyman}) extend its valid regime in the negative side as $m-n$ increases.
%\end{itemize}
%\begin{figure}[ht]
%\centering
%\subfigure[$n_t=5,\;n_r=3$]{
%\includegraphics[width=0.48\columnwidth]{CGF_3_5.eps}}
%\subfigure[$n_t=8,\;n_r=3$]{
%\includegraphics[width=0.48\columnwidth]{CGF_3_8.eps}}
%\subfigure[$n_t=3,\;n_r=3$]{
%\includegraphics[width=0.48\columnwidth]{CGF_3_3.eps}}
%\caption{Comparison of the $n$-asymptotic CGF closed-form (\ref{CGF_converge}), exact high-SNR CGF (\ref{Oyman}) and the
%numerical results  for different antennas settings under ${\rm SNR}= 30 {\rm dB}$. (b) is shown to have wider valid regime than (a) in the left tail since $\Delta=m-n$ is larger.}\label{CGF_comparison}
%\end{figure}

The saddle point (\ref{eq:saddlepoint}) can be computed by
\begin{align}\label{equality}
t
=n\ln P-n\ln m-n+(m+\lambda^\star)\ln (m+\lambda^\star)-(m-n+\lambda^\star)\ln (m-n+\lambda^\star).
\end{align}
Whilst a closed-form solution for (\ref{equality}) is intractable,
it can be trivially computed numerically for any given value of $t$.

By invoking (\ref{SPA_PDF}) and (\ref{eq:CDF_SPA}), we can compute
the distribution (both the PDF and CDF) of the mutual information.
We should point out that $t$ in (\ref{equality}) is a monotonically
increasing function of $\lambda$ (i.e., ${\rm d}t(\lambda)/{\rm
d}\lambda>0$), thus for any $t\geq n\ln \left(P/\beta\right)-n$,
there exists a real root $\lambda^\star$. This, in turn, implies
that the distribution cannot be captured explicitly by
(\ref{equality}) if $t$ is sufficiently small such that $\mu-t>m\ln
\left({\beta}\right)-(m-n)\ln (\beta-1)$ (i.e., when looking
sufficiently far into the left tail region). Nevertheless, the
right-hand side of this inequality is increasing with $m$ (fixed
$n$); thus as $m$ grows, the valid region of (\ref{equality})
extends further into the left tail, allowing smaller outage
probabilities to be calculated. In fact, scenarios for which $m$ is
reasonably large compared with $n$ is quite realistic in many
applications; for example, in cellular systems, for which the base
station may be equipped with a reasonably large number of antennas,
whilst the number of antennas on the mobile device is more
restricted due to limited space constraints.

Fig. \ref{fig:PDF_large_P} depicts the saddle point approximation of
the PDF (\ref{SPA_PDF}) with $I(t)$ computed from
(\ref{CGF_converge}) and (\ref{equality}), comparing with the
Gaussian approximation and Monte Carlo simulations. The saddle point
result is clearly much more accurate than the Gaussian, and lines up
almost perfectly with the simulations when the SNR is sufficiently
large (i.e., at $30$ dB).  Similar observations are made in Fig.
\ref{fig:CDF_large_P}, which shows the corresponding CDF curves
based on (\ref{eq:CDF_SPA}) and the same $I(t)$. Note that for these
results we have chosen $m = 6,\; n=2$, where $m$ is comparatively large enough such
that the validity of (\ref{equality}) extends deep enough into the
left tail region to capture outage probabilities of interest.

%We see that when $m$ is small, the PDF computed by the saddle point
%method (\ref{equality}) provides no values at the left tail with
%small outage probabilities (cf. Fig. \ref{fig:PDF_large_P} (a)),
%whilst for larger $m$, a sufficient region of left tail is captured.
%Also as expected, the large-$P$ asymptotic approximation performs
%better as $P$ increases (cf. Fig. \ref{fig:PDF_large_P} (b), (c)).
%Note that results are shown for reasonably large SNRs (i.e.,
%$P=20\,,30 \,{\rm dB}$).

%Fig. \ref{fig:CDF_large_P} plots the CDF of the mutual information.
%We see that the saddle point approximation keeps close to the
%simulations for reasonably large SNRs (i.e., $P=20\,,30\, {\rm
%dB}$), and become highly accurate as $P$ increases. Results are
%shown for sufficiently and reasonably large $m$.

\begin{figure}[ht]
\includegraphics[width=0.48\columnwidth]{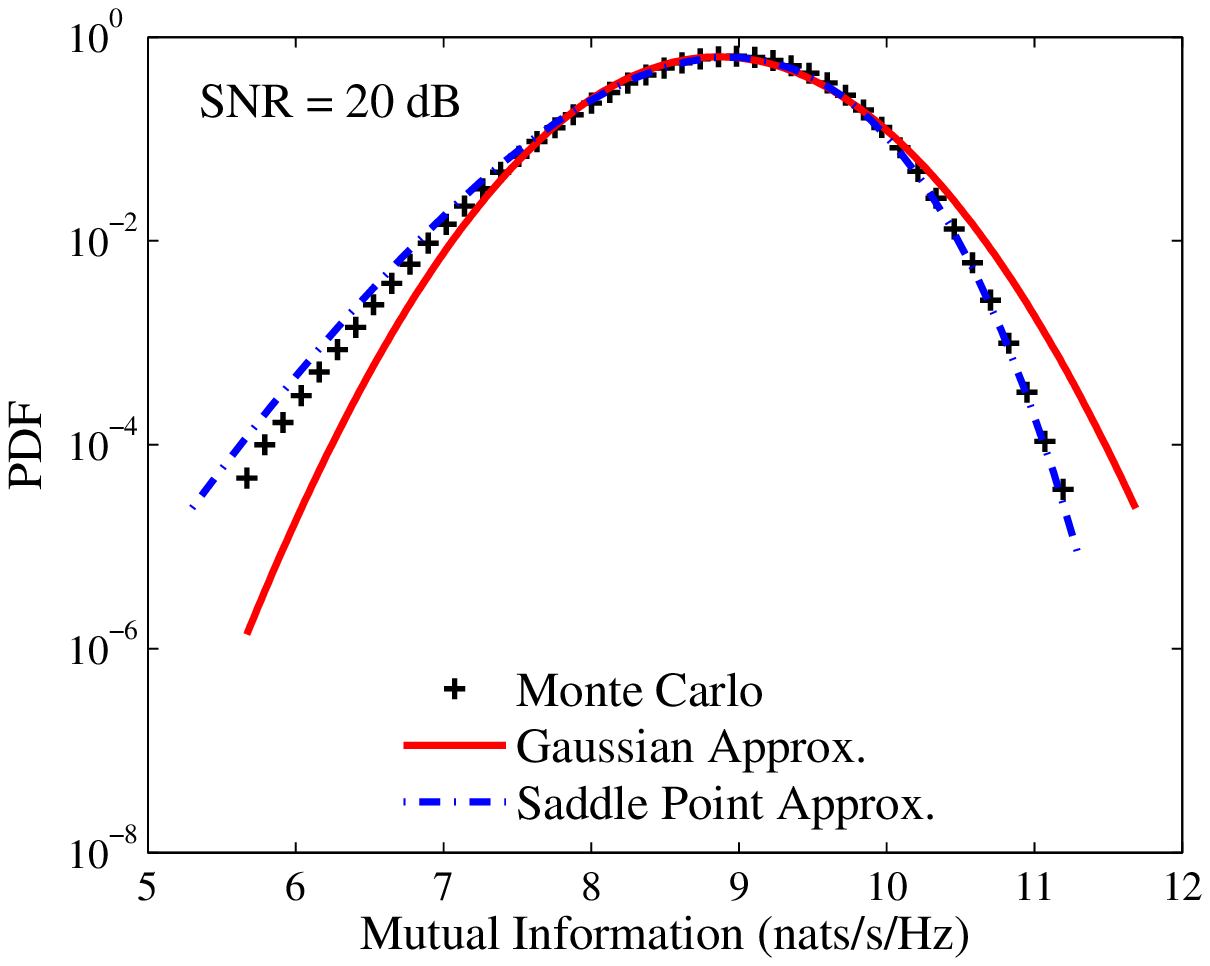}
\includegraphics[width=0.48\columnwidth]{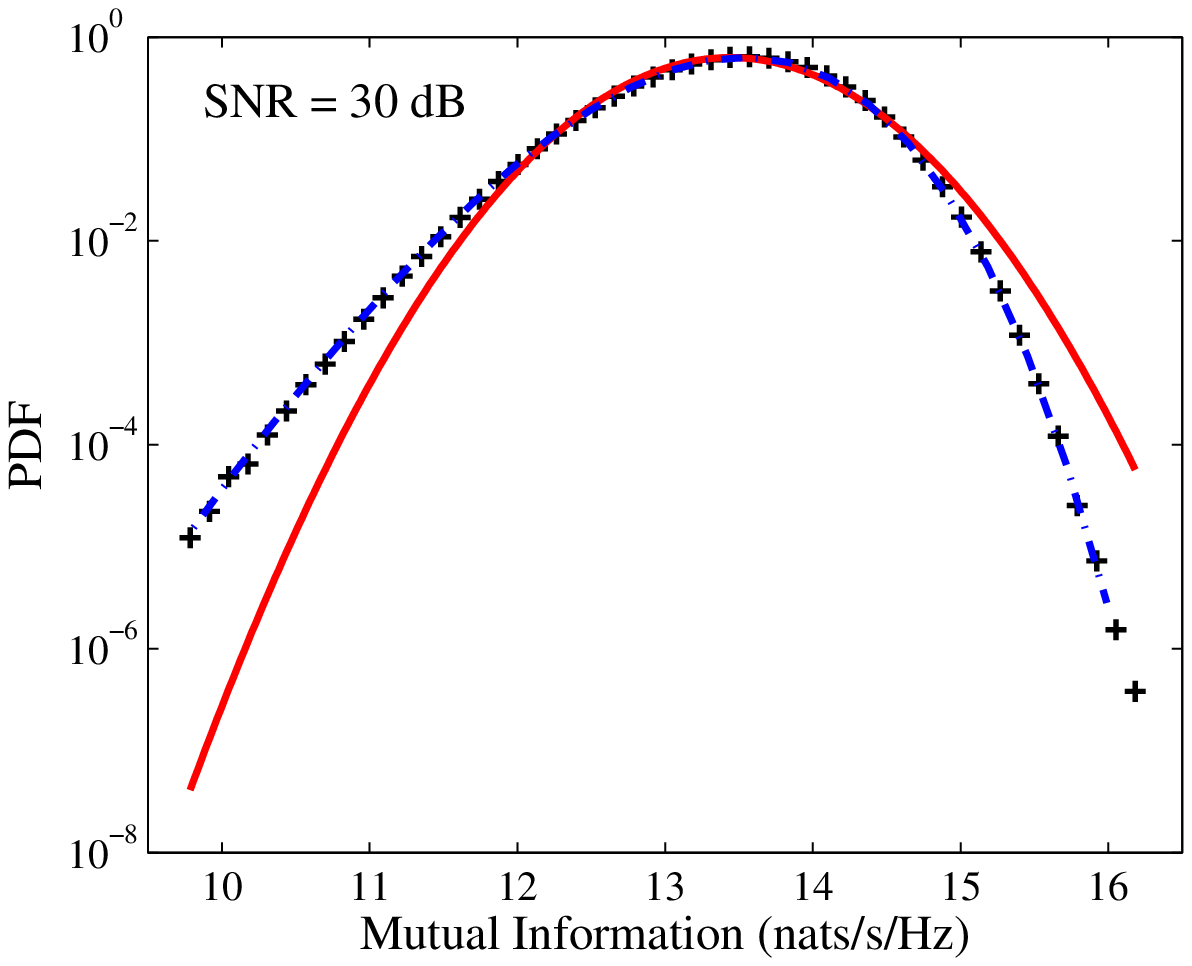}
\caption{PDF of mutual information, comparing the Gaussian
approximation, saddle point approximation, and Monte Carlo
simulations. Results are shown for
$n_t=6,\;n_r=2$ and for different SNRs.}\label{fig:PDF_large_P}
\end{figure}
\begin{figure}
\centering
\includegraphics[width=0.48\columnwidth]{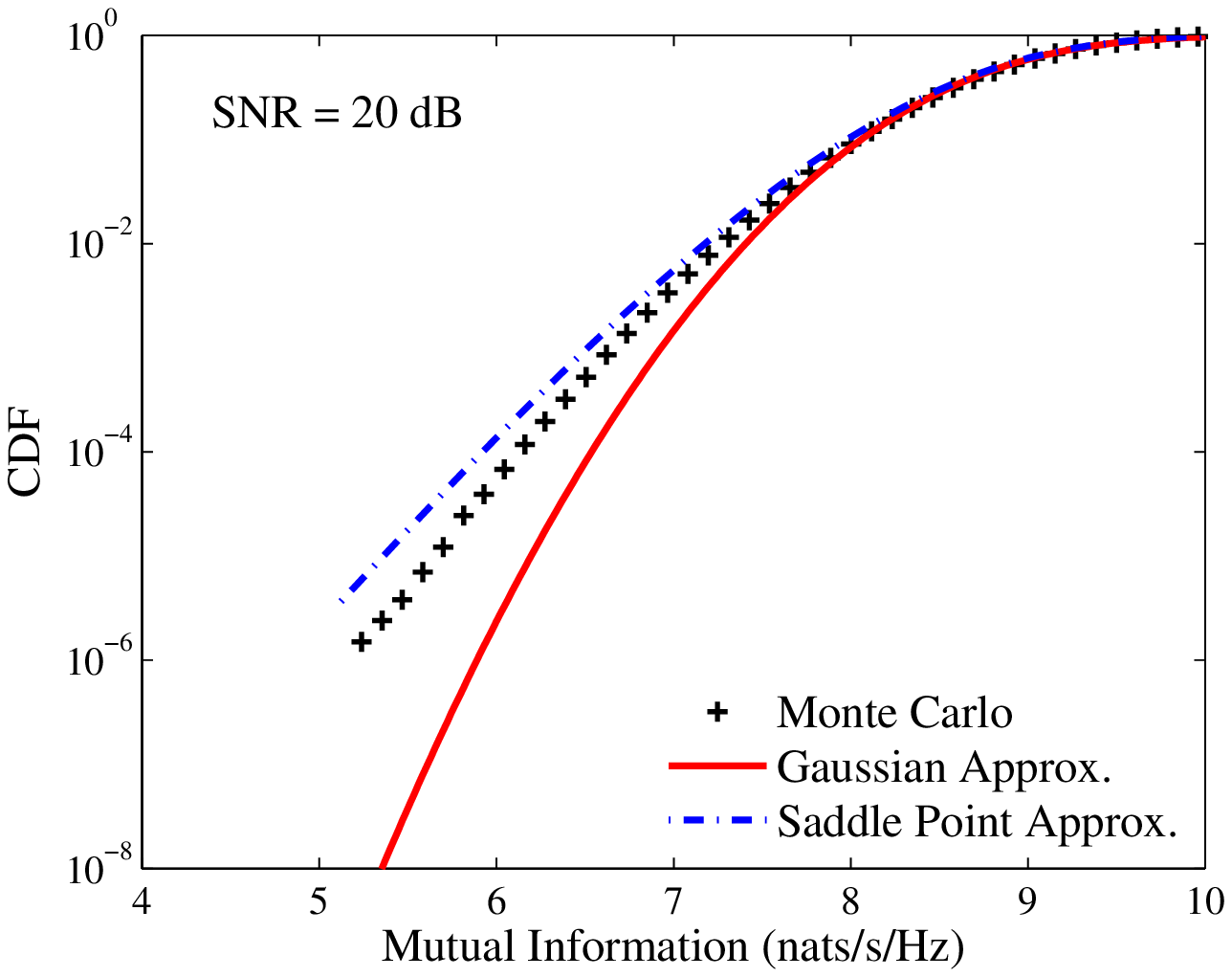}
\includegraphics[width=0.48\columnwidth]{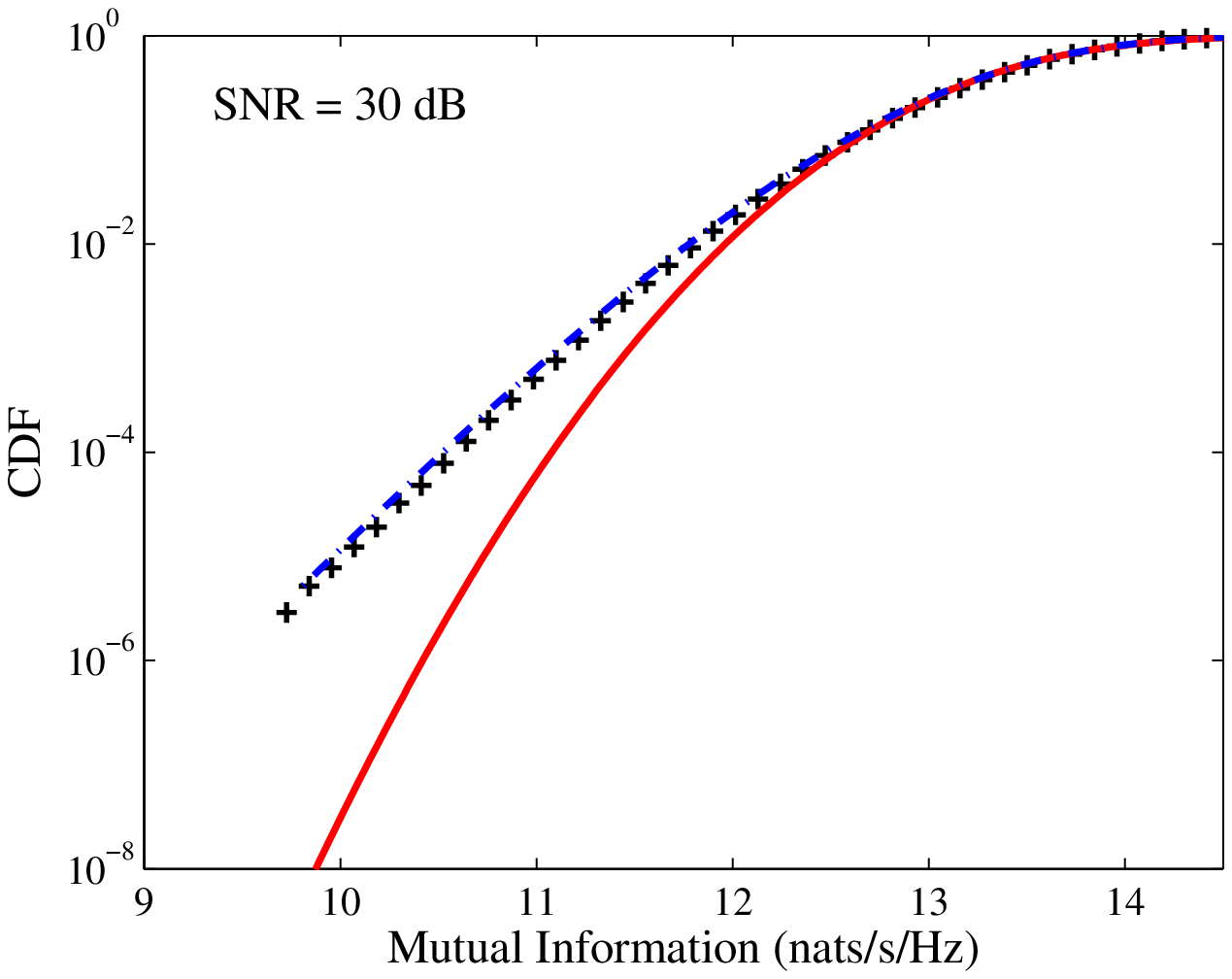}
%\subfigure[$n_t=9, \;n_r=3, \;{\rm SNR} = 30 \;{\rm dB}$]{
%\includegraphics[width=0.48\columnwidth]{Figures/CDF_9_3_30dB.eps}}
\caption{CDF of mutual information, comparing the Gaussian
approximation, saddle point approximation, and Monte Carlo
simulations. Results are shown for $n_t=6,\;
n_r=2$ and for different SNRs.}\label{fig:CDF_large_P}
\end{figure}

\begin{figure}[ht]
\centering
\subfigure[$n_t = n_r = 3$]{
\includegraphics[width=0.48\columnwidth]{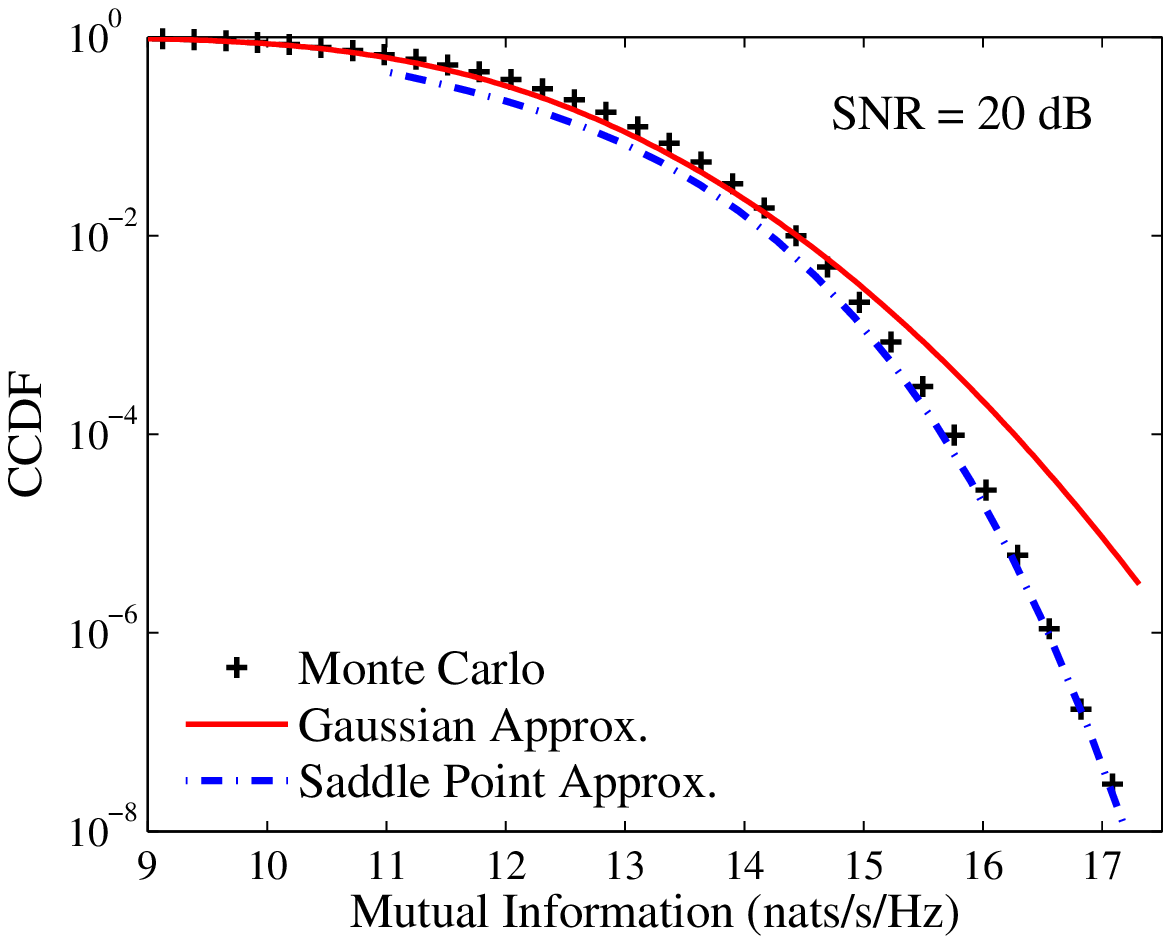}}
\subfigure[$n_t = n_r = 3$]{
\includegraphics[width=0.48\columnwidth]{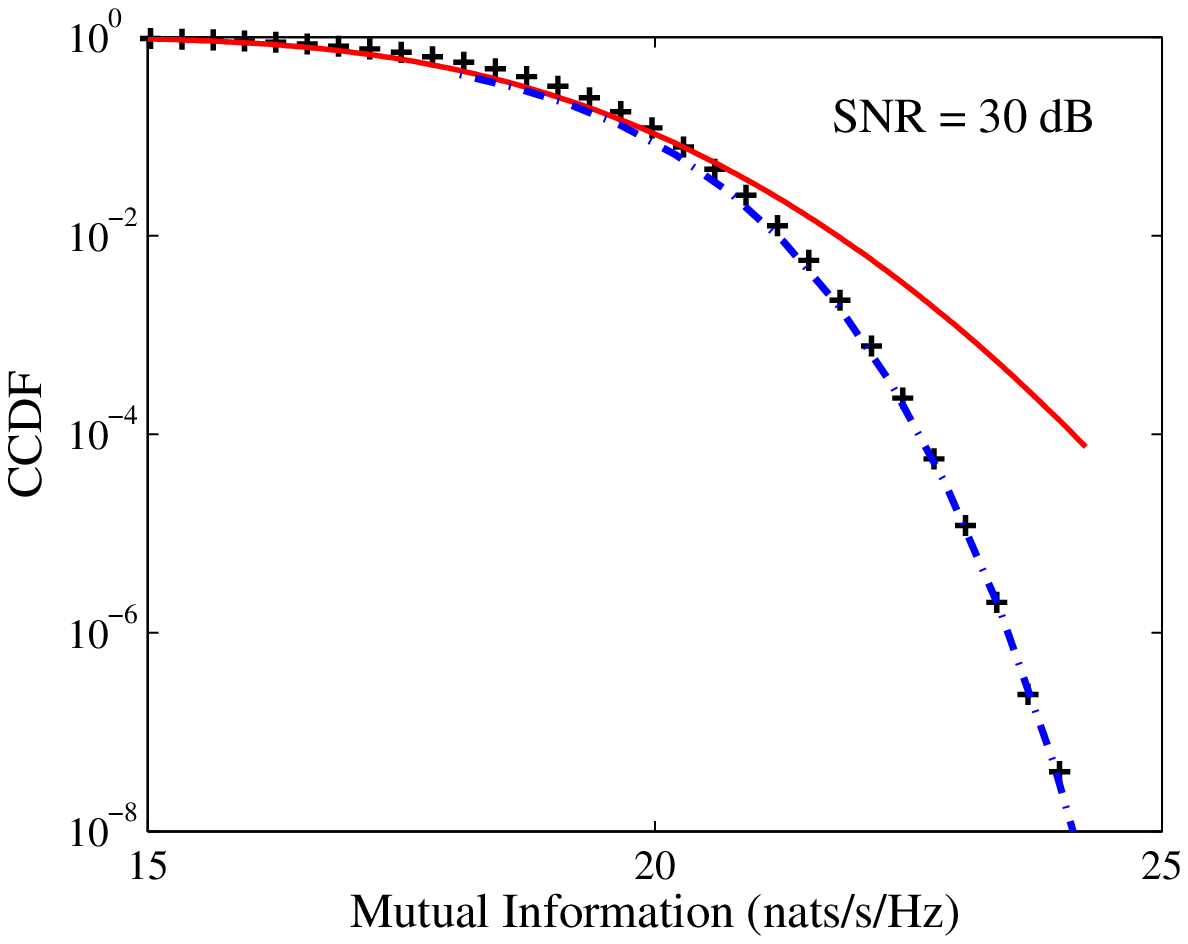}}
%\subfigure[$n_t = 4, n_r = 2, {\rm SNR} = 15 \;{\rm dB}$]{
%\includegraphics[width=0.48\columnwidth]{Figures/CCDF_4_2_15dB.eps}}
\subfigure[$n_t = 4, \; n_r = 2$]{
\includegraphics[width=0.48\columnwidth]{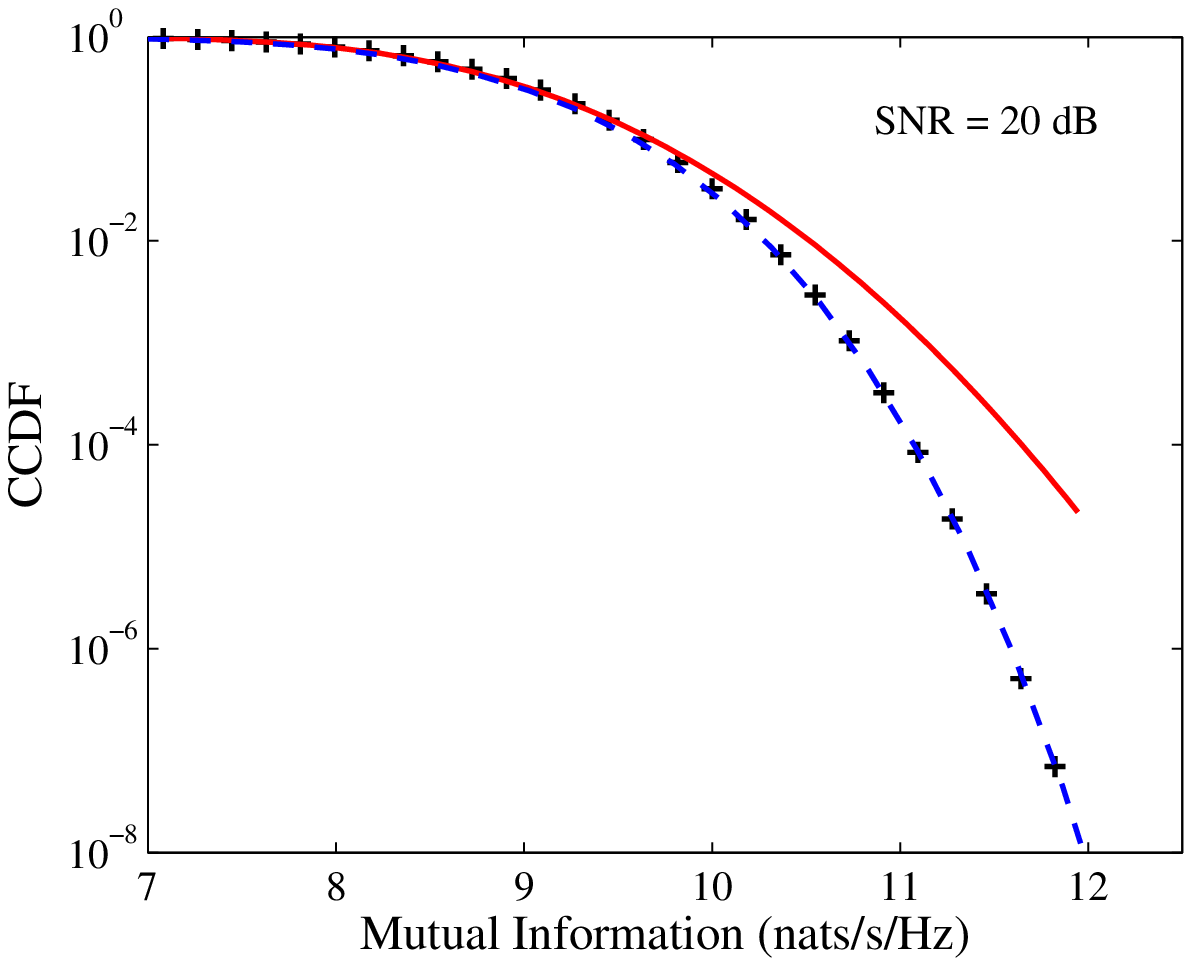}}
\caption{CCDF of mutual information, comparing the Gaussian
approximation, saddle point approximation, and Monte Carlo
simulations. Results are shown for different antenna configurations
and different SNRs.}\label{fig:CCDF_large_P}
\end{figure}

Fig. \ref{fig:CCDF_large_P} depicts the complementary CDF (CCDF) of
the mutual information, comparing the saddle point approximation
based on $I(t)$ computed from (\ref{CGF_converge}) and
(\ref{equality}), the Gaussian approximation, and Monte Carlo
simulations. As for the left tail, we see that the saddle point
approximation becomes extremely accurate when the SNR is
sufficiently high, and significantly outperforms the Gaussian
approximation. In fact, quite surprisingly, even for moderate SNRs
of $20$ dB, the saddle point approximation in the right-hand tail
traces the simulated curve very closely.

As done for the small-SNR scenario, for large-SNR we can also
evaluate the higher order correction terms (in $P$) in order to draw
insight into the accuracy of the leading-order results. To this end,
it is convenient to first introduce the change of variables $x \to
\beta/x$ in the Painlev\'{e} representation of the
CGF (\ref{eq:SU_MIMO_Painleve}) and the large-$n$ equation
(\ref{leading_Equation}). The CGF to leading order in $n$ can then
be written as
\begin{align}\label{eq:Painleve_P}
\mathcal{K}(n\lambda)\sim -n^2\int_{0}^{P} \frac{Y(\beta/x)}{x}{\rm d}x,\qquad n\to \infty,
\end{align}
with $Y(\beta/x)$ satisfying
\begin{align}\label{eq:fundamental_P}
\left[xY'+\left(1+\frac{\lambda+1}{\beta}\right)x^2Y'+Y\right]^2=4\left(-xY'-Y+\beta\right)\left(\frac{x^2}{\beta}Y'-\lambda\right)\left(\frac{x^2}{\beta}Y'\right)
\end{align}
where $Y':={\rm d}Y(\beta/x)/{\rm d}x$. By noting that in
(\ref{CGF_converge}) there exists first-order terms in $P$ which are
$O(\ln P)$ and second-order terms which are $O(1)$, we assume that
the CGF admits the following generic large-$P$ expansion:
\begin{align}\label{eq:largeP_expansion_CGF}
\mathcal{K}(n\lambda)=n^2\left(\lambda \ln P + b_0 + \frac{b_1}{P} + \frac{b_2}{P^2}+\cdots\right)
\end{align}
where the coefficient $b_0$ denotes the constant term in
(\ref{CGF_converge}) and $b_i, i=1, 2, \ldots$ depend on $\beta$ and
$\lambda$. Taking the derivative of $\mathcal{K}(n\lambda)$ w.r.t.
$P$, we obtain the following power series expansion:
\begin{align}\label{eq:largeP_expansion}
Y\left(\frac{\beta}{P}\right)&=-P\frac{\rm d}{{\rm d} P}\left(\frac{\mathcal{K}(n\lambda)}{n^2}\right),\qquad n\to \infty\nonumber\\
&=-\lambda+\frac{b_1}{P}+\frac{2b_2}{P^2}+\cdots
\end{align}
Substituting (\ref{eq:largeP_expansion}) into
(\ref{eq:fundamental_P}) and matching coefficients of the powers
series of $P$ on the left and right-hand sides, we solve the
$b_i$'s:
\begin{align}
&b_1=\frac{\lambda \beta}{\beta +\lambda-1}\;,\nonumber\\
&b_2=-\frac{\beta^2\lambda\left(\beta-1\right)\left(\beta+\lambda\right)}{2\left(\beta+\lambda-1\right)^4}\;,\nonumber\\
&b_3=\frac{\beta^3\lambda(\beta-1)(\beta+\lambda)(\beta+\lambda+1)(\beta-1-\lambda)}{3\left(\beta+\lambda-1\right)^7}\;,\nonumber\\
&\;\,\vdots
\end{align}
%{\bf ** You need to explain why the $n \lambda$ has now changed to
%$\lambda$ etc.  Also, why not present the result below in terms of
%$\beta$?  Would this be more natural in clearly revealing the
%scaling structure with $n$ etc?  Or, is $m - n$ more meaningful in
%the interpretation of when things break down etc... **}
Together with (\ref{eq:largeP_expansion_CGF}), we obtain the
large-$n$--large-$P$ CGF with higher correction terms in $P$:
\begin{align}\label{eq:corrected_largeP}
\mathcal{K}(\lambda)\sim \mathcal{K}_0(\lambda)+\frac{mn\lambda}{m-n+\lambda}\frac{1}{P}-\frac{ m^2(m-n)(m+\lambda)}{2n(m-n+\lambda)^4}\frac{1}{P^2}+\cdots\quad n\to \infty, \quad P\to\infty
\end{align}
where $\mathcal{K}_0(\lambda)$ denotes the leading-order
large-$n$--large-$P$ CGF expression in (\ref{CGF_converge}). Based
on this formula, we draw the following remarks:
\begin{itemize}
\item First, we see that the higher correction terms vanish rapidly
as $m-n+\lambda$ increases (equivalently, $\lambda$ grows for fixed
$m$ and $n$). Meanwhile, as $\lambda$ decreases and approaches
$n-m$, the correction terms become large and eventually invalidate
the expansion. This indicates that the leading-order
large-$n$--large-$P$ approximation for the right-hand tail
(corresponding to positive $\lambda$) is more robust for finite
values of $P$, compared with the approximation for the left-hand
tail (corresponding to negative $\lambda$). This is consistent with
the results shown in Figs. \ref{fig:CDF_large_P} and
\ref{fig:CCDF_large_P}.

%\item Whilst (\ref{eq:corrected_largeP}) provides improvement for moderate $P$ at the right-hand tail, its validity at the left-hand tail should be taken with caution since the correction terms break down as $\lambda$ approaches $n-m$ (equivalently, $m-n+\lambda\to 0$). The underlying limitation here is when we use the asymptotic expansion method, other parameters are not expected to move around.
%A potential method to overcome this hurdle is to rescale other
%parameters to simultaneously change with the asymptotic variable and
%establish a unified expansion. However, if the $m$ (fixed $n$) is
%sufficiently large, (\ref{eq:corrected_largeP}) is able to provide
%enough low outage probabilities.

\item Interestingly, as we keep computing $b_i$'s, it is found that $b_i, i>1$ have the common factor of $m-n$. Assuming this is true for all higher correction terms, then for equal antenna arrays (i.e., $m=n$), we have
    \begin{align}\label{eq:corrected_largeP_beta_1}
\mathcal{K}(\lambda)\sim \mathcal{K}_0(\lambda)+\frac{n^2}{P}\;,\qquad n\to \infty\;.
\end{align}
Note that the single correction term $ n^2 / P$ is independent of
$\lambda$, meaning that the saddle point equation (\ref{equality})
(after setting $m=n$) is unaffected by including correction terms
for finite $P$. This agrees with the formula in \cite[Eq.
(53)]{Moustakas11} and the corresponding argument made therein.
\end{itemize}
%\begin{figure}
%\centering
%\subfigure[$n_t=6,\;n_r=3$]{
%\includegraphics[width=0.48\textwidth]{Figures/CGF_withPcorrection.eps}}
%%\subfigure[$n_t=n_r=3$]{
%%\includegraphics[width=0.48\textwidth]{Figures/CGF_withPcorrection_beta_1.eps}}
%\caption{Comparison of the CGF computed by the leading terms (\ref{CGF_converge}), corrected formula (\ref{eq:corrected_largeP}), and Monte Carlo simulations.}\label{fig:CGF_withPcorrection}
%\end{figure}

Based on (\ref{eq:corrected_largeP}), we plot the saddle point
approximation with first-order correction term (in $P$) in Fig.
\ref{fig:CDF_withPcorrection}. The refinement brought by including
the correction term is clearly evident. Moreover, we find that the
right-hand tail is captured accurately for moderate $P=15\;{\rm dB}$
even without the first-order correction, which is in line with the
discussion in the first point above.
\begin{figure}
\centering
\subfigure[$n_t=10, n_r=2$]{
\includegraphics[width=0.48\textwidth]{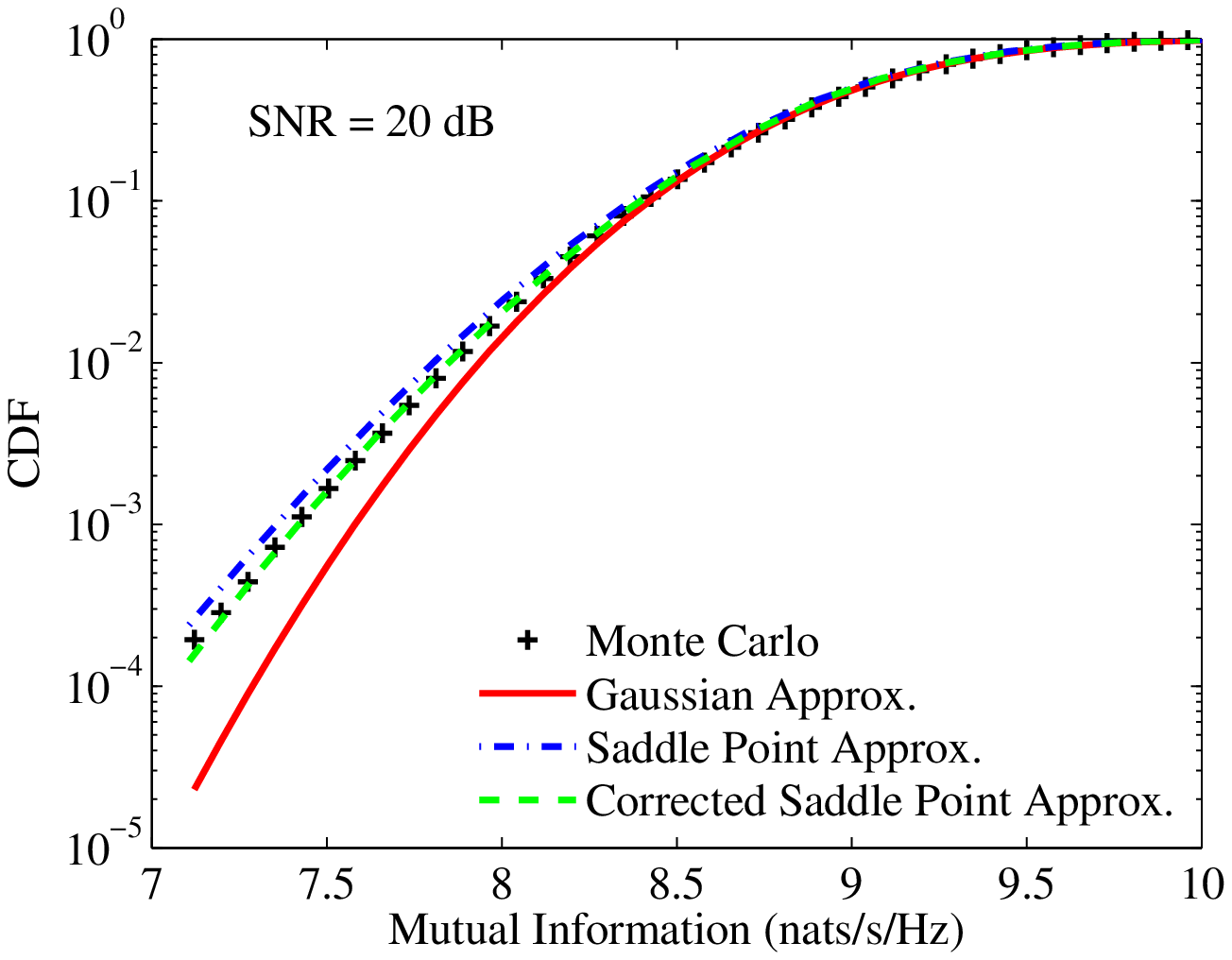}}
\subfigure[$n_t=4, n_r=2$]{
\includegraphics[width=0.48\textwidth]{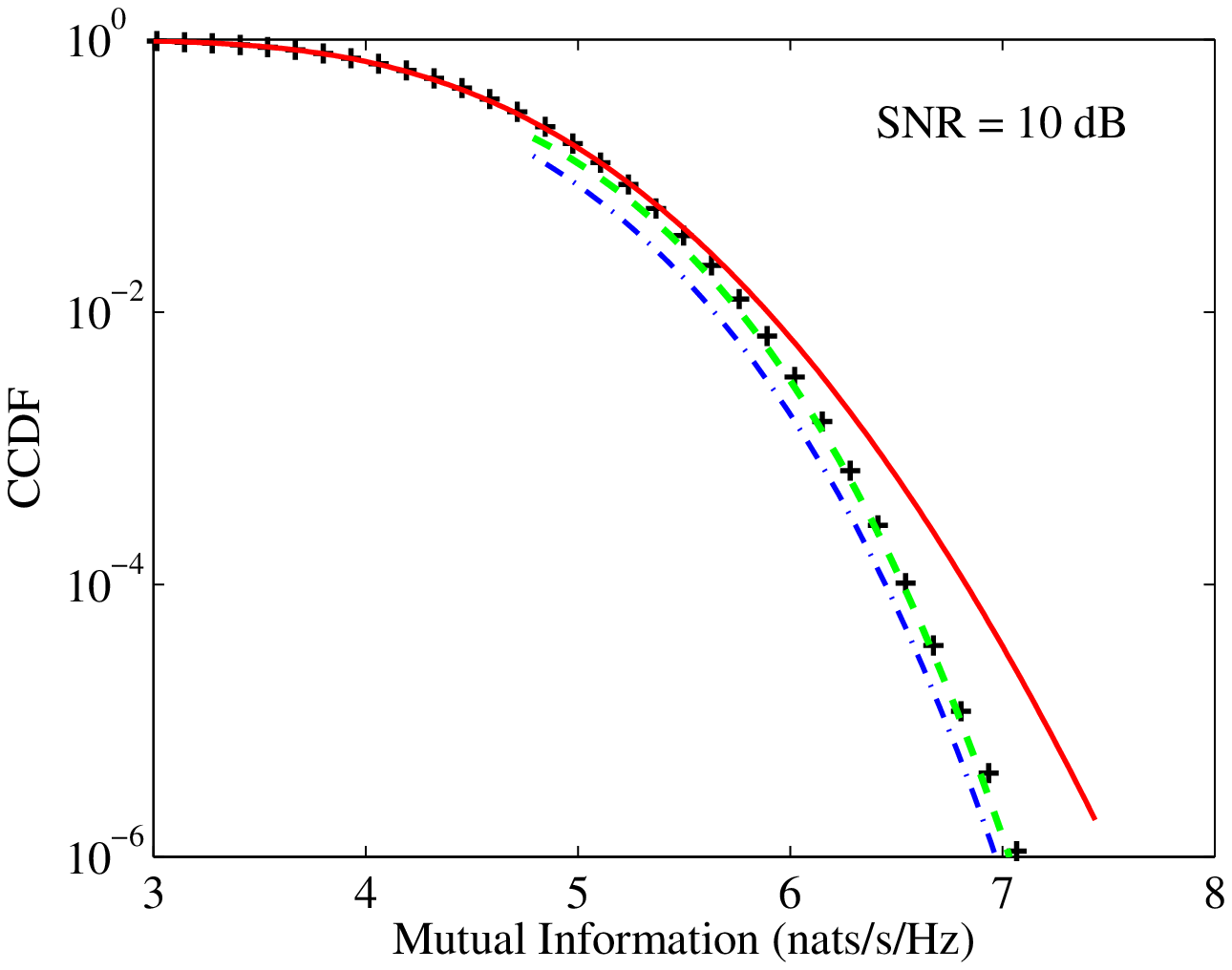}}
\subfigure[$n_t=4, n_r=2$]{
\includegraphics[width=0.48\textwidth]{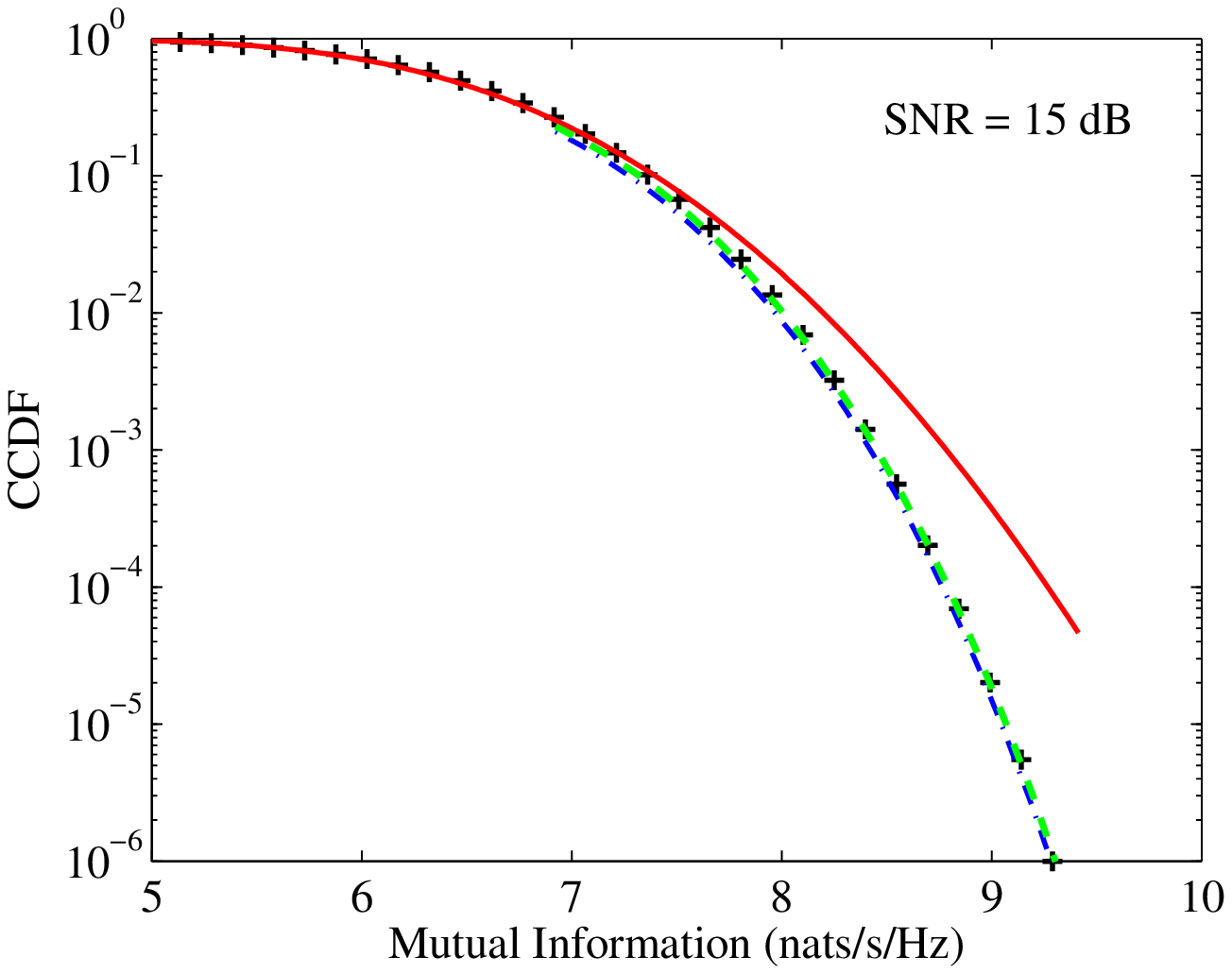}}
%\subfigure[$n_t=n_r=3$]{
%\includegraphics[width=0.48\textwidth]{Figures/CGF_withPcorrection_beta_1.eps}}
\caption{CDF and CCDF of mutual information, comparing the Gaussian
approximation, the saddle point approximation with and without the
first order corrections (in $P$), and Monte Carlo simulations.
Results are shown for different antenna
configurations and different SNRs.}\label{fig:CDF_withPcorrection}
\end{figure}

\subsection{Important Large SNR Behavior in Left and Right Tails (Including DMT)}
In this section, we develop further our analytical results under the
assumption that $t=qn\ln P,\; P\to \infty$, where $q > 0$ is a fixed
constant. This represents the setting where the data rate is
specified to grow as a non-vanishing fraction (i.e., $q$) of the
mean mutual information for large $P$ (i.e., $n\ln P$). This
assumption is important in various contexts; for example, in
specifying the fundamental DMT \cite{Zheng_Tse}, as well as capturing the scheduling gains in
opportunistic multi-user downlink transmissions \cite{Hochwald04}.

We begin with the scenario $q\geq 1$, in which case we are
interested in the mutual information distribution at the right-hand
side of the mean. This scenario is relevant for evaluating the
performance of scheduling algorithms for which the multi-antenna
base-station transmits to the multi-antenna user with the best
channel (i.e., $\mathcal{I}_{\rm best}=\max\{\mathcal{I}_1,
\mathcal{I}_2, \ldots\}$, with $\mathcal{I}_k$ denoting the mutual
information between the base-station and the $k$th user, all of
which are assumed to be independent and to undertake the same
distribution).  In this case, the mutual information achieved by the
system will typically be above the mean mutual information for each
user $\mu$, and the performance gains of such ``best user''
selection algorithms can be characterized by studying the
distribution of the per-user mutual information in the regime
$t\geq\mu$.  See \cite{Hochwald04} and \cite{Moustakas11}.

Substituting $t=qn\ln P, q\geq 1$ into (\ref{equality}), we have the
following asymptotic solution:
\begin{align}
\lambda^\star \sim nP^{q-1}\;,\quad P\to \infty
\end{align}
which further yields
\begin{align}
I(t)&\sim n^2P^{q-1}\ln P^{q}\nonumber\\&=\frac{nt}{P}e^{t/n}\;, \qquad P\to \infty.
\end{align}
Therefore, for high $P$, the CCDF admits
\begin{align}\label{largePright}
{\Pr(\mathcal{I}({\bf x;y})>t})\sim e^{-\frac{nt}{P}e^{t/n}}, \quad t\geq\mu\;.
\end{align}
This agrees with a result derived recently in \cite[Eq.
(79)]{Moustakas11}, by asymptotically solving a set of coupled
equations obtained via a Coulomb fluid formulation. From
(\ref{largePright}), we find the probability of the mutual
information taking greater values than the mean drops very sharply
(doubly exponentially with $t$), indicating that for large SNR, the
best-user scheduling algorithm indicated above will not enhance the
overall data rate significantly.

%{\bf ** Is this result/discussion somewhat contradictory to what is
%said in the previous paragraph,
%about the motivation for selection and so on? **\\ \\
%Samuel: Actually, the motivation of this analysis is to examine how much benefit the best-user selection can bring. However the result above indicates that this selection algorithm is not very advantageous in the large-SNR scenario. That is why the result above seems ``contradictory'' to the best-user selection. In fact, Moustakas has made the same comments as well in \cite{Moustakas11}.\\}

Now we consider the alternative regime, $q<1$, corresponding to the
DMT framework seminally proposed by \cite{Zheng_Tse}. In this case,
however, we find that one cannot simply adopt the direct approach of
substituting $t=qn\ln P$ with $q<1$ into (\ref{equality}), since a
solution for the asymptotic $\lambda^\star$ does not exist. This can
be explained by noting that the solution in fact lies in the range
$\lambda^\star<n-m$ which can not be described by
(\ref{CGF_converge}), because $t=qn\ln P<n\ln (P/\beta)-n$ (i.e.,
the smallest value of $t$ that can be covered by (\ref{equality}))
for $q<1$ and large $P$. Nevertheless, in the following, we are able
to draw upon the exact characterization of the large-$n$--large-$P$
CGF (\ref{leading_Equation}) to derive the DMT formula.

Since in the large deviation regime, $\lambda^\star\sim O(n)$ such
that all the cumulants to leading order in $n$ remain effective for
large $n$, we scale the CGF variable $\lambda\to n\lambda$ before
taking $n\to \infty$ in (\ref{eq:order_CGF}). Further, by recalling
the definition of the CGF, its leading order representation in $P$
should be $O(\ln P)$. Consequently, the asymptotic characterization
of the CGF admits
\begin{align}\label{eq:order_CGF}
\mathcal{K}(n\lambda) \approx -n^2\;
\int_{0}^{P}\frac{Y(\beta/x)}{x}{\rm d}x\approx \;A \;n^2\ln
P,\qquad n\to \infty, \quad P\to \infty
\end{align}
where $Y(\beta/x)$ satisfies (\ref{eq:fundamental_P}) and $A$ denotes a certain function of $\beta$ and $\lambda$. In light of the DMT
formulation \cite{Zheng_Tse}, we require the coefficient of the
$O(n^2\ln P)$ term of the CGF, i.e., the quantity
\begin{align}\label{eq:DMT_2}
A=\lim_{P\to\infty}\left(-\frac{1}{\ln P}\int_{0}^{P} \frac{Y(\beta/x)}{x}{\rm d}x\right)\;.
\end{align}
Note here that we cannot employ the assumption
$Y(\beta/x)=\sum_{k=0}^\infty b_k/x^k$, since integrating this power
series diverges for each term. This motivates us to introduce
suitable variable transformations to (\ref{eq:Painleve_P}) and
(\ref{eq:fundamental_P}) as described below, which are aimed at
scaling $x$ to increase with $P$, whilst keeping the new variable of
integration finite. To this end, observe that
\begin{align}
-\frac{1}{\ln P}\int_{0}^{P} \frac{Y(\beta/x)}{x}{\rm d}x &= -\frac{1}{\ln P}\int_{\epsilon}^{P} \frac{Y(\beta/x)}{x}{\rm d}x -\frac{1}{\ln P}\int_{0}^{\epsilon} \frac{Y(\beta/x)}{x}{\rm d}x \label{transformation_0}\\&=-\int_{\ln \epsilon/\ln P}^{1}Y(\beta/x){\rm d}\left(\frac{\ln x}{\ln P}\right) -\frac{1}{\ln P}\int_{0}^{\epsilon} \frac{Y(\beta/x)}{x}{\rm d}x \label{transformation_1}\\&\sim - \int_{0}^{1} Y(\beta/P^s)\;{\rm d}s\;,\quad P\to \infty \label{transformation}
\end{align}
%{\bf ** Samuel -- explain why the introduction of small $\epsilon$
%is needed.  Why the 2nd line is asymptotic and not exact, and so
%on...  You need describe this step VERY clearly. ** (\emph{Revised})}
where the new variable $s:=\ln x/\ln P$ and $\epsilon$ is an arbitrarily small constant. Here $\epsilon$
is introduced to avoid the singularity at the lower limit when
changing ${\rm d}x/x$ on the right-hand side of
(\ref{transformation_0}) to ${\rm d}(\ln x)$ in
(\ref{transformation_1}). Nevertheless, for fixed $\epsilon$, the
second integral in (\ref{transformation_1}) vanishes as $P\to
\infty$ and the lower limit of the first integral becomes zero, thus
we arrive to the asymptotic formula (\ref{transformation}). With (\ref{transformation}) we have
\begin{align}
A=\lim_{P\to\infty}\left(- \int_{0}^{1} Y(\beta/P^s)\;{\rm d}s\right)
\end{align}
with $Y(\beta/P^s)$ satisfying
\begin{align}\label{eq:leading_Equation_P}
\left[\frac{Y'}{\ln P}+\left(\frac{\beta+\lambda+1}{\beta}\right)\frac{P^s}{\ln P}Y'+Y\right]^2=4\left(-\frac{Y'}{\ln P}-Y+\beta\right)\left(\frac{P^s}{\beta \ln P}Y'-\lambda\right)\frac{P^s}{\beta \ln P}Y'
\end{align}
where $Y':={\rm d}Y(\beta/P^s)/{\rm d}s$. By comparing the
coefficient of $P^s/\ln P$ (i.e., the terms corresponding to the
leading order in $P$) in (\ref{eq:leading_Equation_P}), we have the
equation involving $Y(\beta/P^s)$:
%\begin{align}
%\left[4Y-4\beta+\left(\beta+\lambda+1\right)^2\right]\left(\frac{P^s}{\ln P}Y'\right)^2=0
%\end{align}
\begin{align}
\left(\frac{P^s}{\beta\ln P}Y'\right)^2\left[4Y-4\beta+\left(\beta+\lambda+1\right)^2\right]=0,
\end{align}
which gives a constant solution
\begin{align}
Y(\beta/P^s)=-\frac{(\beta+\lambda+1)^2}{4}+\beta\;.
\end{align}
Thus we obtain the large-$n$--large-$P$ CGF via (\ref{transformation}) and (\ref{eq:order_CGF}):
\begin{align}
\mathcal{K}(n\lambda)\sim n^2\left[\frac{(\beta+\lambda+1)^2}{4}-\beta\right]\ln P\;, \qquad n\to \infty, \quad P\to \infty\;.
\end{align}
By invoking the saddle point equation (\ref{eq:saddlepoint}): $qn\ln
P=\frac{\rm d}{{\rm d}(n\lambda)}\mathcal{K}(n\lambda)$, we have
$\lambda^\star=2q-\beta-1$. This result reconfirms our statement
that the DMT cannot be described by (\ref{equality}), which
implicitly requires $n\lambda^\star>n-m$. With this saddle point,
the rate function $I(t)$ (with $t=qn\ln P,\;q<1$) is evaluated as
\begin{align}
I(qn\ln P)=qn^2\lambda^\star\ln P-\mathcal{K}(n\lambda^\star)=n^2\left[q^2-q(\beta+1)+\beta\right]\ln P\;.
\end{align}
Consequently, the CDF of the mutual information in the left tail for
large $P$ (and large $n$) becomes
\begin{align}
\ln {\rm Pr}(\mathcal{I}({\bf x;y})< qn\ln P)\sim -\left(qn-m\right)\left(qn-n\right)\ln P, \quad x<\mu\;.
\end{align}
This agrees precisely with the well known DMT result in
\cite{Zheng_Tse}.

%{\bf ** This type of discussion is effectively a summary, and fits
%better in the Conclusion. **}
%
%Equipped with this powerful tool, we are able to focus on various regimes of interest, both around the bulk and at the tails.

\section{Conclusion}
Capitalizing upon the exact Painlev\'{e} based representation
for the MGF of the MIMO mutual information in \cite{Chen_McKay}, we
have systematically computed new expansions for the high order
cumulants of the mutual information distribution which apply for
arbitrary SNRs and for asymmetric antenna arrays. In particular,
closed-form expressions were given for the leading order terms (in
$n$), as well as the first-order correction terms which capture
finite-antenna deviations. Based on these new expressions, we
established key novel insights into the behavior of the distribution
under different conditions; for example, explaining why the
$n$-asymptotic Gaussian approximation is more robust to increasing
SNRs for asymmetric systems compared with symmetric systems. This is
an interesting phenomenon which appears difficult to capture with
other methods. In addition, we called upon the Edgeworth expansion
technique along with the high order cumulant formulas to provide
closed-form refinements to the Gaussian approximation for the tail
region corresponding to $O(n^\epsilon)$ ($0 < \epsilon \leq 1$)
deviations from the mean. For deviations of $O(n)$, the so-called
``large deviations'' region, the Edgeworth expansion requires
summing over \emph{all} cumulants and becomes unwieldy; thus, in
this region we employed a saddle point approximation technique along
with asymptotic integration tools to derive very simple and concise
formulas for the CDF for the cases of low and high SNRs. Simulations
showed that our results captured the tail distribution very
accurately for outage probabilities of practical interest. Moreover,
whilst formally derived based on a large-antenna framework, they
were shown to be very accurate even when the antenna numbers are
small. To emphasize the utility of our framework even further, in
the end we recovered well known properties of the tail distribution
of the MIMO mutual information, including the DMT.

We conclude by noting that the key analytical tool underpinning the
analysis in this paper, the Painlev\'{e} based MGF representation in
Proposition \ref{th:Laguerre}, is extremely valuable since, as we
have shown, it facilitates a ``unified'' investigation of the mutual
information distribution under a wider range of scenarios than
appear possible with previous existing tools.  To the best of our
knowledge, together with our recent work \cite{Chen_McKay}, this is
the first time that such tools have been applied to problems in
information theory.  It turns out that these tools are also
applicable to other problems in information theory and wireless
communications, and such topics are currently being pursued.

%%%%%%%%%%%%%%%%%%Appendix%%%%%%%%%%%%%%%%%%%%%%%%%%%%%%%%%%

\section*{Appendix: Relation with Coulomb Fluid Method in \cite{Moustakas11}}

Here we draw the connections between our saddle point results and
those derived based on a Coulomb fluid large deviation approximation
in \cite{Moustakas11}.
%This will give
%us a unified picture of the existing results about the MIMO mutual
%information distribution.
We start by recasting the formulation of \cite{Moustakas11} in terms
of the MGF of the mutual information. To this end, consider the
exact MGF (\ref{MGF}) represented in multi-integral form:
\begin{align} \label{eq:MGF_IntApp}
\mathcal{M}(\lambda)&=\frac{\int_{\mathbb{R}^n_+}\prod_{k=1}^n(1+Py_k)^\lambda
y_k^\alpha e^{-ny_k}\prod_{i<j}(y_i-y_j)^2 {\rm
d}\mathbf{y}}{\int_{\mathbb{R}^n_+}\prod_{k=1}^n y_k^\alpha
e^{-ny_k}\prod_{i<j}(y_i-y_j)^2 {\rm
d}\mathbf{y}}:=\frac{Z(\lambda)}{Z(0)}
\end{align}
where ${\bf y}=(y_1, \ldots, y_n)$ denotes the eigenvalues of ${\bf
HH^\dag}$ and
\begin{align}
Z(\lambda)=\int_{\mathbb{R}^n_+}\exp
\left\{\lambda\sum_{k=1}^n(1+Py_k)+\sum_{k=1}^n\left[(m-n)\ln
y_k-ny_k\right]+2\sum_{i<j}\ln\lvert y_i-y_j \rvert \right\}{\rm
d}\mathbf{y}\;.
\end{align}
Based on the Coulomb fluid interpretation (see
\cite{Dyson,Chen_Manning,Chen_Lawrence,Chen_Manning2,Chen_Ismail}
for details), as $n$ grows large, the CGF associated with
(\ref{eq:MGF_IntApp}) is anticipated to be well approximated with
the following:
\begin{align}\label{Chen_Coulomb}
\mathcal{K}(\lambda):=\ln \mathcal{M}(\lambda)\approx
-\min_{\sigma(y)}F[\sigma(y),\lambda]+\min_{\sigma(y)}F[\sigma(y),0]
\end{align}
where
%the eigenvalues are analogous to the positions of charged particles
%aligned in line. The essence of this heuristic analogy is that, when
%$n\to \infty$, the charged particles are regarded as fluid with a
%density, and $\ln Z(\lambda)\to -F(\lambda)$ (see \cite{Dyson}) with
\begin{align}
F[\sigma,\lambda]=\int_a^b \sigma(y)&\left\{ n^2[y-(\beta-1)\ln
y]\right\}{\rm d}y-\lambda n \int_a^b \ln(1+Py)\sigma(y){\rm
d}y\nonumber\\&\phantom{n^2[y-(\beta-1)}-n^2\int_a^b
\int_a^b\ln\lvert y-z \rvert \sigma(y)\sigma(z){\rm d}y{\rm d}z\;
\end{align}
is the so-called ``free energy'', whilst $\sigma(y)$ is a PDF with
support $[a, b]$.
%associated with solving the optimization
%problem
%\begin{align}\label{prob1}
%&\min_{\sigma(y)} F(\lambda)\nonumber\\
%& s.t. \int_a^b\sigma(y){\rm d}y=1\;.
%\end{align}
Meanwhile, we have the saddle point approximation of the density function:
\begin{align}\label{prob2}
p_{\mathcal{I}({\bf x;y})}(t) \sim \exp \{-\max_\lambda [t\lambda-\mathcal{K}(\lambda)]\}\;,
\end{align}
which, combined with (\ref{Chen_Coulomb}), is equivalent to
\begin{align}\label{McKayChen_LD}
p_{\mathcal{I}({\bf x;y})}(t) \sim \exp
\left\{-\max_\lambda \min_{\sigma (y)}\left\{F[\sigma,\lambda]+t\lambda\right\}+\min_{\sigma (y)}F[\sigma,0]\right\}.
\end{align}
%associated with solving the following optimization problem:
%\begin{align}\label{prob3}
%&\max_\lambda \{\min_{\sigma (y)}[F(\lambda)+x\lambda]-\min_{\sigma (y)}F(0)\}\nonumber\\&s.t. \int_a^b \sigma(y){\rm d}y=1\;.
%\end{align}
%where $E(x):=\max_\lambda \min_{\sigma (y)} [F(\lambda)+x\lambda]$.
%Notice that $E(0)$ can be solved simply by setting $\lambda=0$ in $E(x)$.
%In addition, solving $\max_\lambda(\cdot)$ in (\ref{prob2}) is an %unconstrained convex optimization problem, and we can find the optimal %$\lambda$ by saddle point equation (\ref{eq:saddlepoint}), which leads to the.
Recalling that in the ``large deviations'' regime of interest,
$t\sim O(n), \lambda\sim O(n)$ (see the discussions in Section
\ref{sec:SPA}), we scale $t\to nt, \lambda\to n\lambda$, and
(\ref{McKayChen_LD}) becomes
\begin{align}\label{Moustakas_LD}
p_{\mathcal{I}({\bf x;y})}(nt)\sim \exp\left\{n^2\left\{ -\max_{\lambda} \min_{\sigma (y)}f[\sigma,\lambda,t]+\min_{\sigma(y)}f[\sigma,0,0]\right\}\right\}\;,
 \end{align}
 %with
% \begin{align}\label{prob4}
% \mathcal{E}(r)=\max_{\lambda} \min_{\sigma (y)}f[\sigma (y),\lambda,r], \quad s.t. \int_a^b \sigma(y) {\rm d}y=1
% \end{align}
with
\begin{align}\label{Lagrangian}
 f[\sigma,\lambda, t]=&\frac{F(n\lambda)}{n^2}+t\lambda\nonumber\\=&\int_a^b \sigma(y)
 \left( [y-(\beta-1)\ln y]\right) {\rm d}y+\lambda \left[t-\int_a^b \ln(1+Py)\sigma(y){\rm d}y\right]\nonumber\\
& \qquad\qquad\qquad\qquad-\int_a^b \int_a^b\ln\lvert y-z \rvert
  %& \phantom{\qquad\qquad\qquad n^2[y-(\beta-1)]}-\int_a^b \int_a^b\ln\lvert y-z \rvert
 \sigma(y)\sigma(z){\rm d}y{\rm d}z\;.
 \end{align}
We find that (\ref{Moustakas_LD}) coincides exactly with \cite[Eq.
(23)]{Moustakas11}, where the optimization problems
$\max_\lambda\min_{\sigma(y)}(\cdot)$ are solved \emph{jointly},
eventually resulting in three coupled non-linear equations in
general\footnote{For the special case, $n_t=n_r$, there is one non-linear
equation for the right-hand tail and two non-linear equations for the left-hand tail.}.

In contrast, our method first employs the Painlev\'{e} equation to
obtain the $n$-asymptotic CGF as
$$\mathcal{K}(n\lambda)\sim n^2
\int_\infty^{\beta/P}\frac{G(x)}{x}{\rm d}x, \qquad n\to\infty$$
with $G(x)$ exactly characterized by (\ref{Fundamental}). This
representation corresponds to evaluating (\ref{Chen_Coulomb})
explicitly, without any intuitive Coulomb fluid analogy. Then, armed with
this CGF result, we separately draw upon the saddle point equation
to capture the tail distribution, which corresponds to solving
(\ref{prob2}). Thus, in essence, our saddle point approximation is
solving the equivalent problem to that considered in
\cite{Moustakas11} by \emph{explicitly} finding the asymptotic CGF
(whilst in \cite{Moustakas11} it is implicit). Quite remarkably,
under the high and low SNR regimes considered, it also leads to
simplified results.
%We have shown that these two methods are equivalent and connections can be %drawn in the following aspects:
%\begin{itemize}
%\item $\max_\lambda \{R\lambda-\ln \mathcal{M}(\lambda)\}$ in our method corresponds to $n^2[\mathcal{E}(r)-\mathcal{E}(0)]$ in Moustakas \cite{Moustakas11};
%\item $R=\frac{\rm d}{{\rm d}\lambda}\ln \mathcal{M}(\lambda)$ in our method (i.e.  the equation solving $\max_\lambda \{R\lambda-\ln \mathcal{M}{(\lambda)}\}$ given any $R$) corresponds to Moustakas \cite[Eq.(49)]{Moustakas11};
%\end{itemize}

\bibliographystyle{IEEEtran}
\bibliography{IEEEabrv,references}
\end{document}